# 电 子 科 技 大 学

UNIVERSITY OF ELECTRONIC SCIENCE AND TECHNOLOGY OF CHINA

# 博士学位论文

**DOCTORAL DISSERTATION**

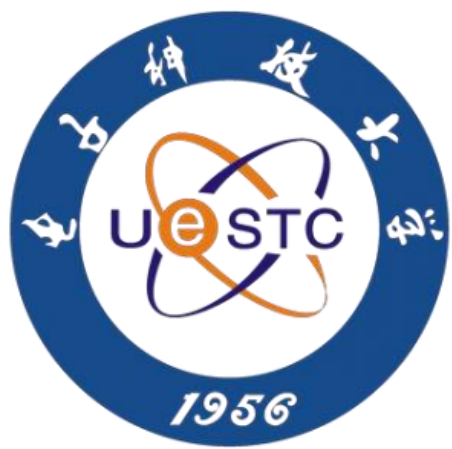

| | |
|---|---|
| 论文题目 | Evolving Text Data Stream Mining |
| 学科专业 | 计算机科学与技术 |
| 学　　号 | 201814080017 |
| 作者姓名 | Jay Kumar |
| 指导教师 | 邵俊明　教授 |

分类号_____________________ 密级_____________________

UDC[注1]_____________________

# 学 位 论 文

## Evolving Text Data Stream Mining

（题名和副题名）

### Jay Kumar

（作者姓名）

| 指导教师 | 邵俊明　　教授 |
| --- | --- |
|  | 电子科技大学　成都 |

（姓名、职称、单位名称）

| 申请学位级别 | 博士 | 学科专业 | 计算机科学与技术 |
| --- | --- | --- | --- |
| 提交论文日期 | __________ | 论文答辩日期 | __________ |
| 学位授予单位和日期 | 电子科技大学　　年　月 | | |
| 答辩委员会主席 | __________ | | |
| 评阅人 | __________ | | |

注 1：注明《国际十进分类法 UDC》的类号。

# Evolving Text Data Stream Mining

**A Doctor Dissertation Submitted to**

**University of Electronic Science and Technology of China**

|  |  |
|---:|:---|
| **Discipline:** | Computer Science and Technology |
| **Author:** | Jay Kumar |
| **Supervisor:** | Prof. Junming Shao |
| **School:** | Computer Science and Engineering |

# 独创性声明

本人声明所呈交的学位论文是本人在导师指导下进行的研究工作及取得的研究成果。据我所知，除了文中特别加以标注和致谢的地方外，论文中不包含其他人已经发表或撰写过的研究成果，也不包含为获得电子科技大学或其它教育机构的学位或证书而使用过的材料。与我一同工作的同志对本研究所做的任何贡献均已在论文中作了明确的说明并表示谢意。

作者签名：__________　　　　　　日期：　　年　　月　　日

# 论文使用授权

本学位论文作者完全了解电子科技大学有关保留、使用学位论文的规定，有权保留并向国家有关部门或机构送交论文的复印件和磁盘，允许论文被查阅和借阅。本人授权电子科技大学可以将学位论文的全部或部分内容编入有关数据库进行检索，可以采用影印、缩印或扫描等复制手段保存、汇编学位论文。

（保密的学位论文在解密后应遵守此规定）

作者签名：__________　　　导师签名：____________
　　　　　　　　　　　　　　日期：　　年　　月　　日

# 摘 要

文本流是随时间生成的有序文本文档序列，如在线社交平台每天都会生成大量此类文本数据。由于文本流具有的独特属性，如无限长度、数据稀疏和演化等，设计文本流挖掘算法是一项极具挑战的任务。因此，在时间和内存的限制条件下，如何从这些文本数据流中学习有用的信息受到越来越多的关注。尽管过去十年提出了许多文本流挖掘算法，但仍存在一些潜在的问题。首先，文本数据的高维特性会严重降低现有学习模型的性能，亟须子空间学习模型或对文本数据进行特征降维。第二个问题是如何有效提取文档的语义文本表示并捕获随时间演变的主题。此外，文本流数据往往存在标签稀缺的问题，而现有方法往往针对都是全标签数据。针对这些问题，本文提出了新的学习模型，用于文本流的聚类和多标签学习。本论文的主要贡献如下。

(1) 传统短文本的 Unigram 表示方法由于没有考虑上下文语境，会导致在计算文档之间的相似度进行聚类时出现词语歧义问题。针对这个问题，本文提出了一种在线语义增强的短文本流聚类狄利克雷（Dirichlet）模型，该模型将单词共现的语义信息用一个概率图模型进行表示，并以在线方式自动对每个到达的短文本进行聚类。为了处理簇演化，使用指数衰减函数自动删除过时的簇。提出的模型在五个数据集上进行了评估，实验表明新提出的方法优于目前最先进的各种聚类模型，达到了平均94.1%的聚类同质性。

(2) 现有方法在跟踪主题演变时往往没有考虑主题关键词分布的变化，极大降低了输出的可解释性并增加了簇的稀疏性。为了解决这些问题，提出了一种用于演化短文本流聚类的在线语义增强概率图模型。该模型动态维护演化中的活跃主题，并使用三角时间函数检测主题相关关键词的变化。与现有方法相比，所提出的模型不仅无需人为确定时间块的大小，还可有效处理大规模数据流。为了降低簇的稀疏性，引入了基于概率的簇合并过程。通过对两个真实数据集和两个人工数据集的实验分析表明，所提出的方法性能不仅优于现有的最先进模型，而且对输入参数具有很强的鲁棒性。

(3) 尽管短文本的语义表示能有效消除词语歧义，但其高维性会成为计算文档簇相似度的瓶颈。为了减少语义表示特征维度，提出了一种基于窗口共现性的新狄利克雷模型。为提高聚类性能，提出的新模型嵌入了一种称为词语特异性的加权分数。与以前的批量推理流程不同，新方法嵌入了情节推理。通过对四

个数据集进行的实验表明，提出的新算法在归一化互信息的衡量指标上平均提高了1.5%。

(4) 针对多标签文本流数据会面临的标签稀缺、标签基数和标签分布随时间演变的问题，提出了一种在线文本流半监督分类算法。通过利用已有少量的标记数据，为每个标签动态维护多个由关键词组成子空间微簇群。然后通过使用 Dirichlet 模型计算概率获取 $k$ 最近的微簇群来进行标签预测。针对关键词组成的子空间存在渐变概念漂移的问题，采用三角时间函数计算词条到达时间与簇寿命的差值进行衰减处理。针对突变的概念漂移：（a）利用指数衰减函数删除过时的微簇群，（b）通过采用基于狄利克雷过程的中餐厅过程创建新的微集群。通过在九个数据集上对十一种基准算法实验表明，新提出的算法在分类性能、运行时间和内存消耗方面均有很好的优势。

**关键词：** 短文本流，概念漂移，多标签学习，半监督分类，概率图模型

# ABSTRACT

A text stream is an ordered sequence of text documents generated over time. A massive amount of such text data is generated by online social platforms every day. Designing an algorithm for such text streams to extract useful information is a challenging task due to unique properties of the stream such as infinite length, data sparsity, and evolution. Thereby, learning useful information from such streaming data under the constraint of limited time and memory has gained increasing attention. During the past decade, although many text stream mining algorithms have proposed, there still exists some potential issues. First, high-dimensional text data heavily degrades the learning performance until the model either works on subspace or reduces the global feature space. The second issue is to extract semantic text representation of documents and capture evolving topics over time. Moreover, the problem of label scarcity exists, whereas existing approaches work on the full availability of labeled data. To deal with these issues, in this thesis, new learning models are proposed for clustering and multi-label learning on text streams. The main contributions of this thesis are given as follows.

(1) Unigram representation of short text does not consider contextual term-space which leads towards term ambiguity issue while calculating the similarity between documents for clustering. To solve this problem, an online semantic-enhanced Dirichlet model for short text stream clustering is proposed, which integrates the word-occurrence semantic information into a new graphical model and clusters for each arriving short text automatically in an online way. To deal with the cluster evolution, it exploits exponential decay function for each cluster which automatically removes outdated clusters. The proposed model is evaluated over five datasets and shows the superiority by achieving an average 94.1% homogeneity.

(2) Most existing approaches track the topic evolution, however, the change in topic related term distribution is not considered which reduces the output interpretability and increases cluster sparsity. To solve these problems, an online semantic-enhanced graphical model for evolving short text stream clustering is proposed which dynamically maintain evolving active topics and detect change of topic related terms using triangular time function. Compared to existing approaches, the proposed model is not only free of determining optimal batch size, but also lends itself to handling large-scale data streams

Abstract

efficiently. To reduce the cluster sparsity, a probability based cluster merging process is introduced. Extensive empirical analysis on two real and two synthetic datasets show that the proposed approach not only outperforms existing stat-of-the-art models in terms of homogeneity and NMI (Normalized Mutual Information) but also less sensitive to input parameters.

(3) Semantic representation of short text is necessary to eliminate term ambiguity issue. However, high dimensional semantic space becomes a bottleneck while calculating the document-cluster similarity. To reduce the semantic term-space, a novel Dirichlet model is proposed where window-based co-occurrence is introduced. To increase the quality of clusters, a novel term weighting score, called the term specificity, is embedded in the proposed model. Unlike previous batch-wise inference procedures, we embed episodic inference methodology. A deep empirical analysis on the four datasets acquired an average of 1.5% improvement in NMI.

(4) Learning on multi-label text stream data poses label scarcity, label cardinality, and evolving label distribution over time. To learn and predict such text stream, an online semi-supervised classification algorithm is proposed. By leveraging a few labeled instances, the proposed approach dynamically maintains multiple subspaces of terms for each label with a set of evolving micro-clusters. For labels prediction, $k$ nearest micro-clusters are employed by using the calculated probability of the Dirichlet model. To handle the gradual concept drift in term space, a triangular time function is adopted to calculate the difference between term arriving time and cluster life-span. Whereas, the abrupt concept is dealt with by considering two procedures: (a) deleting outdated micro-cluster by exploiting an exponential decay function, and (b) creating new micro-clusters by adopting Chinese restaurant process based on the Dirichlet process. The conducted experimental study has demonstrated the superiority of the proposed method by comparing with eleven state-of-the-art algorithms on nine datasets in terms of classification performance, runtime, and memory consumption.

**Keywords:** Short-text stream, Concept drift, Multi-label learning, Semi-supervised Learning, Probabilistic Graphical Model

# Contents

# List of Figures

# List of Tables

# List of Acronyms

| | |
|---|---|
| LDA | Latent Dirichlet allocation |
| PGM | Probabilistic graphic model |
| DP | Dirichlet process |
| CRP | Chinese restaurant process |
| ML | Multi-label learning |
| NMI | Normalized mutual information |
| ICF | Inverse cluster frequency |
| VSM | Vector space model |
| SB | Sumblr |
| CF | Cluster feature |
| KNN | $k$-Nearest neighbors |
| AMR | Adaptive model rules |
| AHOT | Adaptive Hoeffding Option Tree |
| OZBAML | Oza Bag Adwin Multi-label |
| SCD | Single Classifier Drift |
| OCB | Online Coordinate Boosting |
| ISOUPT | Structured Output Prediction Tree |
| KNNPA | $k$-Nearest Neighbors adaptive with ADWIN+PAW |
| SAMkNN | Self-Adjusting Memory $k$-NN |
| MLSAMPkNN | ML Self-Adjusting Memory Punitive $k$-NN |
| MASS | ML Classification with specific features |
| TRANS | ML large-margin subspace learning |

# Chapter 1  Introduction

## 1.1 Significance of the Research

A *text stream* is a continuous series of ordered documents[1]. A vast range of scientific and social applications generate sequential text data with high velocity every day, such as Weblogs, Facebook, Twitter, and many more[2][3]. Such tremendous data growth is possible due to rapid advancements in digital hardware and software in recent years. Dealing with such real-time data has expanded the boundaries of business and changed the way of research[4][5]. For example, over 3 billion searching queries are processed by Google every day[6]. The objective of text mining task is to extract meaningful patterns or predictions from the given data[7]. Thereby, the driven patterns or predictions can assist to take better decisions to boost the organization. Mining such data not only grows business revenue but also provides enormous power to stabilize the multi-billion dollars industry.

The need for text mining for many applications has drastically increased during the past decade which directly attracted research industries to work on optimized solutions. With this digital era, stream of data is not confined to only big web industries but also small units of society, such as web browsing, smartphones, home appliances and traveling data can be utilized to expand commercial knowledge. To handle such text streams, recent machine learning approaches, such as HPStream[8], ConSTREAM[9], GSDPMM[10] and ML-FSL[11], have proven significant for better decision making.

## 1.2 Text Stream Mining

During the past two decades, designing algorithms for mining from text stream is among most active research problems which motivate to extract useful information from massive data. A general process of data stream mining is shown in Figure 1–1. A series of different models in literature for this task have proven vital to solve many real-world problems such as spam detection [12][13], event detection[14], disaster management[15][16], scientific document analysis[17], information retrieval[18], and recommendation systems[19]. Designing the mining algorithm for any data may serve different objectives like classification, clustering, time-series analysis and pattern mining.

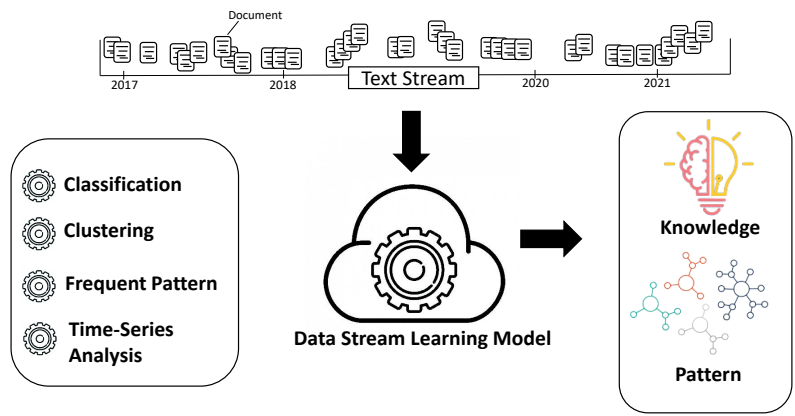

Figure 1–1 The overview of data stream learning procedure.

Mining and analyzing such text streams brings a variety of challenges to research community due to the following unique properties[20]. (a) Volume: The massive amount of data is potentially infinite in length, (b) Velocity: The speed of arriving data is high and processing of instances must be lower or equal to the data velocity, (c) Evolution: The intrinsic data distribution or extrinsic characteristics of data may change over time. In the light of the aforementioned unique characteristics of data streams, designing algorithms is a non-trivial task and brings the following constraints[21].

(1) **Finite Memory.** In the stationary data mining task, generally, the size of data to be processed is confined to the memory limit. In some cases, if the size of data exceeds the memory limits, various techniques have been developed to split the data into chunks to iterative process. However, unlike stationary data mining, mining stream poses limited memory challenge as the data is potential of infinite length. Therefore, only essential data needs to be stored in the limited memory.

(2) **Limited Processing Time.** The static data mining can console prediction while processing all data multiple times. However, the continuous arrival of data on high speed in stream demands the real-time prediction and allows processing the instance only once. The constraint for an algorithm to be ready for prediction at any time make this task more difficult to solve.

(3) **Change Adaptivity.** The learning algorithms designed for static data assume a stationary distribution (e.g., i.i.d) and constant nature of data. However, this assumption do not hold for streaming data where distribution of data and attributes of instances may change at any time. This change is generally divided into two categories, concept drift and concept evolution. Usually, change in any kind of distribution in data is referred as concept drift, such as label distribution, feature distribution over label, and label boundary. The concept evolution refers to emergence and disappearance of either labels or features.

Although dealing with these problems rely upon the nature of data and real-time application, thus gradual improvements are reported in the series of previous models.

## 1.3 Recent Research Progress in Text Stream Mining

Knowledge discovery from text data stream and stream analytics bring variety of methods, tools and technologies. These approaches include data visualization[22], business intelligence[23] and statistical analysis[24]. Machine learning is the core of each methodology and technology for text data analysis. Substantial improvements have become possible in data mining due to machine learning in the past two decades. However, machine learning tools have become confined to static data and are unable to deal with dynamic nature of data. Therefore, text stream mining has become a hot research problem for data mining community for last several years. In the projection of demanded applications, text stream mining covers many task which include clustering and classification. The objective and recent progress related to both tasks are discussed in following sections.

### 1.3.1 Clustering of Text Documents

The task of clustering is to construct the groups of similar objects from a corpus. The clustering of text documents in static corpus has been focused for more than two decades due to its versatile applicability. However, the same task of grouping became more challenging where a system needs to update itself to consider newly generated documents everyday. During the last decade, a variety of algorithms and frameworks have been proposed to deal with different subset of challenges[25]. Particularly, clustering of text stream is dealt with two major methodologies: similarity based and model based.

**Similarity based approaches.** It represents each instance with unified vector length[26]. Initially, a general framework to deal with streaming data is proposed by Aggarwal et al.[27]. The framework is composed of online and offline clustering modules. The online component maintains a fixed number of clusters while creating new clusters and deleting old clusters. Both creation and deletion follows different scheme of distribution. It is supposed to assign a much higher number of micro-clusters than the number of actual

Doctoral Dissertation of University of Electronic Science and Technology of Chinamicro-clusters. The deleted clusters from the model are stored in permanent storage. The offline component then analyze the stored component with merging and tracking of cluster. Later on, many popular clustering algorithms are mapped to this general framework. Afterwards, Zhong et al.[28] apply an adaptive spherical K-means clustering on streaming data. The proposed approach creates unit-length vectors for all the micro-cluster centroid. However, scalability and sparsity issues still exist[29]. Later on, OSKM[28] is proposed by combining adaptiveness and scalable algorithm on stream text data together. An additional decay mechanism is also designed to analyze the impact on the clustering process. Likewise, HPStream[8] clustering algorithm is designed by adopting projected clustering in stream data to tackle the sparsity issue. On the other side, FW-KMeans[30] is also embedded with a generic framework that can assign feature weights to reduce the feature space. Another challenge to exploit the feature weights is to utilize the module for text data. Previous algorithms could not handle the high-dimensional vector space, and the text data contains thousands of unique terms in the active stream. A factor is introduced to adjust the term distribution in the cluster. The popular algorithm ConSTREAM[9] is exploited to deal with text data in stream. A *cluster droplet* is introduced as cluster feature, which represents only those terms related to local space. By using the decay mechanism, outliers are also detected by calculating the cosine similarity between micro-clusters. DenStream[31] is introduced by designing DBSCAN[32] algorithm for streaming data. Unlike other algorithms, this approach does not require a static number of clusters. However, the minimum number of data points and radius are still required. In general, the limitation of similarity-based clustering approaches is that they require a manually fixed threshold for assigning a document to an old or new cluster. Moreover, sparsity is another issue, specifically dealing with short length text document, when adopting threshold-based approaches. For such cases, few works consider feature subspace for creating clusters.

**Model based approaches.** This methodology represent each group with different subset of feature space. The early classical attempt for text clustering is Latent Dirichlet Allocation (LDA)[33]. However, it cannot handle the temporal data for text streams. For this purpose, many LDA variants have been proposed to consider the text streams such as dynamic topic model (DTM)[34], dynamic mixture model (DMM)[35], temporal LDA (T-LDA)[36], streaming LDA (S-LDA)[37], and Dirichlet mixture model with feature partition (DPMFP)[38]. Later, Dirichlet multinomial mixture model-based dynamic clus-tering topic (DCT) model was designed to deal with short text streams by assigning each document with single topic[39]. Very soon, GSDMM[40] was proposed to extend DMM with collapsed gibbs sampling to infer the number of clusters. However, most of these models did not investigate the evolving topics (clusters) in text streams where the number of topics usually evolves over time.

To automatically detecting the number of clusters, [41] proposed a temporal Dirichlet process mixture model (TDMP). It divides the text stream into many chunks (batches), and assumes that the documents inside each batch are interchangeable. Later, GSDPMM[10] was proposed with collapsed gibbs sampling to infer the number of clusters in each batch. In contrast to LDA, GSDPMM not only converges faster but also dynamically assigns the number of clusters over time . However, both TDMP and GSDPMM models do not examine the evolving topics, and, these models process the text stream for multiple times. Thereafter, MStreamF[42] was thus proposed by incorporating a forgetting mechanism to cope with cluster evolution, and allows processing each batch only one time.

### 1.3.2 Classification of Text Documents

The task of predicting new observed documents from a pre-defined set of categories is known as *Classification*[43]. The learning for known categories makes classification task different from clustering. In stationary data situation the model is initially trained on samples and then it is utilized to predict the unseen instances. Whereas, in streaming environment, the task of learning and prediction go parallel over time. The learning task majorly depends on category assignment for each instance and can be usually divided into two types of task: multi-class learning and multi-label learning.

**Multi-class classification.** The task of multi-class learning involve more than two pre-defined categories (classes) where every instance can have only one class[44]. Initially for streaming task, many static learning algorithms are transformed to deal with continuous sequence of data. For example, [45] proposed an streaming decision tree using the Hoeffding bound. It learns incrementally and builds a decision tree from continuously arriving samples. Similarly, [46] adapts traditional Support vector machines to propose CVM (Core Vector Machine). It reduces traditional SVM[47] complexity by using novel data point representation named Minimum Enclosing Ball. Furthermore, the naïve Bayes[48] classifier is also exploited to work in streaming data by [49]. The defined approach represents the whole data set into a hierarchical Gaussian-mixture tree where each node contains the statistical summary of data points. Many researchers also employed assemble approach to work with streaming data. For example, [50] proposed an online version of Bagging and boosting for streaming data where each sample can be repeated multiple times.

**Multi-label learning.** The task of multi-label learning involve more than two predefined categories (labels) where each instance may associated with more than one label[51]. For example, a news article may relate to *politics*, *election* and *government* themes simultaneously. Multi-label classification problem is solved through two methodologies, (i) problem transformation and (ii) algorithm adaptation. The former technique transforms the data to fit into mutli-class classifier, such as 'label powerset' where subsets of labels are treated as individual class. Whereas, the latter approach transforms the algorithm for multi-label data such as ML-KNN[52]. During the past decade, many multi-label learning models have been proposed to address the classification task from different perspective. Initially, classic multi-label algorithms were transformed for streams such as ML-HoeffdingTree[53], iSOUP-Tree[54], and ML-KNN[55], to mention a few. Unlike these techniques, some research works focused on reducing the feature space to fit into the model such as [56] exploits count-min sketch for dimension reduction in multi-label text streams. Also, later studies analyzed that these approaches still exploit global feature space whereas tracking label related features is important to improve classification performance. For this reason a variety of approaches such as ProSVM[57], FRS-LIFT[58], S-CLS[59], and ML-FSL[11] have been proposed to consider feature subspace.

## 1.4 Thesis Content and Contributions

In the recent years, many state-of-the-art text stream mining models have been developed. However, there are still some limitations in existing approaches that are not handled. This dissertation is mainly intended to solve short text stream clustering and multi-label classification task.

For short text representation in stream environment, most existing short text clustering algorithms exploit independent word representation in their cluster models which tend to cause ambiguity. Additionally, in social media stream, a topic is a group of similar texts that is active for a certain period of time. Whereas, each topic have its own life-span for which a model needs to maintain until its activation period. Apart from these discussed issues, most clustering approaches for text stream process batch by batch way.

Processing batch way assume that there exist no concept evolution inside batch. However, this assumption needs to approximate the optimized batch-size. Usually, a topic is about a geographically occurred event and due to its long time-span the distribution of words change over time as the event. Additionally, each topic generally revolve around some core topic-representative terms, thus tracking the evolution of such terms for each topic is difficult. Additionally, to deal with velocity of text stream and unknown point of evolution, previous algorithms suffer from cluster sparsity. In other words, clustering process of state-of-the-art models create multiple clusters for each topic. To solve this issue their approach require additional cost of inference procedure which iterates each batch multiple times. Although with approximating the optimal batch-size, inference procedure brings more complication to deal with the clustering task. Generally, representing the terms of text data require a high dimensional term-space. However, the term-space becomes squared when the model needs a semantic representation for n-gram vector space. A very high dimensional term-space not only requires a large amount of memory, but also participate to degrade the model performance. Apart from memory, processing high dimensional data requires more computation time. However, specifically for stream clustering task, a model should consume processing time at most equal to the stream velocity. Additionally, term weight score for better cluster quality is important to distinguish between noisy and useful terms.

In many real-world scenarios, classification of text document is required which deals with predefined number of document categories. However, it become more complex task when each document may be associated with many mutually non-exclusive categories. The efficient and robust analysis of multi-label documents in data streams has become a complex task due to different types of challenges related to multi-label learning and concept evolution (drift) such as high dimensionality, label scarcity and label cardinality. Additionally, representation of text for such machine learning algorithms is itself a non-trivial task while considering evolving stream environment with limited labeled instances.

**Contributions.** This research work proposes some efficient models to solve aforementioned problem to process text stream. This thesis presents four main contributions focusing on different perspective of text stream processing task. Figure 1–2 shows the relationship among different research content. The main contributions are summarized as follows.

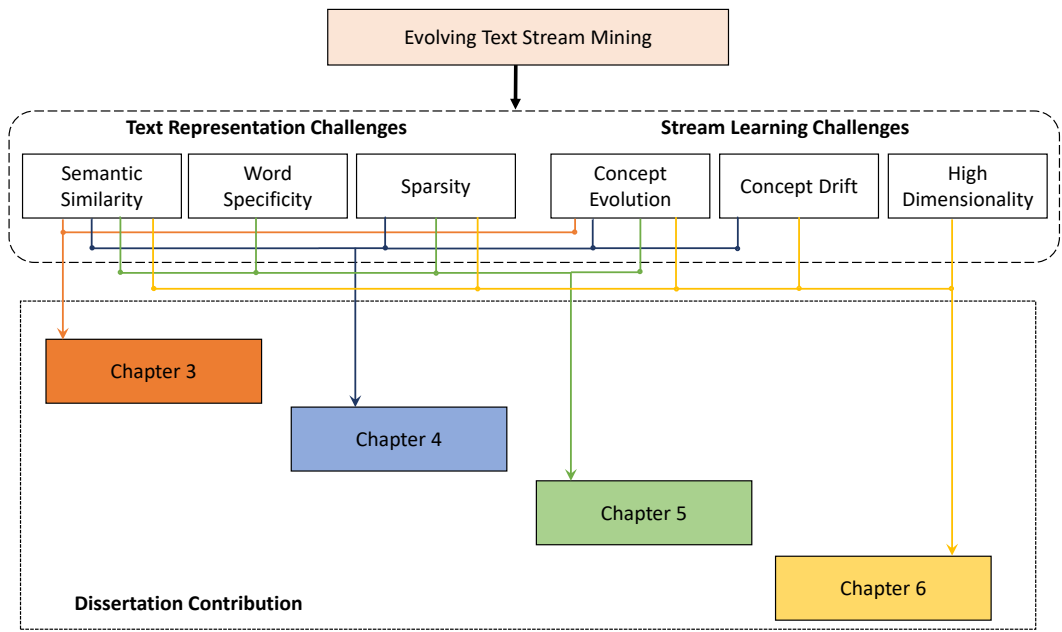

Figure 1–2 The relationship among different research content.

(1) An online semantic-enhanced Dirichlet model for short text stream clustering is proposed. Compared to existing approaches, It allows processing each arriving short text in an online way. The online model is not only free of determining the optimal batch size, but also lends itself to handle large-scale data streams efficiently. Additionally, it is able to handle "term ambiguity" problem effectively using term co-occurrence matrix. Whereas, equipped with Poly Urn Scheme, the number of clusters (topics) are determined automatically in the cluster model.

(2) An enhanced online graphical model for evolving short text stream clustering is proposed. Compared to the state-of-the-art algorithms, the proposed model can capture semantics by embedding evolving term co-occurrence matrix, find the topic related term-subspace in online way, and track cluster term evolution using feature-level triangular time decay. Additionally, it allows merging highly similar clusters to infer the topic related clusters, which yields the number of topics near to actual ones automatically.

(3) An non-parametric Dirichlet model for clustering short text is presented which reduces the semantic term-space by considering sentence window of each document. It also considers word specificity problem by observing ratio between term frequency and its co-occurred neighbors. Additionally, an episodic inference procedure is introduced to infer the number of clusters for each topic.

(4) A novel non-parametric Dirichlet model for online semi-supervised classification of multi-label text is proposed. An evolving micro-cluster representation is introduced which dynamically allows maintaining multiple subspaces of terms for each label. Additionally, it is capable of handling both the gradual and sudden concept drifts in term space by employing the triangular time function and Chinese restaurant process based on Dirichlet process. To capture the first-order label dependency, the presented model embeds label co-occurrence probability with cluster similarity for next label prediction, which supports high prediction performance.

## 1.5 Thesis Organization

This dissertation is organized as follows.

Chapter 1. first discusses the significance, motivation and overall contributions of this research with versatile applicability in different domains.

Chapter 2. provides a brief discussion about key terminologies useful to build the background knowledge.

Chapter 3. presents an online semantic-enhanced Dirichlet Model for short text stream clustering which also automatically detects new topics.

Chapter 4. proposes a novel approach for short text stream clustering which detects the gradual concept drift whilst preserving semantic similarity.

Chapter 5. proposes an algorithm which solves semantic-space sparsity problem for clustering short text stream. In addition, it introduces episodic inference procedure to reduce the cluster sparsity.

Chapter 6. focuses on multi-label text streams and presents an online semi-supervised classification algorithm.

Chapter 7. summarizes the overall significance and contributions of this research work. In addition, some brief directions regarding further investigation of related problems are discussed.

# Chapter 2  Fundamental Concepts and Methods

This chapter provides the key concepts and methods useful to understand this research work. Initially, basic definitions of key terms are described with formal representation. The research work of the dissertation is specifically targeted for text data, for this reason, methods of preprocessing of raw text document are discussed. Then the representation of preprocessed data and ways of processing in stream environment with task oriented perspective are briefly described with some examples. Lastly, the evaluation criteria and respective measures are discussed with formal representation.

## 2.1 Definitions

**DEFINITION 1** (**TEXT DOCUMENT**) A document is a set of ordered sequence of words, formally we denote as $d = \{w_1, w_2, \ldots, w_N\}$. The $N$ represents the total number of words in a document and as each document has its own length therefore $N$ varies from document to document.

**DEFINITION 2** (**TEXT DATA STREAM**) The text data stream ($S_t$) is sequence of continuous arrival of text documents over infinite time. Let us define data stream as $S_t = \{d_t\}_{t=1}^{\infty}$. The speed of arriving documents over a time unit is referred as velocity of stream.

**DEFINITION 3** (**CLUSTERING**) In machine learning, the task of gathering similar objects (text documents) is referred as *clustering* [60]. A group of similar documents is called as *Cluster*. The formal representation of unknown number of clusters in stream is denoted as $Z = \{z_t\}_{t=1}^{\infty}$, and each cluster $z_t$ contains documents represented as $z_t = \{d_1^{z_t}, d_2^{z_t}, \ldots, d_n^{z_t}\}$. For non-overlapping text clustering, each document is the member of only one cluster, so $z_i \cap z_j = \phi$, where $i \neq j$.

**DEFINITION 4** (**CLASSIFICATION**) Suppose, there is a set of $M$ discrete labels $\mathcal{L} = \{l_1, l_2, \ldots, l_M\}$ and each document $d_t$ is associated with one or more labels, denoted as $(d_t \to L_{dt})$. The objective of classification task is to build a model $f(D \to \mathcal{L})$ using $x$ number of labeled documents to classify future unlabeled documents[61].

**DEFINITION 5** (**CONCEPT DRIFT**) A concept drift refers to distributional change of any kind in arriving input data over time. This distributional change may cause due to change of input sources or environmental changes which triggered the distributional change. The decision boundaries of machine learning task totally rely over data distribution. In this context[62][63], concept drift is categorized into real and virtual concept drift. An example is shown in Figure 2–1. Mathematically, a concept drift from timestamp $t_0$ to $t_1$ can be expressed as follows.

$$\exists d : \mathcal{G}_{t0}(d,\mu,\sigma) \neq \mathcal{G}_{t1}(d,\mu,\sigma) \tag{2-1}$$

Here, $\mathcal{G}$ is the distribution with $\mu$ and $\sigma$ representing its mean and variance, respectively. For the sake of simplicity, we assume $\mathcal{G}$ as normal distribution. In the context of clustering task, the concept drift refers to change in cluster boundary or cluster-feature distribution. Whereas, in the context of classification, change in prior probability of a label $P(l)$ or in posterior probability $P(l|w)$ indicate the concept drift.

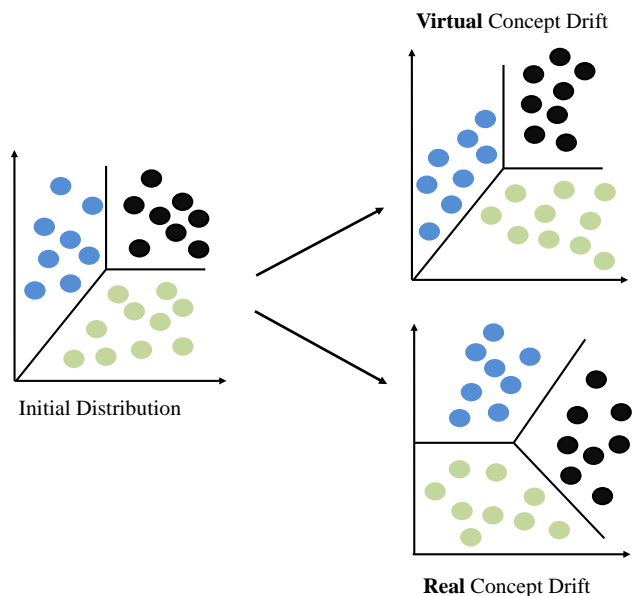

Figure 2–1 An example of two types of concept drifts in terms of decision boundary.

**DEFINITION 6** (**CONCEPT EVOLUTION**) The concept evolution usually concerns with change in feature space and change in label space. With the flow of streaming data, new features (new words in arriving document) may appear and old features may disappear, formally we denote this scenario as $W_{t0} \neq W_{t1}$ where $W_{t0}$ is feature space at timestamp $t_0$. Likewise, new labels may occur and old labels may disappear in streaming environment, formally it is denoted as $\mathcal{L}_{t0} \neq \mathcal{L}_{t1}$. Here, $\mathcal{L}_{t0}$ is the label space at timestamp $t_0$.

**DEFINITION 7** (**DIRICHLET PROCESS**) The Dirichlet Process (DP) is a non-parametric type of stochastic process, used to model the random generation of a distribu-tion represented as $G_0 \sim \text{DP}(\alpha, \mathcal{G})$. Here, $\mathcal{G}$ is the base distribution, and $G_0$ is the drawn sample from the distribution. Interestingly, the drawn sample itself is a distribution, thus also referred to as distribution over distribution. The drawing concentration is controlled by $\alpha$ parameter.

**DEFINITION 8** (**POLYA URN SCHEME**) It is the procedure to draw a sequence of samples $G_1, G_2, \ldots$ from a distribution [64]. The scheme is summarized as follows.

$$G_n | G_{1:n-1} \sim \frac{\alpha}{\alpha + n - 1} + \frac{\sum_{k=1}^{n-1} \delta(G_n - G_k)}{\alpha + n - 1} \qquad (2\text{--}2)$$

Here, $\delta(x) = 1$ if $x = 0$, and $\delta(x) = 0$ otherwise. Initially, we assume that there exists an empty urn and predefined distribution of colors. As we draw a color from a base distribution (i.e. $G_1 \sim \mathcal{G}$), we put a ball of drawn color into the empty urn. At this defined point the urn is no longer empty. In the next turn, we either draw a color that already exist in urn with probability of $\frac{n-1}{\alpha+n-1}$ or draw a new color with probability of $\frac{\alpha G_0}{\alpha+n-1}$. The same color may appear more than once due to the repetitive drawing procedure. This defines that we have $n$ number of draws and $K$ unique colors. The procedure of partitioning $n$ draws into $K$ clusters is defined by a well-known process called the Chinese Restaurant Process [65].

**DEFINITION 9** (**CHINESE RESTAURANT PROCESS**) Suppose, there exists a restaurant with infinite number of tables and each table surrounds infinite number of empty chairs for the customers. Initially, all tables are empty therefore first arriving customer has to sit on the first table. Later on, the $n^{th}$ customer either chooses to sit on any occupied table with probability of $\frac{n_k}{\alpha+n-1}$ or chooses an empty table with probability of $\frac{\alpha}{\alpha+n-1}$. Here, $n_k$ is the number of customers sitting on a specific table. A new customer tends to be attracted towards a highly crowded table which is called richer gets rich characteristics. The samples of CRP are drawn from the distribution $\mathcal{G}_{crp}$ and the stick-breaking process shows the property of the distribution where the procedure ranges towards infinity:

$$\mathcal{G}_{crp}(G) = \sum_{k=1}^{\infty} \theta_k \delta(G - G_k), \quad G_k \sim \mathcal{G} \qquad (2\text{--}3)$$

The mixture weights $\theta = \{\theta_k\}_{k=1}^{\infty}$ can be formalized by $\theta \sim GEM(\gamma)$ [66]. We exploit Equation (2–3) for the generative process of the Dirichlet process multinomial mixture model (DPMM) [10] as follows :

$$z_d | \theta \sim \text{Mult}(\theta) \quad d = 1, \ldots, \infty$$

$$G_k|\beta \sim \text{Dir}(\beta) \quad k = 1, \ldots, \infty$$

$$d\,|z_d, \{G_k\}_{k=1}^{\infty} \sim p\,(d|G_{z_d})$$

where, $z_d$ is the assigned documents to the cluster, which are multinomial distributed. The probability of document $d$ generated by topic $z$ is summarized as:

$$p\,(d|G_z) = \prod_{w \in d} \text{Mult}\,(w|G_z) \tag{2–4}$$

Here, the naive Bayes assumption is considered where words in a document are independently generated by the topic. In other words, the position of words in a document is not considered while calculating the probability.

By following the CRP to design a generative process for infinite number of tables (topics), we combine two probabilities (1) probability of topic popularity (defined by CRP) and (2) similarity between document and topic in terms of word distribution.

$$p(z_d|G_z) = p(G_z).p(d|G_z) \tag{2–5}$$

Here, $p(G_z)$ defines the probability of topic popularity (in CRP it is table popularity), which is $\frac{n_k}{\alpha+n-1}$ for choosing existing topic and $\frac{\alpha}{\alpha+n-1}$ for choosing new topic. The term $p(d|G_z)$ represents the similarity between topic (table) and document (customer), multinomial distribution with Dirichlet process is used for existing topic and for the new topic as well.

## 2.2 Text Preprocessing

A raw document needs to be transformed to feed into machine learning algorithms. It includes text cleansing and text representation[67].

### 2.2.1 Raw Text Cleansing

The purpose of this step is to extract the useful term(s) from the unstructured text document. The main steps for this task are discussed as follows.

**Tokenization**: is the process to split document text into small chunks[68]. In most cases, a chunk can be a word. However, in *n*-gram scheme, a group of *n* consecutive words make a single chunk. Additionally, for the language grammar learning task, a sentence is considered as a chunk.

**Noise Elimination**: is the process to remove general (e.g., comma and special characters) as well as language specific terms (e.g., stop-words).

**Lemmatization**: is language dependent process which transforms the stem word into root word. For example, '*give*' is stem word of '*given*' or '*flew*' is stem word of '*fly*'. This transformation helps feature representation for better learning.

### 2.2.2 Text Representation

The representation of words in the documents is used to transform into a matrix for machine learning algorithms. For this reason, a variety of word representation techniques have been introduced which include feature representation and their statistical properties.

For each document, the unique tokens of all the documents are represented into vector and this representation is referred as *vector space modeling*. The size each vector is equal to number of unique tokens (terms) in whole collection of documents. In other words, each dimension of the vector represent a unique chunk. The calculation of each dimension value for a document is based on different statistical measure, here some common scores are discussed.

**Binary**: stores 1 if this token exist in the document, 0 otherwise.

**Term Frequency**: stores the frequency count of the token in the document.

**TF. IDF**: is a popular representation and formally for a word *w* in document *d* it is defined as,

$$TF.IDF(w|d) = count(w|d) \times 1 + log\left(\frac{D}{count(d|w)}\right) \qquad (2\text{--}6)$$

Here, $count(w|d)$ is number of occurrence in a particular document, *D* is total number of documents in the collection, and $count(d|w)$ is number of documents containing word *w*. This score combines the local value as well as global value of a particular word[69].

## 2.3 Stream Learning

Like machine learning task in stationary environment, the learning in stream is also divided on the basis of label availability. A hierarchy is shown in Figure 2–2.

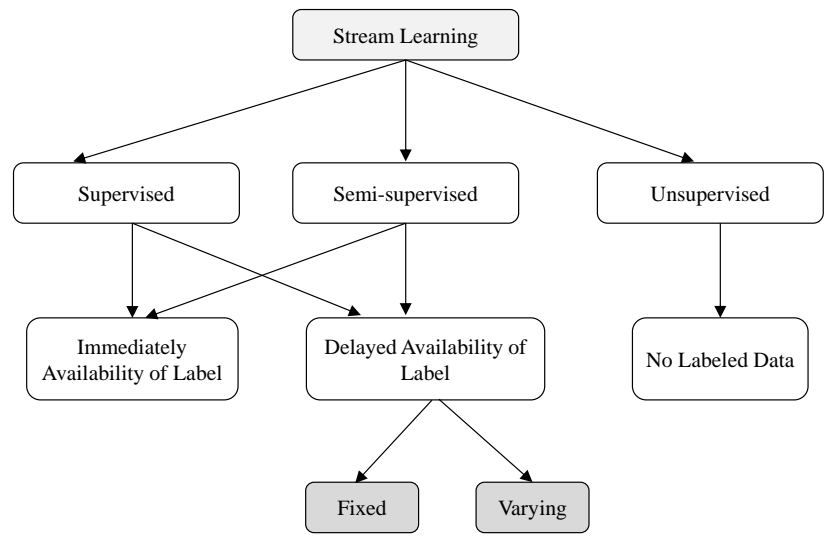

Figure 2–2 Data stream learning task based on label availability.

**Unsupervised Learning**: Learning in the total absence of labeled data is referred as *unsupervised learning*. Mostly, unsupervised learning task deal with clustering of similar instances over stream. Unlike static learning, a small portion of data is available for processing and the model needs to maintain the current concept or distribution. Whereas for the static data processing, all the data is available at once.

**Supervised Learning**: The objective of this learning is to build a model over a given labeled data and then the task is to predict the unlabeled arriving instances. In most of the cases, a small percentage of data in stream in manually labeled compared with unlabeled instances. The assumption of label availability in supervised task is two fold, either labels of the data are available after prediction (immediately or delayed) or will never be available.

**Semi-Supervised Learning**: It leverages both supervised and unsupervised task where a large volume of data is unlabeled and the task is to cluster the data (labeled and unlabeled) and then train the model on the data to predict the unseen arrival instances in the stream. In this type of learning, a small percentage of label instance will be available in the future to update the model.

### 2.3.1 Type of Classification Approaches

(1) **Single Classifier**: This methodology builds single classifier for learning from data where model is frequently updated over time. The simplicity of single classifier enables it to handle concept drift very efficiently. The major drawback of models based on single classifier is that their complex inner structure is very difficult to change thus unable to induce robust and accurate model. Usually, the model incrementally learn from recent data while old data is forgotten. However, the forgotten concepts may still be useful with current or upcoming concepts. Due to this reason, the performance and accuracy is compromised. Figure 2–3a shows the processing flow of single classifier.

(2) **Ensemble Learning**: This methodology combines more than one classifier work for single objective[70]. The power of multiple classifier lies behind the integrity so that weakness of one classifier is dealt by another one. Although complexity (in terms of computation) of ensemble approaches is much more higher than single one but it generally ensures the performance stability for streaming data[71]. Figure 2–3 shows difference between single and ensemble classification approaches.

In literature[72][73], the ensemble methodology either place classifier in stack or parallel manner. The parallel method usually split the prediction task into sub-tasks where each classifier is able to predict the subset of label set. Whereas, the classification in stack manner depends on voting system where each incoming instance is labeled from major classifiers vote.

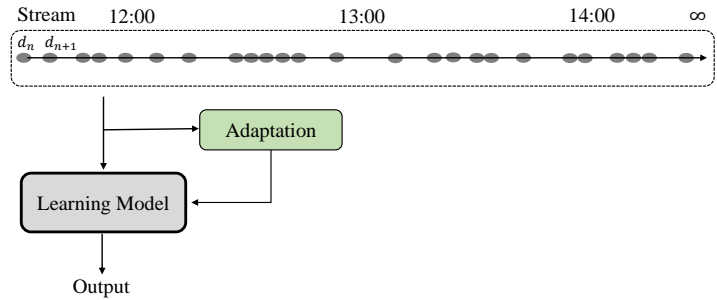

(a) Single Classifier

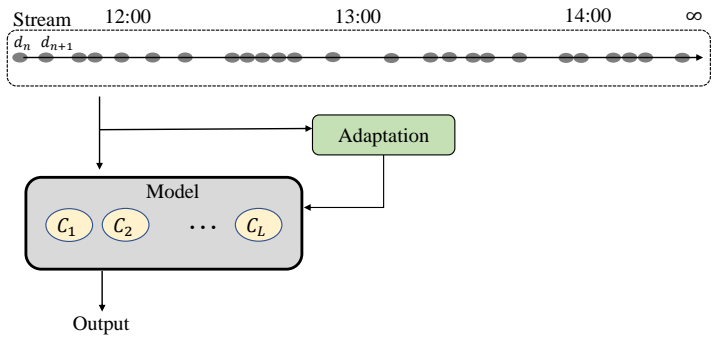

(b) Ensemble Model

Figure 2–3 Single versus ensemble learning.

## 2.3.2 Data Processing Strategy

In the stream environment, due to the infinite and continuous arrival of instances the data is generally processed in the following ways by algorithms. Each way has its own significance and applicability for different environments.

(1) **Time Slice**: For many applications, the recent data is more important. The time slice window split the data for processing is based on particular time unit[74]. For example, in tweeter, over the time unit of *one hour* (see Figure 2–4) the tweets of recent hour will be processed at once, thus in this way number of tweets to process will vary from hour to hour.

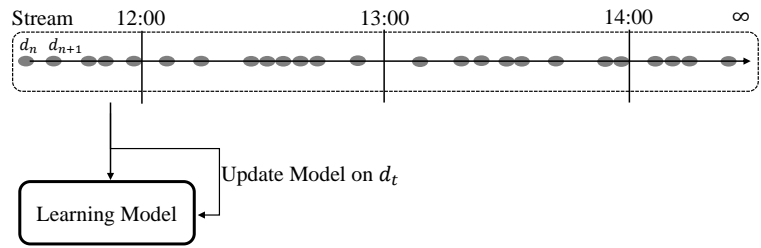

Figure 2–4 Time window processing of arriving data in stream.

(2) **Batch**: In the continuous arrival of instances, batch-wise processing split the stream into fixed sized batch. The instances will not be processed until the batch is full. An example is shown in Figure 2–5. The batch-wise processing does not allow revisiting the previous old batches.

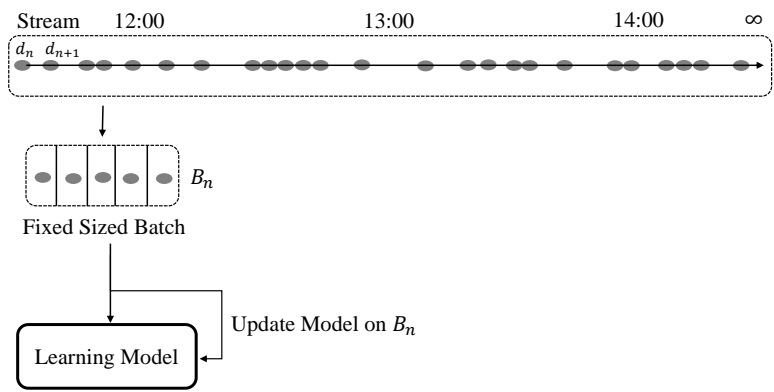

Figure 2–5 Batch-wise processing of arriving data in stream.

(3) **Online**: It only considers processing in instance by instance manner. In other words, it is the batch-wise processing where size of the batch is one.

### 2.3.3 Adaptivity Mechanism

The concept drift in streaming environment demands the model to adapt the changes and transform into new model. The model must detect the concept drift or concept evolution and retain its performance; otherwise it would lead towards devastating situation in many applications. The incremental learning enables streaming algorithms to update the learned model whenever the new instances are available. This allows models to decrease the chance of performance degradation over time. The adaptation of any model depends on detecting new concepts and forgetting outdated concepts. The metadata of the model enables it to identify new concepts whereas forgetting mechanism facilitates removing old concepts.

(1) **Data Management**: This technique describe which data is useful and vital to detect the concept drift or concept evolution for algorithm to update the model. The strategy is to use the most recent data for learning. The algorithms which follow online processing fashion continuously refine the model. Whereas, models which follow batch-wise processing or time-window only maintain current chunk of data or statistical properties of data. The significant change in the properties help the learning model to identify concept drift or concept evolution for model updation.

(2) **Forgetting Mechanism**: This technique assist the learning model to remove the outdated data or outdated distribution from the model so that it can predict according to new learned concepts. Generally, a function (e.g., exponential decay) is attached either to chunk of data or individual instance. The life of any chunk or instance depend to approximated parameter provided by domain experts. A sliding window is a common example where data is selected based on time or batch-size, where the model only keeps the recent $n$ chunks or time slices.

### 2.3.4 Data Maintenance on Data Streams

The constraint of limited memory in streaming task (i.e., clustering), where all flowing data in stream can be stored, confines the algorithm to design significantly optimized data structure[75]. Therefore, employing special data structure which can store a summary of incremental data is important.

(1) **Feature Vector**: For summarizing the large volume of data, BIRCH [76] introduced feature vector data structure. It stores the statistical summary of instances. It uses the cluster feature vector notion to maintain the statistical summaries of data clusters. Commonly, three statistical properties are stored to represent the cluster instances i.e., ($N$, $LS$, $SS$). Here, $N$ is total number of instances belong to a particular cluster, $LS$ is a vector containing linear sum of attribute values, and $SS$ is the squared sum of attribute values of all the instances inside the cluster. These three features are used to calculate the centroid and radius of cluster.

(2) **Grid Representation**: It represent the dense feature subspace for data abstraction and usually each cluster contains subset of the global space. An example is shown in Figure 2–6. Here, ten dimensions are considered as global space where subset of features are considered as subspace. A cluster may represent by global space or any subspace.

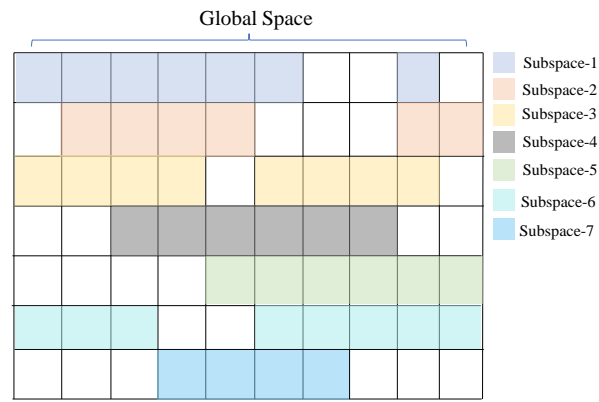

Figure 2–6 An example of grid representation.

## 2.4 Evaluation Metrics

To assess the performance of different task in machine learning, a variety of measures are available targeting different perspective of results. This dissertation provide clustering as well as classification algorithms for text stream, therefore, we discuss related evaluation measures with formal definition.

### 2.4.1 Clustering Performance Metrics

The assessment of clustering algorithms is commonly measured using following scores which help to direct the approaches for enhancement.

(1) **Cluster Purity**: is the most popular and commonly used evaluation measure in machine learning for clustering [77]. The purity of cluster $z \in Z$ is defined by,

$$pur(z) = \frac{1}{|z|} \max(n_{c,z}|c \in C) \quad (2\text{--}7)$$

Here, $|z|$ is the size of cluster, $C$ represent all classes, and $n_{c,z}$ reflects the number of instances in $z$ related to class $c$.

(2) **Homogeneity**: measures that each cluster should have only members of a single class [77], formally defined as:

$$Homo = 1 - \mathrm{H}(C|Z) \times H(C). \quad (2\text{--}8)$$

Here, $H(C|Z)$ is the conditional entropy and $H(C)$ represent the coverage of classes, defined as,

$$H(C|Z) = \sum_{z,c} \frac{n_{c,z}}{N} \cdot \log\left(\frac{n_{c,z}}{\sum_c n_{c,z}}\right)$$

$$H(C) = \sum_c \frac{\sum_z n_{c,z}}{N} \cdot \log\left(\frac{\sum_z n_{c,z}}{N}\right).$$

Here, the total number of documents is represented by $N$.

(3) **V-measure**[78]: calculates how successfully the criteria of completeness and homogeneity are satisfied, defined as,

$$V\text{-}M = \frac{2 \cdot (Homo \times Co)}{(Homo + Co)}. \quad (2\text{--}9)$$

Here, $Co = 1 - \mathrm{H}(Z|C) \times \mathrm{H}(Z)$ and each term is defined as,

$$H(Z|C) = \sum_{c,z} \frac{n_{c,z}}{N} \log\left(\frac{n_{c,z}}{\sum_z n_{c,z}}\right)$$

$$H(Z) = \sum_z \frac{\sum_c n_{c,z}}{N} \cdot \log\left(\frac{\sum_c n_{c,z}}{N}\right).$$

(4) **Normalized Mutual Information**: measure calculates overall clustering quality, defined as,

$$NMI = \frac{\sum_{c,z} n_{c,z} \log\left(\frac{N \cdot n_{c,z}}{n_c \cdot n_z}\right)}{\sqrt{\left(\sum_c n_c \log \frac{n_c}{N}\right)\left(\sum_z n_z \log \frac{n_z}{N}\right)}} \quad (2\text{--}10)$$

Here, $n_z$ is number of documents in $z$ and $n_c$ is number of documents of class $c$.

## 2.4.2 Classification Performance Metrics

There are different measures for binary, mutli-class and multi-label classification task. This dissertation propose multi-label classification approach (in Chapter **6**) therefore the definition of related scores are provided which include example-based Accuracy, Hamming loss, and micro-average Recall [79]. Let us assume $\hat{y}(x) \in \{0, 1\}^m$ denote predicted set of labels and $y(x) \in \{0, 1\}^m$ represent ground truth labels of instance $x$. Here, $m$ is the size of label space.

(1) **Hamming loss**: of instance $x$ is defined as the percentage of the wrong labels to the total number of labels, defined as,

$$\text{Hamming loss}(x) = \frac{1}{m}|y(x) \Delta \hat{y}(x)| \qquad (2\text{--}11)$$

Here, $\Delta$ denotes the symmetric difference of two sets.

(2) **Accuracy**: is the proportion of correctly predicted labels to the total number of predicted and ground truth labels, defined as,

$$\text{Sample-based Accuracy}(x) = \frac{1}{N}\sum_{i=1}^{N} \frac{|y_i(x) \cap \hat{y}_i(x)|}{|y_i(x) \cup \hat{y}_i(x)|} \qquad (2\text{--}12)$$

where $N$ is the number of samples (examples).

(3) **Micro-average recall**: evaluates the proportion of correctly predicted label out of predicted set of label and for each label $l$ it is defined as follows.

$$\text{Micro-average Recall} = \sum_{i=1}^{N} \frac{|\hat{y}_i \cap y_i|}{|\hat{y}_i|} \qquad (2\text{--}13)$$

## 2.4.3 Evaluation in Streaming Environment

Machine learning task for data stream settings must consider some main constraints. Therefore, apart from task specific evaluation measure, the following measures are also very important and should be considered.

(1) **Memory**: Stream oriented algorithms must abide by the memory limitation as the arriving instances are infinite, thus it demands stable memory consumption over the entire stream. Therefore, it is essential to observe the memory requirement of algorithms to process the data as well as to store the model.

(2) **Decision Time**: As the velocity of data stream confine the algorithm to consume very less time to process an instance. The processing speed should behave parallel at most with data stream velocity. Otherwise the delay in decision time will become the bottleneck of any effective model. Therefore, time to process single instance needs to considered while evaluation.

(3) **Adaptation Time**: The time consumed by the algorithm to update or refine the model structure is referred as adaptation time. The occurrence of concept drift or concept evolution enforce any model to update. Whereas, to produce more quality patterns the model needs to refine its structure. Therefore, in the ideal situation, the adaptation time of any streaming algorithm should be less than the speed of arriving batch or instances.

## 2.5 Summary

This chapter gives the background knowledge to understand data stream learning. Definition of data stream and related learning tasks including clustering, classification, concept evolution and concept drift detection are given. To prepare the text data of stream of learning task, a variety of text cleansing and text representations techniques are briefly discussed. The prepared data can be used for different learning objective which include unsupervised, supervised, and semi-supervised task. In this context, processing manner and adaptivity of learning model are also discussed. To assess the overall model performance, a variety of evaluation measures and relative formal definitions are briefly described.

# Chapter 3  An Online Semantic-enhanced Dirichlet Model for Short Text Stream Clustering

Mining text stream data to extracting useful patterns brings many challenges. Representation of text becomes non-trivial task when text instances arrive at high velocity and model needs to maintain the active term space while consuming limited time and space. This chapter discusses about short text representation problems for clustering in stream environment and proposes a new Dirichlet model to solve term ambiguity and tracking active topics issues.

## 3.1 Introduction

A massive amount of short text data is constantly generated with online social platforms such as microblogs, Twitter and Facebook. Clustering of such short text streams has thus gained increasing attention in recent years due to many real-world applications like event tracking, hot topic detection, and news recommendation[80]. However, due to the unique properties of short text streams such as infinite length, evolving patterns and sparse data representation, short text stream clustering is still a big challenge[27, 60].

During the past decade, many approaches have been proposed to address the text stream clustering problem from different points of view, and each method comes with specific advantages and drawbacks. Initially, traditional clustering algorithms for static data were enhanced and transformed for text streams[28]. Very soon, they are replaced by model-based algorithms such as LDA[81], DTM[34], TDPM[41], GSDMM[82], DPMFP[83], TM-LDA[36], NPMM[84] and MStream[42], to mention a few. However, for most established approaches, they often work in a batch way, and assume the instances within a batch are interchangeable. This assumption usually cannot hold for topic-evolving text data corpus. Determining an optimal batch size is also a non-trivial task for different text streams[85].

Additionally, unlike long text documents, short text clustering further suffers from the lack of supportive term occurrence to capture semantics[86]. For most existing short text clustering algorithms like Sumblr[87], DCT[39] and MStreamF[42], exploiting independent word representation in their cluster models tends to cause ambiguity. Let us show the following four tweets, for example:

  T1: *"A regular intake of an Apple can improve your health and muscle stamina."*

  T1: *"A glass of fresh apple juice is recommended for breakfast."*

  T2: *"New Apple Watch can monitor your health."*

  T2: *"Apple is launching iPhoneX this December."*

Tweets of these two topics share few common terms, i.e., '*health*' or '*apple*'. It creates an ambiguity if the model deals with only single term representation to calculate the similarity. However, the co-occurring terms representation (i.e., context) helps a model to identify the topic[①] correctly.

## 3.2 Related Work

  Latent Dirichlet Allocation (LDA)[81] is the most popular approach introduced to model the generation of topics from the static corpus. LDA is able to extract *k* number of topics by identifying subspace of terms from the ocean of term features through constructing a document-topic matrix. However, it cannot deal with the temporal data for text streams. That is the reason many variants of LDA are proposed for temporal data, such as dynamic topic model (DTM)[34], dynamic mixture model (DMM)[35], online clustering of text streams (OCTS)[88], temporal LDA (T-LDA)[36], and streaming LDA (S-LDA)[37]. These models assume that each document contains rich content, and thus they are not suitable to deal with the short text streams. Later, Dirichlet multinomial mixture model-based dynamic clustering topic model (DCT) is proposed to deal with short text streams by assigning each document with a single topic[39]. However, these models need a high value of *k* (number of clusters) to generate clusters through Dirichlet process. Later on, GSDMM[82] extends the DMM model to infer the number of clusters by integrating collapsed Gibbs sampling. However, most of these models do not explore the evolving topics (clusters) in text streams where the number of topics usually evolves over time[89].

  To automatically detect the number of clusters, Ahmed et al.[41] propose a temporal Dirichlet process mixture model (TDMP). It divides the text stream into many chunks (batches), and assumes that the documents inside each batch are interchangeable. Like other models, it follows the fine-grain clustering model to generate many small clusters. To infer the number of clusters in each batch, GSDPMM[10] is proposed with collapsed Gibbs sampling. In contrast to other models, GSDPMM not only converges faster than

---

① Topic and cluster will be interchangeably used in this dissertation

LDA but also dynamically assigns the number of clusters over time. However, both TDMP and GSDPMM models do not examine the evolving topics by maintaining the active topics. Thereafter, MStreamF[42] is thus proposed by incorporating a forgetting mechanism to cope with cluster evolution and allows processing each batch only one time. However, estimating an optimal batch size is a non-trivial task. Along with the disadvantage of fixed batch size, it also cannot deal with the term ambiguity problem. Recently, NPMM model[84] is introduced by using the word-embeddings to eliminate a cluster generating parameter of the model. Word-embedding is useful to remove term ambiguity, therefore it can represent the relationship between different terms. However, word-embeddings of a particular language model needs to be pre-trained, hence cannot capture evolving semantics for generating quality clusters. Therefore, capturing semantics by removing term ambiguity is a non-trivial task while dealing with online clustering of text streams. Our proposed approach can solve term ambiguity issue in online way while dealing with evolving clusters.

Additionally, Zhang et al. argue that a text document is often full of general (topic-independent) words and short of core (topic-specific) words. For this purpose, Liu et al.[90] propose term smoothing to reduce the dimensions in model-based approaches which helps to highlight core terms. NPMM[84] maintains a bit vector to represent terms of the topics. The bits in the vector change on arriving documents by exploiting word-embeddings. However, most previous model-based approaches do not apply feature reduction on cluster feature set, instead, each cluster is represented by terms of its documents. Due to the multinomial assumption of Dirichlet process, these approaches cannot track evolving topics for online text streams.

## 3.3 Proposed Approach

This section gives a brief discussion about the representation and formulation of the proposed algorithm.

To solve these aforementioned issues, we propose an online semantic-enhanced Dirichlet model for short text stream clustering. Compared to existing approaches, it has following advantages. (1) It allows processing each arriving short text in an online way. The online model is not only free of determining the optimal batch size, but also lends itself to handling large-scale data streams efficiently; (2) To the best of our knowledge, it is the first work to integrate semantic information for model-based online clustering, which is able to handle "term ambiguity" problem effectively and finally support high-quality clustering; (3) Equipped with Poly Urn Scheme, the number of clusters (topics) are determined automatically in our cluster model.

### 3.3.1 Model Representation

We build our model upon the DPMM[10], which is an extension of the DMM model to deal with evolving clusters. We call our model as OSDM (Online Semantic-enhanced Dirichlet Model), aiming at incorporating the semantic information and cluster evolution simultaneously for short text stream clustering in *an online way*. The graphical model of OSDM is given in Figure 3–1.

We show two major differences in our model to highlight the novelty. First, for word-topic distribution, we embed semantic information by capturing the ratio of word co-occurrence. Thereby, independent word generating process and word co-occurrence weight are well considered in topic generation. Secondly, our model works instance by instance fashion to cluster the documents, instead of batch by batch. For comparison, Figure 3–1a further show the MStreamF[42] model. At initial stage before clustering documents of a batch, MStreamF update vocabulary set (active terms) from all the documents in a batch, then it starts the clustering each document of the batch. However, OSDM does not consider fixed number of documents to create vocabulary set, instead it incrementally updates with each arriving document.

### 3.3.2 Model Formulation

Defining the relationship between documents and clusters is the most crucial task while dealing with the text stream clustering problem. The threshold-based methodology[91] adapts similarity measures to define the homogeneity threshold between a cluster and a document. If the dissimilarity between the exiting clusters and a new arriving document is above the threshold, then a new cluster is created. However, due to the dynamic nature of the stream, it is very hard to define the similarity threshold manually.

In contrast, we assume that documents are generated by DPMM (see Section 2.1). Most recent algorithm MStreamF improved DPMM to cluster short text documents in the stream. As a further study, we integrate the semantic component in DPMM model. Additionally, we integrate term importance on the basis of cluster frequency. The derived equation for calculating the probability of a document $d$ choosing existing cluster $z$ is given in Equation (3–1).

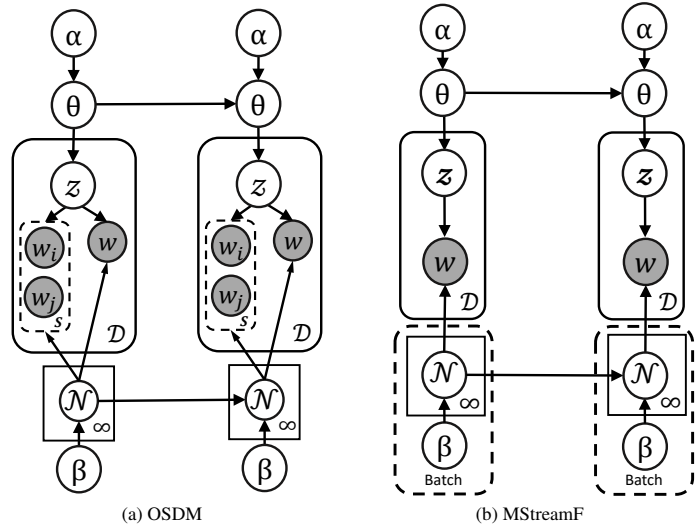

Figure 3–1 The graphical representation of OSDM and MStream.

$$p\left(z_d = z | \vec{z}, \vec{d}, \alpha, \beta\right) = \left(\frac{m_z}{D-1+\alpha D}\right) \cdot \left(\frac{\prod_{w \in d} \prod_{j=1}^{N_d^w} (n_z^w \cdot lCF_w) + \beta + j - 1}{\prod_{i=1}^{N_d} n_z + V\beta + i - 1}\right) \cdot \left(1 + \sum_{w_i \in d \wedge w_j \in d} cw_{ij}\right) \quad (3-1)$$

The first term of this Equation $\left(\frac{m_z}{D-1+\alpha D}\right)$ represents completeness of the cluster. Here, $m_z$ is the number of documents contained by the cluster $z$ and $D$ is the number of current documents in active clusters[①]. Whereas, $\alpha$ is the concentration parameter of the model. The middle term of the equation based on multinomial distribution (see Equation (2–4)) with psuedo weight of words $\beta$ defines the homogeneity between a cluster and a document. $N_d$ and $N_d^w$ represents total number of words and term frequency of word $w$ in document $d$, respectively. The symbol $n_z^w$ is the term frequency of the word $w$ in the cluster $z$. The current vocabulary size of the model is represented by $V$. $n_z$ is the number of words in the cluster $z$. $ICF_w$ calculates the term importance over the active clusters in the model, which is defined as follows.

$$ICF(w \in d) = \log\left(\frac{|Z|}{|w \in Z|}\right) \quad (3-2)$$

Here, $|Z|$ represents the number of active clusters in the model. The denominator part of Equation ((3–2)) is the number of those cluster which contains the word $w$. The term $\left(1 + \sum_{w_i \in d \wedge w_j \in d} cw_{ij}\right)$ defines the semantic weight of term co-occurrence between the

---

① Active clusters refer to those clusters which are not yet deleted from the model.

cluster and a document. Formally, we define a value of an entry $cw_{ij}$ in the co-occurrence matrix as follows.

$$cw_{ij} = \frac{\sum\limits_{d' \subseteq z} n_{d'}^{w_i}}{\sum\limits_{d' \subseteq z} n_{d'}^{w_i} + \sum\limits_{d' \subseteq z} n_{d'}^{w_j}} \quad (w_i, w_j) \in d' \tag{3-3}$$

Here, $n_z^{d'}$ is frequency count of word $w_i$ in document $d'$. The ratio between $w_i$ and $w_j$ must satisfy the property $cw_{ij} + cw_{ji} = 1$. We calculate the term co-occurrence weight of those terms which are common in the cluster $z$ and document $d$. Term co-occurrence matrix is constructed where two terms are co-occurred in a single document. Therefore, if the size of cluster feature set (discussed in Section 3.3.3) is $|V_z|$, then it is not necessary that the co-occurrence matrix would be $|V_z| \times |V_z|$. Let us take a example, suppose, we have a document $d = \{w_1 : 3, w_2 : 2, w_3 : 3\}$, we calculate each entry of matrix as,

$$cw_{w_1,w_2} = \tfrac{3}{3+2} = 0.6$$

$$cw_{w_2,w_1} = \tfrac{2}{2+3} = 0.4$$

$$cw_{w_1,w_3} = \tfrac{3}{3+3} = 0.5$$

$$cw_{w_3,w_1} = \tfrac{3}{3+3} = 0.5$$

$$cw_{w_2,w_3} = \tfrac{2}{2+3} = 0.4$$

$$cw_{w_2,w_3} = \tfrac{3}{3+2} = 0.6$$

So far, we have defined the probability of a document choosing existing cluster, then we have to define the probability for a document to creating a new cluster. By following the DPMM for infinite number of clusters, which transform $\theta \sim GEM(\gamma)$ into $\theta \sim GEM(\alpha D)$, because the hyper-parameter for the mixture model should be dynamically change over time (see Section 2.1). Therefore, the probability of creating a new cluster is as follows.

$$p\left(z_d = z | \vec{z}_{\neg d}, \vec{d}, \alpha, \beta\right) = \left(\frac{\alpha D}{D - 1 + \alpha D}\right) \cdot \left(\frac{\prod_{w \in d} \prod_{j=1}^{N_d^w} \beta + j - 1}{\prod_{i=1}^{N_d} V\beta + i - 1}\right) \tag{3-4}$$

Here, the pseudo number of clusters related documents in the model is represented as $\alpha D$, and $\beta$ is the pseudo term frequency of each word (exist in document) of the new cluster.

### 3.3.3 The cluster feature set

The similarity-based text clustering approaches usually follow vector space model (VSM) to represent the cluster feature space[92]. However, a topic needs to be represented as the subspace of global feature space. Here, we use a micro-cluster feature set to represent each cluster. Namely, a cluster is represented as the summary statistics of a set of words of related documents. In our model, a cluster feature (CF) set is defined as a 6-tuple $\{m_z, n_z^w, cw_z, len_z, l_z, u_z\}$, where $m_z$ is the number of documents in the cluster $z$, $n_z^w$ is the number of frequency of the word $w$ in the cluster, $cw_z$ is the word to word co-occurrence matrix, $len_z$ is the number of words in the cluster $z$ which is sum of all frequencies of words, $l_z$ is the cluster weight, and $u_z$ is the last updated time stamp.

The desirable addition property of cluster feature allows updating each micro-cluster in an online way.

**Definition 3.3.1** A document $d$ can be added to a cluster $z$ by using the *addition property*.

$$m_z = m_z + 1$$
$$n_z^w = n_z^w + N_d^w \quad \forall w \in d$$
$$cw_z = cw_z \cup cw_d$$
$$len_z = len_z + len_d$$

Here, $cw_d$ is word to word co-occurrence of the document, and $len_d$ represents the number of total words in the document. The complexity of updating a cluster by adding a document is $O(\mathcal{L})$, where $\mathcal{L}$ is the average length of the document. This property is useful to update evolving micro-clusters in the text stream clustering procedure.

**Algorithm 1:** OSDM

1   $S_t : \{d_t\}_{t=1}^{\infty}$, $\alpha$ : concentration parameter, $\beta$ : pseudo weight of term in cluster, $\lambda$ : decay factor  Cluster assignments $z_d$
2   $K = \phi$
3   **while** $d_t$ in $S_t$ **do**
4      $t = t+1$
5      $K = removeOldZ_i(K)$
6      $K = reduceClusterWeight(\lambda, K)$
7      **foreach** $z_i \in K$ **do**
8          $P_{Z_i} = prob(z_i, d_t)$ using Eq. (3–1)
9      **end**
10     $i = \arg\max_i (P_{Z_i})$
11     $P_{Z_n} =$ calculate the probability of new cluster using Equation (3–4)
12     **if** $P_{Z_i} < P_{Z_n}$ **then**
13         $m_{z_n} = 1$
14         $n_{z_n}^w = N_{d_t}^w$
15         $cw_{z_n} = cw_{d_t}$
16         $len_{z_n} = len_{d_t}$
17         $l_{z_n} = 1, u_{z_n} = t$
18         $K = K \cup z_n$
19     **else**
20         $m_{z_i} = m_{z_i} + 1$
21         $n_{z_i}^w = n_{z_i}^w + N_{d_t}^w$
22         $cw_{z_i} = cw_{z_i} \cup cw_{d_t}$
23         $len_{z_i} = len_{z_i} + len_{d_t}$
24         $l_{z_i} = 1, u_{z_i} = t$
25     **end**
26   **end**

### 3.3.4 OSDM Algorithm

We propose a semantic-enhanced non-parametric Dirichlet model to cluster the short text streams in an online way, called OSDM. The proposed algorithm allows processing each instance incrementally and updates the model accordingly.

The procedure of OSDM is given in Algorithm 1. Initially, it creates a new cluster for the first document and the document is assigned to the newly created CF set. Afterward, each arriving document in the stream either choose an existing cluster or generate a new cluster. The corresponding probability for choosing either of an existing cluster or a new cluster is computed using Equation (3–4) and (3–1), respectively. The CF vector with the highest probability is updated using the addition property.

To deal with the cluster evolution (i.e., evolving topics) in text streams, many ex-isting approaches often delete the old clusters by using some of the forgetting mechanisms (e.g., decay rate)[9, 28, 29]. Instead of deleting old clusters, MStreamF[42] deletes old batches. In this study, we investigate the importance of each micro-cluster to handle the cluster evolution problem. Specifically, the importance of each micro-cluster is decreased over time if it is not updated. $l_z$ in CF stores weight of each cluster. If the weight is approximately equals to zero, then the cluster is removed from the model, i.e., it cannot capture recent topics in the text stream. For this purpose, we applied the exponential decay function, $l_z = l_z \times 2^{-\lambda \times (\triangle t)}$. Here, $\triangle t$ is the elapsed time from the last update, and $\lambda$ is the decay rate. The decay rate must be adjusted depending upon the applications at hand. The initial value of $l_z$ (See Line 17 of Algorithm 1) is set to 1. Afterward, the importance of micro-cluster is exponentially decreases over time. We can also store the deleted clusters in a permanent disk for offline analysis.

**Complexity Analysis.** The OSDM algorithm always maintains the average $\bar{K}$ number of current topics (CF sets). Every CF set store average $\bar{V}$ number of words in $n_z^w$ and at most $|\bar{V}_z| \times |\bar{V}_z|$ in $cw_z$. Thus the space complexity of OSDM is $O(\bar{K}(\bar{V} + \bar{V}^2) + VD)$, where $V$ is the size of active vocabulary and $D$ is the number of active documents. On other side, OSDM calculates the probability of arriving document with each cluster (see Line 7 of Algorithm 1). Therefore, the time complexity of OSDM is $O(\bar{K}(\mathcal{L}\bar{V}))$, where $\mathcal{L}$ is the average size of arriving document.

## 3.4 Experimental Study

This section provides the detailed description of experimental setup and results analyses.

## 3.5 Datasets

To evaluate the performance of the proposed algorithm, we conduct experiments on three real and two synthetic datasets. These datasets were also used in [10, 39, 42, 84, 93, 94] to evaluate short text clustering models. In the preprocessing step, we removed stop words, converted all text into lowercase, and stemming. The description of the datasets is as follows.

- **News:** This dataset is collected by [40], which contains 11,109 news title belong to 152 topics.

- **Reuters:** Similar to [82] we skip the documents with more than one class and obtained the dataset consists of 9,447 documents from 66 topics.
- **Tweets:** This dataset contain 30,322 tweets which are relevant to 269 topics in the TREC [①] microblog.

Naturally, we may find a situation where topics in social media appear only for a certain time period and then disappear. However, the documents of each topic in original dataset is observed for long period of time. Therefore, we construct two synthetic datasets (i.e., *Reuters-T* and *News-T*) for which we first sorted the documents by topic. After sorting, for each dataset we divide documents into sixteen equal chunks and shuffled them.

### 3.5.1 Evaluation metrics

We adopted five different evaluation metrics for deep analysis of all algorithms, which include Normalized Mutual Information (NMI), Homogeneity, V-Measure, Accuracy and cluster Purity. We utilized sklearn[②] API to implement these metrics. We compute the measures on overall clustering results[40]. Homogeneity measures that each cluster should have only members of a single class. Whereas, V-measure calculates how successfully the criteria of completeness and homogeneity are satisfied. Cluster purity measures the true positive instances in each cluster. The typical NMI measure calculates the overall clustering quality.

### 3.5.2 Baselines

We have selected four state-of-the-art representative algorithms for stream text clustering to compare OSDM. A brief description of these algorithms are given as follows.

(1) **DTM**[34] is an extension of Latent Dirichlet Allocation which traces the evolution of hidden topics from corpus over time. It was designed to deal with the sequential documents.

(2) **Sumblr (Sb)**[87] is an online stream clustering algorithm for tweets. With only one pass, it enables the model to cluster the tweets efficiently while maintaining cluster statistics.

---

① http://trec.nist.gov/data/microblog.html
② http://scikit-learn.org

(3) **DMM**[40] is a Dirichlet multinomial mixture model for short text clustering, which does not consider temporal dependency of instances.

(4) **MStreamF**[42] is the latest model to deal with infinite number of latent topics in short text while processing one batch at a time. Two models of MStreamF were proposed, one with one-pass clustering process, and another with Gibbs sampling. We refer to the former algorithm as MStreamF-O (MF-O) and the latter as MStreamF-G (MF-G).

We try to find the optimal parameter values of all baseline algorithms with grid search. Finally, we set $\alpha = 0.01$ for DTM, $\beta = 0.02$ for Sumblr. For MStreamF-O and MStreamF-G, we set $\alpha = 0.03$ and $\beta = 0.03$. As defined in [42], we set the number of iterations to 10 and *saved-batches* = 2 for MStreamF-G. We set $\alpha = 0.3$ and $\beta = 0.3$ for DMM. The DTM, DMM and Sumblr needs fixed number of cluster as input therefore we set $K = 300, K = 170$ and $K = 80$ for Tweets, News and Reuters datasets, respectively. We set $\alpha = 2e^{-3}, \beta = 4e^{-5}$ and $\lambda = 6e^{-6}$ for OSDM. The source code of OSDM is publicly available at: https://github.com/JayKumarr/OSDM.

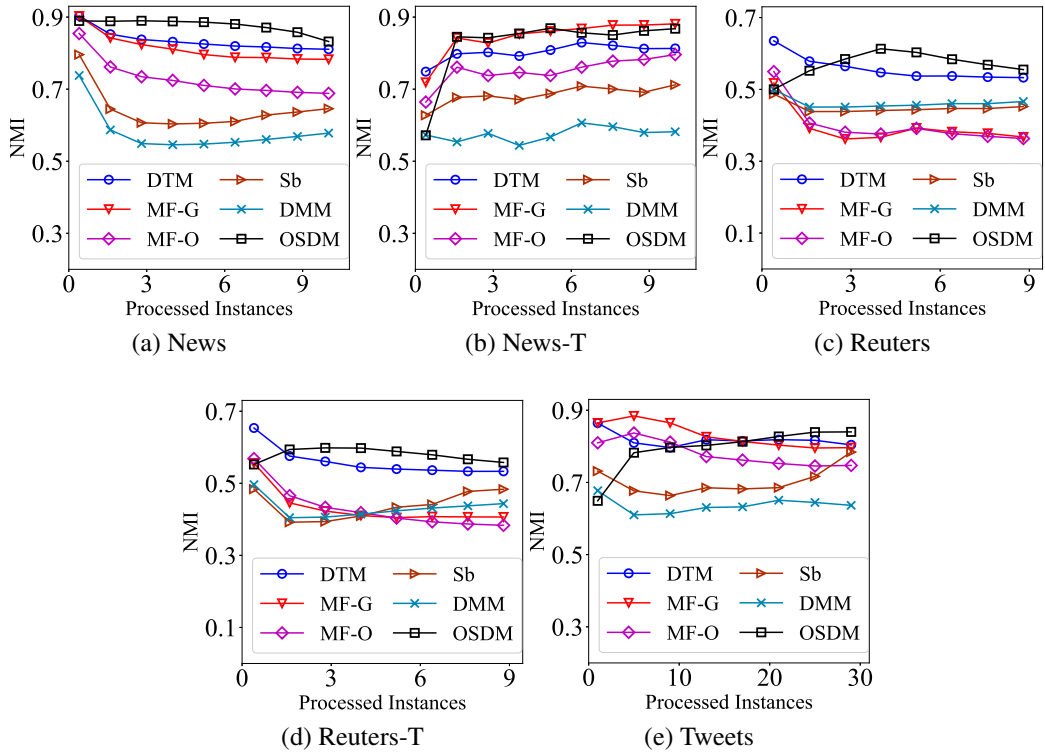

Figure 3–2 The NMI score of all algorithms over time (in thousand points).

### 3.5.3 Comparison with state-of-the-art methods

In this section, we provide a detailed comparative analysis of OSDM with state-of-the-art algorithms. The overall results are summarized in Table 3–1. We report NMI, Homogeneity, v-measure, purity and accuracy of each algorithm. Additionally, we also evaluate the performance of each algorithm over different time-stamps of the stream (see Figure 3–2). Further, we studied the parameter sensitivity and runtime of OSDM, respectively.

Table 3–1 The performance of different algorithms on five data sets in terms of different measures.

| Algorithm | Evaluation | News | Tweets | Reuters | News-T | Reuters-T |
|---|---|---|---|---|---|---|
| OSDM | NMI | **0.815** | **0.836** | **0.552** | 0.858 | **0.554** |
| MF-O |  | 0.685 | 0.746 | 0.361 | 0.803 | 0.381 |
| MF-G |  | 0.780 | 0.795 | 0.364 | **0.888** | 0.405 |
| Sb |  | 0.575 | 0.698 | 0.464 | 0.723 | 0.494 |
| DTM |  | 0.808 | 0.800 | 0.537 | 0.810 | 0.537 |
| DMM |  | 0.586 | 0.636 | 0.448 | 0.582 | 0.476 |
| OSDM | Homogeneity | **0.951** | **0.936** | **0.954** | **0.900** | **0.964** |
| MF-O |  | 0.654 | 0.695 | 0.374 | 0.778 | 0.385 |
| MF-G |  | 0.751 | 0.738 | 0.319 | 0.900 | 0.343 |
| Sb |  | 0.547 | 0.758 | 0.402 | 0.747 | 0.574 |
| DTM |  | 0.833 | 0.822 | 0.659 | 0.837 | 0.657 |
| DMM |  | 0.588 | 0.622 | 0.466 | 0.565 | 0.497 |
| OSDM | V-Measure | 0.805 | **0.831** | **0.479** | 0.857 | 0.478 |
| MF-O |  | 0.684 | 0.744 | 0.361 | 0.803 | 0.380 |
| MF-G |  | 0.779 | 0.793 | 0.361 | **0.888** | 0.400 |
| Sb |  | 0.575 | 0.696 | 0.458 | 0.723 | 0.436 |
| DTM |  | **0.808** | 0.800 | 0.526 | 0.810 | **0.527** |
| DMM |  | 0.586 | 0.636 | 0.448 | 0.582 | 0.476 |
| OSDM | Purity | **0.907** | **0.890** | **0.962** | **0.851** | **0.972** |
| MF-O |  | 0.552 | 0.529 | 0.602 | 0.636 | 0.608 |
| MF-G |  | 0.653 | 0.801 | 0.530 | 0.835 | 0.606 |
| Sb |  | 0.414 | 0.609 | 0.609 | 0.580 | 0.770 |
| DTM |  | 0.767 | 0.749 | 0.793 | 0.765 | 0.795 |
| DMM |  | 0.456 | 0.473 | 0.673 | 0.398 | 0.694 |
| OSDM | Accuracy | **0.880** | 0.665 | **0.927** | **0.769** | **0.952** |
| MF-O |  | 0.420 | 0.246 | 0.577 | 0.584 | 0.447 |
| MF-G |  | 0.517 | **0.707** | 0.452 | 0.606 | 0.461 |
| Sb |  | 0.606 | 0.539 | 0.652 | 0.653 | 0.620 |
| DTM |  | 0.647 | 0.246 | 0.669 | 0.294 | 0.644 |
| DMM |  | 0.334 | 0.150 | 0.649 | 0.073 | 0.500 |

From Table 3–1, we can see that OSDM outperformed all baseline algorithms on almost every dataset in terms of all measures. Here, MStreamF-G yielded much better results on the News-T data in terms of NMI measure. The reason behind might be the multiple iterations of each batch in the stream. However, MStreamF-G requires more execution time to process the data. In contrast, our proposed algorithm OSDM processes the data only once. And we can also observe that OSDM achieves the highest NMI in other data sets. In addition, the crucial part of evaluating the cluster similarity is mea-sured by the homogeneity measure. We can see that OSDM outperformed all previous algorithms. It also shows the same statistics except for v-measure of DTM. Likewise, our model generates more pure clusters. Furthermore, to investigate the performance over time, we plot the performance of all algorithms over time in Figure 3–2.

### 3.5.4 Sensitivity Analysis

We perform sensitivity analysis for OSDM with respects to three input parameters: concentration parameter $\alpha$, $\beta$, and decay function parameter $\lambda$ on the *Tweets* dataset. From Figure 3–3a, we can observe the effect of $\alpha$, which ranges from $9e^{-3}$ to $9e^{-1}$. The performance in terms of all evaluation measures is stable over the different values of parameters. The $\alpha$ parameter is responsible for finer clustering, that is why we can observe a little fluctuation in initial values. Figure 3–3b shows the performance on different values of $\beta$, which ranges from $1e^{-4}$ to $1e^{-2}$. As we already defined that we modified homogeneity part of the clustering model (see Equation (3–1)), and $\beta$ is the related hyper-parameter. We can observe that after a certain range, the values of all the evaluation measure become stable. The crucial point to be observed is the stability of homogeneity on different values of $\beta$. Figure 3–3c shows effect of $\lambda$ ranges from $9e^{-4}$ to $9e^{-6}$. Our model follows the forgetting mechanism on decay factor $\lambda$ and the clusters are deleted from model when the value is approximately equals to zero. We can observe the performance of OSDM on different decay factors. It can be observed that the behavior of a given evaluation measure is stable over time.

### 3.5.5 Runtime

To compare the runtime of different algorithms, we performed all experiments on a PC with *core i5-3470* and 8GB memory. Figure 3–4 shows the runtime of all algorithms on the tweets dataset. We can observe that Sumblr required the highest execution time to cluster the instances. Whereas, the runtime of other algorithms are comparable. Due to simple execution process of each instance MStreamF-O took least time because it does not need to maintain semantic similarity. Comparatively, MStreamF-G required much higher time than OSDM. The reason is that it needs to execute each batch data multiple times. Due to online nature, the overall speed of OSDM is more efficient than most existing algorithms, and the benefit is strengthened with more and more arriving instances.

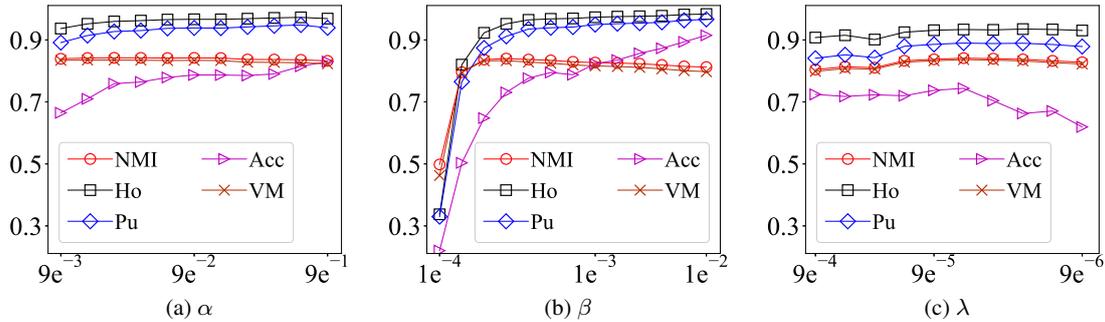

Figure 3–3 The sensitivity analysis with different parameters, including $\alpha, \beta$ and $\lambda$.

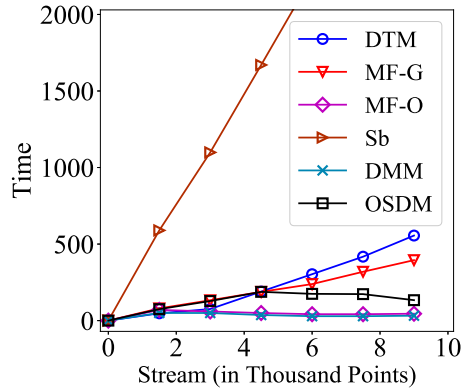

Figure 3–4 The runtime of different text stream clustering algorithms.

## 3.6 Summary

In this chapter, we propose a new online semantic-enhanced Dirichlet model for short text stream clustering. In contrast to existing approaches, OSDM does not require to specify the batch size and the dynamic number evolving clusters. It dynamically assigns each arriving document into an existing cluster or generating a new cluster based on the poly urn scheme. More importantly, OSDM tried to incorporate semantic information in the proposed graphical representation model to remove the term ambiguity problem in short-text clustering. Building upon the semantic embedding and online learning, our method allows finding high-quality evolving clusters. Extensive results further demonstrate that OSDM has better performance compared to many state-of-the-art algorithms.

Apart from the successfully capture the small subspaces of topic, there are still some shortcomings. The proposed model successfully detects the new evolving topics while capturing the semantic feature space with co-occurrence matrix. However, sparsity of clusters in model increased as size of cluster feature space increase over time. If the value of input parameters are not in defined range then the model generate too many cluster leading very fine-grain representation, thus fails to analyze the topics. The value of all parameters should depend on particular domain. Such as, $\lambda$ depends on the estimated velocity of a particular domain given specific unit (model time unit). Likewise, values of $\beta$ and $\alpha$ also belong to domain topic population and term space.

# Chapter 4  A Non-parametric Clustering Model for Evolving Short Text Stream

The proposed model in previous chapter successfully detects the new evolving topics while capturing the semantic feature space with co-occurrence matrix. However, sparsity of clusters in model increased as size of cluster feature space increase over time. Additionally, for some real world topics, the topic related core-terms evolve over time which represent the topic term space evolution. This chapter gives a detailed discussion about distributional change in term space and model sparsity for clustering short text streams. A new model is also proposed with deep analysis to solve these issues.

## 4.1 Introduction

A short text stream is the continuous arrival of short-length documents over time. Nowadays, more and more short text data has been generated on social media and other platforms every day, e.g., Twitter, Facebook, QA blogs, and News blogs. Clustering such continuous short text documents has gained increasing attention in recent years due to its diverse applications, i.e., news recommendation[95], topic tracing[96], and hot topic detection[97], etc. However, due to the unique properties of short text streams such as infinite length, sparse word representation and topic evolution, clustering short text streams is still a big challenge. During the past decade, many approaches have been introduced to deal with different problems of short text stream clustering.

**Challenge 1: Evolving Number of Topics on Text Streams.** For handling text streams with infinite length, initially, static text clustering approaches were transformed to deal with streaming data by considering temporal dependency to discover latent topics [26, 91, 98]. By transforming the traditional text clustering algorithms, similarity-based approaches were introduced such as HPStream[8], Feature Weighting K-Means (FW-KMeans)[30], DenStream[31], Spherical K-Means (SPKM)[28], COBWEB[99] and ConStream[9]. With a pre-specified threshold, similarity-based approaches calculate the similarity scores between the arriving document and existing clusters. It was not long after that statistical model-based approaches were introduced to deal with text streams such as Dynamic Topic Model (DTM)[34], Temporal Text Mining (TTM)[100], Temporal Dirichlet process mixture model (TDPM)[41], Trend analysis model (TAM)[101], Temporal latent Dirichlet allocation (TM-LDA)[36], DPMFP[83], and Gibbs sampling for Dirichlet mixture model (GSDMM)[82], to mention a few. These approaches often require a pre-specified number of latent topics, and cannot deal with evolving unknown number of topics in text streams. However, due to the dynamic velocity of streams, determining evolving number of topics over streams is a non-trivial task.

**Challenge 2: "Term Ambiguity" of Short-text Word Representation.** Additionally, unlike long text documents, short text clustering further suffers from the lack of supportive term occurrence to capture semantics[86] due to the sparse word representation. For most existing short text clustering algorithms like Sumblr[87], Dynamic cluster topic model (DCT)[39], MStreamF[42] and SIF-AE[80], exploiting independent word representation in their cluster models tends to cause "term ambiguity"[86] (i.e., the same word has different meanings in different contexts). To deal with this problem, previous works often enrich short text representations by incorporating features from external resources[102][84].

**Challenge 3: Online Evolving Topic Modeling.** Another potential problem of most existing approaches is the batch-wise processing. The batch-wise processing assumes that the instances are interchangeable within a batch①[97, 103]. Determining an optimal batch size is also a non-trivial task for different text streams. These algorithms also need to process each batch multiple times to infer the clusters. For online processing, NPMM[29] is proposed to discover evolving number of topics over time and allows finding core terms with the help of external word-embeddings. However, due to the context of terms changes over time, pre-trained word embeddings are thus not efficient in real-time streams[104].

To clearly describe the mentioned problems, Figure 4–1 shows an example where the three challenges are depicted. The first problem is the number of topics at timestamp 1 varies from Timestamp 2. The second problem is the "term ambiguity" issue, where "*Apple*" is a context depending term, enriched in different topics. The third challenging problem is the change in term distribution over time for the topic "Trump Election Campaign".

## 4.2 Proposed Approach

To tackle the aforementioned issues, we propose an online semantic-enhanced graphical model (OSGM) for evolving short text stream clustering. Compared to the state-of-the-art

---

① A batch is a chunk of instances, objects or documents.

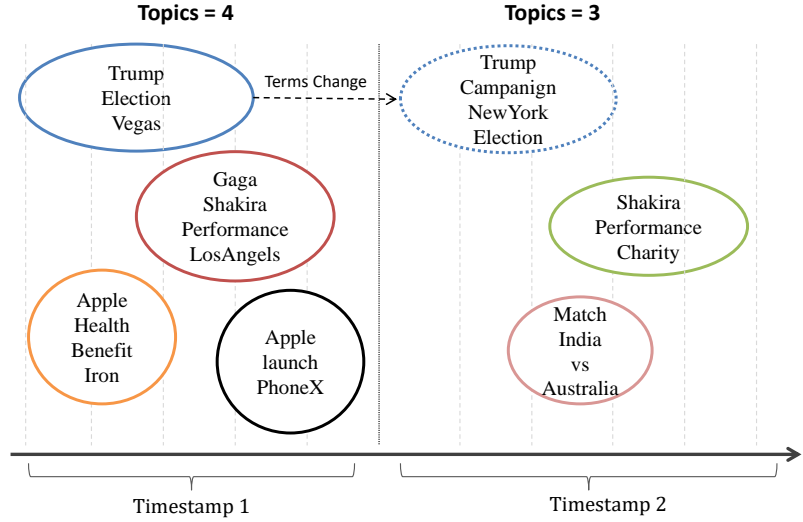

Figure 4–1 Illustration of challenges of short text stream clustering.

algorithms, our proposed model can: (a) automatically detect the number of topics over time using the Polya Urn scheme, (b) capture semantics by embedding evolving term co-occurrence matrix, (c) find the topic related term-subspace in online way, and (d) track cluster term evolution using feature-level triangular time decay. The main contributions of this chapter are highlighted as follows.

- **Evolving Topic Modeling with Online Term Distribution Exploring:** Instead of batch-wise processing and eliminating the constraint of external embeddings, OSGM works in an online way to automatically detect the number of clusters and exploit cluster-document population (see Section 4.2.2.3) to identify evolving topics in changing term-subspaces.

- **Semantics Enrichment:** To deal with the "term ambiguity" issue in short text clustering, an evolving term co-occurrence matrix is introduced to capture the relationship between terms for improving the cluster purity. We further exploit the inverse cluster frequency for singular terms as semantic smoothing.

- **Robustness:** OSGM allows merging highly similar clusters, which yields the number of topics near to actual ones automatically. Moreover, unlike traditional Dirichlet based non-parametric models, OSGM is more robust to the concentration parameter (Dirichlet Process parameter, cf. Section 2.1) and background term noise parameter.

We first give detailed description of cluster feature set and then introduce our online semantic-enhanced graphical model (OSGM) for evolving short text stream clustering.

## 4.2.1 Cluster Feature Set

A micro-cluster in stream is represented by cluster feature set which contains different statistical values about the instances within a micro-cluster. Most similarity-based and model-based methods follow the vector space model (VSM) to represent the cluster feature space[92]. However, a topic needs to be represented in the term-changing subspaces. Here, an extended cluster feature (CF) is employed, where each cluster is represented by words of related documents. In our model, a CF set is defined as a 7-tuple $\{m_z, n_z^w, cw_z, n_z, l_z, u_z, ta_z\}$. The description of each tuple is given in Table 4–1. The CF has the important addable property which allows a cluster to update incrementally over time.

**Definition 4.2.1**  A document $d$ can be added to a cluster $z$ by using the *addable property* to update cluster.

$$m_z = m_z + 1$$
$$n_z^w = n_z^w + N_d^w \quad \forall w \in d$$
$$cw_z = cw_z \cup cw_d$$
$$n_z = n_z + N_d$$
$$ta_z = ta_z \uplus ta_d \quad \forall w \in d$$

Here, $cw_d$ is co-occurrence of each word with every word of same document, and $N_d$ represents the number of total words in the document. $\uplus ta_d$ merges the arrival time of each document term. Let us take an example, suppose, we have cluster containing two documents $z = \{d_1, d_2\}$ and its CF,

$$d_1 = \{w_1 : 3, w_2 : 2\}$$
$$d_2 = \{w_1 : 1, w_3 : 1, w_4 : 1\}$$

$$m_z = 2$$
$$n_z^w = \{w_1 : 4, w_2 : 2, w_3 : 1, w_4 : 1\}$$
$$n_z = 8$$

$$cw_z = \begin{bmatrix} & w_1 & w_2 & w_3 & w_4 & \\ & - & 0.4 & 0.5 & 0.5 & w_1 \\ & 0.6 & - & 0.0 & 0 & w_2 \\ & 0.5 & 0.0 & - & 0.5 & w_3 \\ & 0.5 & 0 & 0.5 & - & w_4 \end{bmatrix}$$

$ta_{w_1} = \{timestamp : 1, timestamp : 2\}$

$ta_{w_2} = \{timestamp : 1\}$

$ta_{w_3} = \{timestamp : 2\}$

$ta_{w_4} = \{timestamp : 2\}$

$ta_z = \{ta_{w_1}, ta_{w_2}, ta_{w_3}, ta_{w_4}\}$

and suppose we have a new document $d_i$ to add in cluster $d_i = \{w_1 : 1, w_2 : 1\}$, then the CF of $z$ will be,

$m_z = \mathbf{3}$

$n_z = \mathbf{10}$

$n_z^w = \{w_1 : \mathbf{5}, w_2 : \mathbf{3}, w_3 : 1, w_4 : 1\}$

$ta_{w_1} = \{timestamp : 1, timestamp : 2, \mathbf{timestamp:3}\}$

$ta_{w_2} = \{timestamp : 1, \mathbf{timestamp:3}\}$

As we can observe that a new timestamp is added, which is the size of cluster. In other words, the life-span of terms is measure over the scale of cluster size. Another point to highlight, that new document contains $w_1$ and $w_2$, therefore, the new score will be updated by adding document score of word to word co-occurrence, which will be

$$cw_z = \begin{bmatrix} & w_1 & w_2 & w_3 & w_4 & \\ & - & \mathbf{0.9} & 0.5 & 0.5 & w_1 \\ & \mathbf{1.1} & - & 0.0 & 0 & w_2 \\ & 0.5 & 0.0 & - & 0.5 & w_3 \\ & 0.5 & 0 & 0.5 & - & w_4 \end{bmatrix}.$$

The complexity of updating a cluster by adding a document is $O(\mathcal{L}^2)$, where $\mathcal{L}$ is the average length of the documents. This property is useful to update the evolving microcluster in the text stream clustering.

Table 4–1 Symbols and notations.

| Symbol | Definition |
| --- | --- |
| $D$ | total number of active documents |
| $V_d$ | unique terms / vocabulary in document $d$ |
| $V_z$ | unique terms / vocabulary of cluster $z$ |
| $\mathcal{L}$ | average length of document in stream |
| $\bar{V}_z$ | average vocabulary size of active clusters ($V_z \cap V_d \neq \{\}$) |
| $cw_{ij}$ | co-occurrence of words $w_i$ and $w_j$ |
| $N_d^w$ | occurrence of word $w$ in document $d$ |
| $N_d$ | total number of words in document $d$ |
| $n_z^w$ | term frequency of $w$ in cluster $z$ |
| $cw_z$ | word to word co-occurrence matrix |
| $z_d$ | assigned cluster of document $d$ |
| $m_z$ | number of documents in cluster $z$ |
| $n_z$ | number of words in cluster $z$, $\sum_{w \in z} n_z^w$ |
| $l_z$ | decay weight of cluster $z$ |
| $u_z$ | last updated time stamp of cluster $z$ |
| $ta_z$ | arriving timestamps of the words in clusters $z$ |
| $V_{z \cup d}$ | vocabulary size of $V_z \cup V_d$ |

## 4.2.2 Online Semantic-enhanced Graphical Model

The probabilistic graphical model (PGM) is used to represent the stochastic generation of objects (short text document in our case) with Bayesian probabilistic modeling. To model the topic generation for the text stream clustering problem, the most crucial task is to define the relationship between documents and topics (clusters) in terms of a probability distribution. Figure 4–2 shows our graphical model: OSGM. We extend and transform the MStreamF model from batch-wise to online process. In the graphical model, $\theta$ represents the distribution of topics $z$ drawn with $\alpha$ as concentration parameter and $\mathcal{N}$ is the distribution of terms over generated topics with $\beta$ as term coefficient. The $\varphi$ controls the change of topic related terms over time with $\Gamma$ as term recency threshold. The decay process to remove the outdated topics from the model is controlled by $\lambda$ as a decay factor. The generative process for topic generation in stream is presented as follows:

1. Sample $\theta_t \sim Dirichlet(\alpha, \theta_{t-1})$
2. For each document $d$:
   (a) draw $\mathcal{N}_{t,z}|\beta_{t,z} \sim Dirichlet(\beta_{t-1,z}, \mathcal{N}_{t-1,z})$
   (b) sample $z_d \sim Multinomial(\theta_t)$
   (c) For each word or co-occurrence of word:

i. $w_d \sim Multinomial(\mathcal{N}_{t,z}|z_d)$

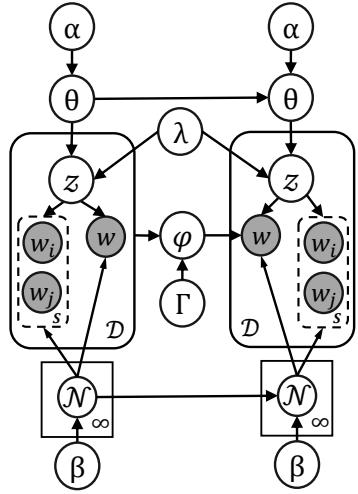

Figure 4–2 The graphical model of OSGM. The shaded are observed and unshaded are latent entities.

The characteristics of OSGM are four-fold: eliminate term ambiguity with calculating document-cluster similarity, handle evolving number of topics, determine the evolving topics in changing term-subspaces, and remove outdated topics. The pseudo code of our proposed approach is given in Algorithm 2.

#### 4.2.2.1 Document-Cluster Similarity

Initially the first arriving document creates its own cluster. For next arriving document, either it should be added in any active cluster of model ($\mathcal{M}$) or a new cluster is created for it. For calculating similarity between a document and an existing cluster, we transform Equation (2–5) and derive the probability of an arriving document $d$ to choose an existing cluster $z$ as we combine two probabilities (1) probability of topic popularity (defined by CRP) and (2) similarity between document and topic in terms of word distribution.

$$p(z_d|G_z) = p(G_z).p(d|G_z) \tag{4–1}$$

Here, $p(G_z)$ defines the probability of topic popularity (in CRP it is table popularity), which is $\frac{n_k}{\alpha+n-1}$ for choosing existing topic and $\frac{\alpha}{\alpha+n-1}$ for choosing new topic. The term $p(d|G_z)$ represents the similarity between topic (table) and document (customer), multinomial distribution with Dirichlet process is used for existing topic and for the new topic as well. By deriving the equation for the probability of topic popularity we define

$$p(G_z) = \left(\frac{m_z}{D-1+\alpha D}\right). \tag{4–2}$$

Here, $m_z$ is the number of documents in a particular cluster (customers on existing table in CRP) and $D$ reflects the number of active documents. For capturing new distribution, as controlled by the $\alpha$ concentration parameter, we define

$$p(G_{New}) = \left(\frac{\alpha D}{D-1+\alpha D}\right), \tag{4-3}$$

where $\alpha D$ defines pseudo portion of active documents for new cluster (topic). To simulate the homogeneity in generative process we derive probability of a document generated by a particular topic as:

$$p(d|G_z) = p(w_z|w_d).p(w_z^i w_z^j | w_d^i w_d^j) \text{ where } i \neq j$$

Here, it is assumed that topic related terms are generated by mixture of independent terms and related-terms generation (for capturing term semantics). The first part of equation reflect the probability of independent terms and the predecessor part for term relativeness. We derive the probability for independent terms as,

$$p(w_z|w_d) = \left(\frac{\prod_{w \in d} \prod_{j=1}^{N_d^w} p(w_z^j) + \beta + j}{\prod_{i=1}^{N_d} p(w_z^j) + (\beta V_{z \cup d}) + i}\right) \tag{4-4}$$

The nominator part calculate the probability of words contain by cluster and document $w \in d$, The denominator part normalizes the equation by multiplying probability of all words of cluster which uses the global term space $V_{z \cup d}$ by taking union of unique term of both document and cluster. The correlation of terms in topic generation is derived as,

$$p(w_z^i w_z^j | w_d^i w_d^j) = \sum_{(w_i,w_j) \in \mathcal{V}_{d \cap z}} \frac{p(w_i w_j|z)}{p(w_i|z) + p(w_j|z)}$$

Here, $p(w_i w_j|z)$ is probability of two words co-occurring. As we know that $p(w_z|w_d) \leq 1$ therefore we combine them by adding 1 in $p(w_i w_j|z)$ as,

$$p(d|G_z) = p(w_z|w_d).\left(1 + \sum_{(w_i,w_j) \in \mathcal{V}_{d \cap z}} \frac{p(w_i w_j|z)}{p(w_i|z) + p(w_j|z)}\right)$$

By extending DPMM, we propose novel probability score defined as,

$$p\left(z_d = z|\vec{z},\vec{d},\alpha,\beta\right) =$$
$$\left(\frac{m_z}{D-1+\alpha D}\right) \cdot \left(\frac{\prod_{w \in d} \prod_{j=1}^{N_d^w} (n_z^w \cdot lCF_w) + \beta + j - 1}{\prod_{i=1}^{N_d} n_z + (\beta V_{z \cup d}) + i - 1}\right) \cdot \left(1 + \sum_{w_i \in d \wedge w_j \in d} cw_{i,j}\right) \tag{4-5}$$

The first term $\left(\frac{m_z}{D-1+\alpha D}\right)$ represents $p(G_z)$ of Equation (2-5). Here, $D$ is the number of current documents in active clusters[①]. Whereas, $\alpha$ is the concentration parameter of the model. We designed two probabilities to define $p(d|G_z)$ of Equation (2–5). The first part captures similarity in singular term space (the middle term of the Equation (4–5)) and second part captures similarity in semantic term space (co-occurrence of terms) to solve term-ambiguity problem. The middle term is based on the multinomial distribution (see Equation (2–4)) with a pseudo weight of words $\beta$ defines the homogeneity between a cluster and a document. Previously, occurrence of a term in cluster $n_z^w$ is used while calculating the homogeneity (similarity). However, in the static environment inverse document frequency is used to calculate the term importance in global space. Here, in the dynamic environment we define similar weight measure called inverse cluster frequency $ICF_w$ to calculate term importance, defined as:

$$ICF(w \in d) = \log\left(\frac{|Z|}{|\{Z|w \in V_z\}|}\right). \tag{4–6}$$

The denominator part of Equation (4–6) is the number of those clusters which contains the word $w$, and $|Z|$ is total number of active clusters. Occurrence of a term $w$ exists in more clusters, lesser the importance it has. If a term $w$ is found in very few clusters, it signifies its high importance. To deal with the "term ambiguity" issue in short text clustering, an evolving term co-occurrence matrix is introduced to capture the relationship between terms for improving the cluster purity. For example, suppose we have two clusters $C1$ and $C2$.

$$n_{C1}^w = \{w_1 : 3, w_2 : 4, w_1 : 3, w_3 : 1, w_4 : 2\}$$

$$n_{C2}^w = \{w_1 : 3, w_2 : 4, w_1 : 3, w_5 : 1, w_6 : 2\}$$

$$cw_{C1} = \begin{bmatrix} & w_1 & w_2 & w_3 & w_4 & \\ & - & 0.5 & 0 & 0 & w_1 \\ & 0.5 & - & 0.25 & 0 & w_2 \\ & 0 & 0.75 & - & 0.5 & w_3 \\ & 0 & 0 & 0.5 & - & w_4 \end{bmatrix}$$

---

① Active clusters refer to those clusters which are not yet deleted from the model.

$$cw_{C2} = \begin{bmatrix} & w_1 & w_2 & w_5 & w_6 & \\ - & 0 & 0 & 0.5 & w_1 \\ 0 & - & 0.75 & 0 & w_2 \\ 0 & 0.25 & - & 0.5 & w_5 \\ 0.5 & 0 & 0.5 & - & w_6 \end{bmatrix}$$

We assume $d' = \{w_1 : 1, w_2 : 1, w_7 : 1\}$ as an arriving document, and calculate the similarity of $d'$ with both clusters. As we can observe that the $w_1$ and $w_2$ are common between both clusters. If we do not consider co-occurrence then the probability of both clusters are same, however, $w_1$ and $w_2$ co-occurred in $C1$. Therefore, the third part $\left(1 + \sum_{w_i \in d \wedge w_j \in d} cw_{i,j}\right)$ in Equation (4–5) defines the weight of term co-occurrence between the cluster and a document. Formally, we define a value of an entry $cw_{i,j}$ in the co-occurrence matrix as follows.

$$cw_{i,j} = \frac{\sum_{d' \subseteq z} N_{d'}^{w_i}}{\sum_{d' \subseteq z} N_{d'}^{w_i} + \sum_{d' \subseteq z} N_{d'}^{w_j}} \quad \text{s.t.} \ (w_i, w_j) \in d' \tag{4–7}$$

Here, $n_{d'}^{w_i}$ is the frequency count of word $w_i$ in the document $d'$. The ratio between $w_i$ and $w_j$ must satisfy $cw_{i,j} + cw_{j,i} = 1$, where $i \neq j$. We calculate the term co-occurrence weight of those terms which are common in the cluster $z$ and document $d$. Therefore, if the size of cluster feature set (discussed in Section 4.2.1) is $|V_z|$, then it is not necessary that the co-occurrence matrix would be $|V_z| \times |V_z|$.

### 4.2.2.2 Automatic Cluster Creation

Initially the first arriving document creates its own cluster. To identify new topics over time, we need to define the probability that can automatically create new cluster for the new topic if the incoming document in the stream does not relate to any existing active cluster. For automatic cluster creation, we transform Equation (2–5) and derive probability as follows.

$$p\left(z_d = z_{new} | \vec{z}_{\neg d}, \vec{d}, \alpha, \beta\right) = \left(\frac{\alpha D}{D - 1 + \alpha D}\right) \cdot \left(\frac{\prod_{w \in d} \prod_{j=1}^{N_d^w} \beta + j - 1}{\prod_{i=1}^{N_d} (\bar{V}_z \beta) + i - 1}\right) \tag{4–8}$$

Here, $\bar{V}_z$ is the average vocabulary size of active clusters. The $\alpha D$ represents the pseudo number of documents and $\beta$ helps to calculate pseudo term similarity with a new cluster. If this small probability of an arriving document is greater than the probability of existing clusters, then the model creates a new cluster containing the arrival document (see Line 16 Algorithm 2).

**Algorithm 2:** OSGM

1    $S_t : \{d_t\}_{t=1}^{\infty}$, $\alpha$ : concentration parameter, $\beta$ : pseudo weight of term in cluster, $\lambda$ : decay factor, $\Gamma$ : feature decay threshold Cluster assignments $z_d$
2    initialize $\mathcal{M} = \phi$      // empty model
3    $t_c = 0$      // model timestamp
4    **while** $d_t$ in $S_t$      // start streaming
5    **do**
6      $t_c = t_c + 1$
7      $\mathcal{M} = updateActiveClusters(\lambda, \mathcal{M})$      // Algorithm 4
8      $\mathcal{M} = updateTermsSubspace(\Gamma, \mathcal{M})$      // Algorithm 3
9      **foreach** $z_i \in \mathcal{M}$ **do**
10        **if** $V_d \cap V_z \neq \phi$ **then**
11          $P_{Z_i} = p(z_d = z_i, d_t)$ using Equation (4–5)
12        **end**
13      **end**
14      $i = \arg\max_i (P_{Z_i})$
15      $P_{Z_n} = p(z_d = z_{new}, d_t)$ using Equation (4–8)
16      **if** $P_{Z_i} < P_{Z_n}$ **then**
         // create new cluster
17        $m_{z_n} = 1, n_{z_n}^w = N_{d_t}^w$
18        $cw_{z_n} = cw_{d_t}$, $n_{z_n} = N_{d_t}$
19        $l_{z_n} = 1, u_{z_n} = t_c, ta_{z_n} = ta_{d_t}$
20        $\mathcal{M} = \mathcal{M} \cup z_n$
21      **else**
22        update $z_i$ using Definition 4.2.1
23        $l_{z_i} = 1, u_{z_i} = t_c$
24      **end**
25    **end**

#### 4.2.2.3 Online Evolving Topic Extraction in Changing term-subspaces

The previous study [105] argues that a topic related text document has few core terms to represent a topic. Whereas, for some topics, distribution of core terms also changes over time. To highlight core terms and term subspace change in online way, we first maintain arriving timestamps of each term. We then calculate age of the cluster by adopting triangular number of its size (see Line 5 Algorithm 3). Afterwards, we compare arrival timestamps of each term with age of cluster to reduce noisy terms. The triangular time is originally employed to fade out outdated micro-clusters[106], defined as,

$$\text{Triangular Number } \Delta f(T) = \left((T^2 + T)/2\right). \tag{4–9}$$

Here, $T$ is the timestamp number. The recency score of a term is measured by ratio of the sum of arrivals and age of cluster (see Line 8 Algorithm 3). For example, cluster $z$ contains 9 documents then the age of cluster $z$ will be:

$$Age_z = \Delta f(9) - \Delta f(1) = ((9^2 + 9)/2) - ((1^2 + 1)/2)$$

Let us suppose $D1$, $D2$ and $D8$ belong to $z$ which are:

$D_1$: "entry performance of Lady gaga."

$D_2$: "Lady gaga stole the heart by exploding performance."

$D_8$: "Gaga and shakira rocked!."

and their cluster timestamps[①] are 1, 2 and 8, respectively. The arrivals of terms ($ta$) in clusters is stored as:

$$ta_{\text{gaga}} = \{timestamp : 1, timestamp : 2, timestamp : 8\}$$

and the sum of time arrivals is $sum(ta_{\text{gaga}}) = \sum_{t \in ta_{\text{gaga}}} t$ To calculate recency score of $gaga$ which will be

$$recency_{\text{gaga}} = ((1 + 2 + 8) \times 100) / Age_z$$

if the *recency* is less than our defined threshold $\Gamma$ then we consider that this term cannot represent current concept of cluster. In other words if a term is frequently observed in cluster with the passage of cluster time-span then the term is useful to represent the current concept of terms (see Line 9 Algorithm 3). It can be analyzed that more terms become less important when new documents are added in a cluster. In this way, the parameter $\beta$ of multinomial distribution heavily affects the probability due to the increasing number of unimportant terms. With the triangular number time, important terms are highlighted and noisy terms are filtered out gradually over time. The whole procedure is given in Algorithm 3.

### 4.2.2.4 Deleting Outdated Clusters

Followed by clusters creating process, the model should maintain only active clusters (current concept) by deleting outdated clusters (old concept). Usually in real time environment, each topic may have different active time-span. For example, *"USA-election"*

---

① cluster timestamp is equal to number of documents within respective cluster

**Algorithm 3:** Update Cluster Term-subspace

/* Update weight of terms in each active cluster                                                  */

1  **Function** $updateTermsSubspace(\Gamma, \mathcal{M})$
2      $S_z = \Delta f(1)$ using Equation (4–9)
3      **foreach** $z \in \mathcal{M}$ **do**
4          $E_z = \Delta f(m_z)$ using Equation (4–9)
5          $Age_z = E_z - S_z + 1$
6          **foreach** $(w, ta_w) \in ta_z$ **do**
            // sum arrival timestamps of w
7              $sumta_w = \sum_{t \in ta_w} t$
8              $recency_w = (sumta_w \times 100)/Age_z$
9              **if** $recency_w < \Gamma$ **then**
10                 $remove(w, z)$
11             **end**
12         **end**
13     **end**
14     return $\mathcal{M}$

may be active on social media for two months, whereas, tweets on a particular "football-match" may active for two days. Many existing approaches often delete old clusters using some forgetting mechanisms (e.g., decay rate), some deletes instances of old batches [28][9][29][42]. The life-span of each document depends on the active time period of a topic, therefore, in contrast to deleting old batches we adopt forgetting mechanism cluster importance score based on stream velocity. Specifically, the importance of each cluster is decreased over time if it is not updated. For this purpose, we apply an exponential decay function,

$$l_z = l_z \times 2^{-\lambda \times (t_c - u_z)} \quad (4\text{--}10)$$

Here, $l_z$ represents the importance score of a cluster $z$ and $t_c$ is current timestamp of model. Initially, we set $l_z = 1$ of each new cluster (see Line 19 Algorithm 2). Importance score of a cluster decreases over time if it is not receiving any document. If the score of a cluster approximately reaches to zero (see Line 7 Algorithm 4), then the cluster is processed to be removed from the model, i.e., it cannot capture recent topic in the text stream. Algorithm 4 shows the step by step procedure.

### 4.2.2.5 Merging Clusters

As we discussed that $\beta$ parameter of multinomial distribution is responsible for calculating the homogeneity (similarity). However, due to the exceeding noisy terms over time in a cluster, the probability of noisy terms may become dominant over core terms. This leads the model to create many micro-clusters for a single topic. Previous approaches iterate the stream multiple times to infer the number of clusters. In contrast, we incorporate the cluster merging process to check if an outdated cluster can be merged with active cluster by calculating the probability using Equation (4–5) between two clusters. Algorithm 4 shows the step by step procedure to merge an outdated cluster with an active cluster. The two clusters are merged using Definition 4.2.1.

---

**Algorithm 4:** Update Active Clusters

/* Update weight of all active clusters                                                */
1 **Function** $updateActiveClusters(\lambda, \mathcal{M})$
2     $Old = \phi$
3     **foreach** $z \in \mathcal{M}$ **do**
4        calculate $l_z$ using Equation (4–10)
5        **if** $l_z \approx 0$ **then**
6           $Old = Old \cup z$
7        **end**
8     **end**
9     **foreach** $z_i \in Old$ **do**
10        **foreach** $z_{act} \in (\mathcal{M} - Old)$ **do**
11           **if** $V_{z_{act}} \cap V_{z_i} \neq \phi$ **then**
12              $P_{Z_i} = p(z_d = z_{act}, z_i)$ using Equation (4–5)
13           **end**
14        **end**
15        $max = \arg\max_i(P_{Z_i})$
16        $P_{Z_n} = p(z_d = z_{new}, z_i)$ using Equation (4–8)
17        **if** $P_{Z_{max}} < P_{Z_n}$ **then**
18           $\mathcal{M} = \mathcal{M} - z_i$
19        **else**
20           $merge(z_i, Z_{max})$ using Definition 4.2.1
21        **end**
22     **end**
23 **return** $\mathcal{M}$
24 **End Function**

---

**Complexity Analysis:** The OSGM algorithm always maintains an average $\bar{K}$ number of current topics. Every CF set stores an average $\bar{V}$ number of words in $n_z^w$, $2 \times |\bar{V}|$ timestamps of each word in $ta_z$, and at most $|\bar{V}_z| \times |\bar{V}_z|$ in $cw_z$. Thus the space complexity of OSGM is $O(\bar{K}(3\bar{V} + \bar{V}^2) + VD)$, where $V$ is the size of active vocabulary and $D$ is the number of active documents. On other side, OSGM calculates the probability of arriving document with almost each active cluster (see Line 9 of Algorithm 2). Additionally, it also eliminates outdated vocabulary by checking each CF. Therefore, the time complexity of OSGM is $O(\bar{K}(\mathcal{L}\bar{V}))$, where $\mathcal{L}$ is the average size of arriving document.

## 4.3 Experiments

This section provides the selection of datasets and state-of-the-art algorithm for experimental setup, results analysis and parameter sensitivity of proposed model.

### 4.3.1 Datasets

To evaluate the performance of the proposed algorithm, we conduct experiments on four dataset: two real and two synthetic. Two real datasets include News and Tweets which are already discussed in Section 3.5. The synthetic dataset construction is different from previous chapter. For synthetic datasets, we sorted documents by topics then divide the dataset into sixteen equal chunks and shuffled them. Afterwards, we analyzed 5 main topics whose term distribution changed in sub-topics over time. To depict change in term distribution, we combined sub-topics and placed them at continuous position. The Figure 4–3 shows the instance position of 5 and 4 topics in stream for News and Tweets, respectively. Here, black point instances show the initial distribution of terms of a new topic, whereas, the green points depicts instances of the term distributional change (concept drift) of same topic. An example from this dataset is given in Figure 4–4.

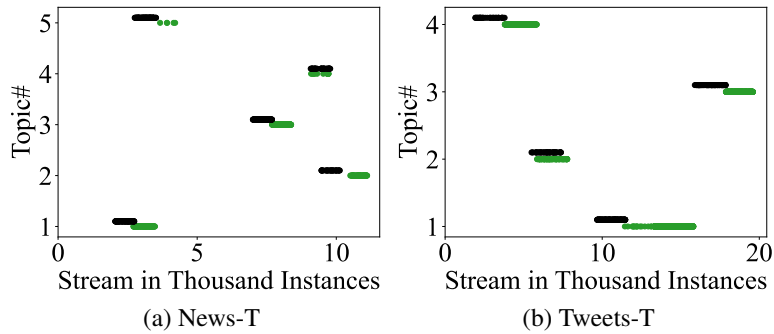

(a) News-T  (b) Tweets-T

Figure 4–3 In News-T and Tweets-T dataset, there are five and four main topics, respectively.

The statistics of the datasets are reported in Table 4–2 where $|D|$ represents total number of documents, $|V|$ is number of unique terms, $|T|$ is total number of topics, and $\mathcal{L}$ is average document length. Both datasets contain less than 10 average document length ($\mathcal{L}$) which assure datasets fit for short text analysis. Specifically, the News dataset contains lower average document length, compare to Tweets dataset, because the dataset was constructed by collecting headlines of news streams.

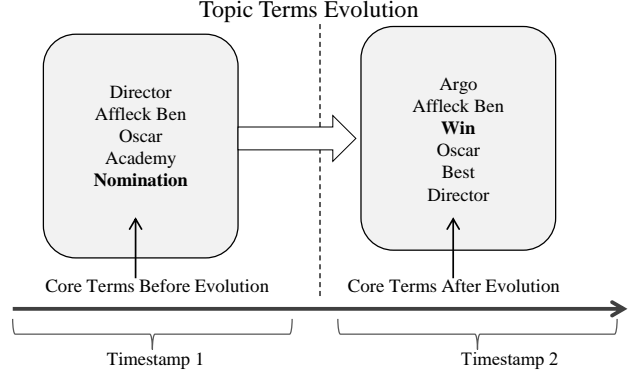

Figure 4–4 The changes of core terms for a topic from News dataset.

Table 4–2 Statistics of Each dataset.

| Dataset | $|D|$ | $|V|$ | $|T|$ | $\mathcal{L}$ |
|---|---:|---:|---:|---:|
| News | 11109 | 8110 | 152 | 6.23 |
| Tweets | 30322 | 12301 | 269 | 7.97 |
| News-T | 11109 | 8110 | 147 | 6.23 |
| Tweets-T | 30322 | 12301 | 265 | 7.97 |

### 4.3.2 Methods of Comparison

For the deep analysis, we compare two variations of our proposed approach i) OSGM and ii) OSGM-ES with baselines. The former approach includes semantic smoothing of inverse clustering frequency, whereas latter approach excludes it. We have selected five state-of-the-art representative algorithms for text stream clustering to comparsion.

(1) **DMM**[40] is an initial model based on multinomial Dirichlet modeling to simulate CRP.

(2) **OSDM**[107] is the most recent model to deal infinite number of topics in online way.

(3) **DTM**[34] is an extension of Latent Dirichlet Allocation which traces the evolution of hidden topics from corpus over time. It was designed to deal with the sequential documents.

(4) **GSDMM**[40] is a Dirichlet multinomial mixture model for short text clustering, which does not consider the temporal dependency of instances. Whereas, the number of clusters is inferred with iterative Gibbs sampling procedure.

(5) **MStreamF**[42] is a recent model to deal with the infinite number of latent topics in short texts while processing one batch at a time. Two models of MStreamF were proposed, one with one-pass clustering process, and another with Gibbs sampling. We refer to the former algorithm as MStreamF-O (MF-O) and the latter as

MStreamF-G (MF-G).

(6) **NPMM**[84] is the latest model to deal with the infinite number of latent topics in short texts while exploiting pre-trained Glove word-embeddings to capture semantics. Authors proposed two variations of their algorithm, one with iterative process referred as NPMM-G, and another without it which is NPMM-O.

### 4.3.3 Evaluation Measures

We adopted four highly used evaluation metrics for deep analysis of all algorithms, which include Purity (Pur), V-Measure (V-M), Homogeneity (Homo), and Normalized Mutual Information (NMI). We utilized sklearn[①] API to compute the measures on overall clustering results[40].

### 4.3.4 Parameter Setting

We try to find the optimal parameter values for best results for most algorithms with grid search. Finally, we obtain best results on $\alpha = 0.01$ for DTM, $\alpha = 0.3$ and $\beta = 0.3$ for DMM. The DTM and DMM need fixed number of cluster as input therefore we set $K = 300$ and $K = 170$ for Tweets and News, respectively. For MStreamF-O, and MStreamF-G, we set $\alpha = 0.03$ and $\beta = 0.03$. By following [42], we set the number of iterations to 10 and *saved-batches* $= 2$ for MStreamF-G. As defined in [84], we set $\alpha = 0.03$, $\beta = 0.03$, $\delta = 0.03$, and $\varepsilon = 3e^{-7}$ for NPMM. For OSDM, we follow the given parameter setting where $\alpha = 0.002$, $\beta = 0.0004$ and $\lambda = 6e^{-6}$. We select $\alpha = 0.05$, $\beta = 0.004$ and $\alpha = 0.09$, $\beta = 0.006$ for OSGM and OSGM-ES, respectively. We set $\Gamma = 10$ and $\lambda = 1e^{-6}$ for both variants of proposed algorithm.

### 4.3.5 Comparison to state-of-the-art approaches

Here we compare OSGM with the selecting state-of-the-art algorithms. Table 4–3 gives the overall clustering results on different real-world and synthetic data streams. We report NMI, V-measure, Homogeneity, and purity for each algorithm. The results show the superior performance of the proposed approach with and without semantic smoothing. The analysis of both variations is discussed in Section 4.3.8. In total, OSGM yields the best performance over almost all the datasets. NPMM-G gives the best results in the

---

① http://scikit-learn.org

Table 4–3 The performance of different algorithms on four datasets in terms of different measures.

| Algorithm | Dataset | Evaluation Measures | | | |
|---|---|---|---|---|---|
| | | NMI | V-Measure | Homogeneity | Purity |
| OSGM | | 0.815 | 0.815 | 0.893 | 0.824 |
| OSGM-ES | | 0.822 | 0.822 | 0.900 | 0.841 |
| OSDM | | 0.814 | 0.805 | **0.950** | **0.907** |
| MF-G | | 0.780 | 0.779 | 0.751 | 0.653 |
| MF-O | News | 0.685 | 0.684 | 0.654 | 0.552 |
| DTM | | 0.800 | 0.800 | 0.822 | 0.749 |
| NPMM-O | | 0.809 | 0.809 | 0.802 | 0.710 |
| NPMM-G | | **0.843** | **0.843** | 0.834 | 0.745 |
| DMM | | 0.592 | 0.592 | 0.595 | 0.474 |
| GSDMM | | 0.821 | 0.821 | 0.751 | 0.605 |
| OSGM | | 0.822 | 0.822 | 0.879 | 0.819 |
| OSGM-ES | | **0.836** | **0.836** | **0.910** | **0.855** |
| OSDM | | 0.822 | 0.831 | 0.890 | 0.829 |
| MF-G | | 0.795 | 0.793 | 0.738 | 0.801 |
| MF-O | Tweets | 0.746 | 0.744 | 0.695 | 0.529 |
| DTM | | 0.800 | 0.800 | 0.822 | 0.749 |
| NPMM-O | | 0.769 | 0.769 | 0.758 | 0.634 |
| NPMM-G | | 0.821 | 0.821 | 0.813 | 0.696 |
| DMM | | 0.392 | 0.623 | 0.549 | 0.141 |
| GSDMM | | 0.798 | 0.798 | 0.728 | 0.581 |
| OSGM | | **0.854** | **0.854** | 0.919 | 0.878 |
| OSGM-ES | | 0.853 | 0.853 | **0.958** | **0.936** |
| OSDM | | 0.854 | 0.854 | 0.901 | 0.868 |
| MF-G | | 0.848 | 0.848 | 0.829 | 0.705 |
| MF-O | News-T | 0.833 | 0.833 | 0.810 | 0.669 |
| DTM | | 0.819 | 0.819 | 0.844 | 0.773 |
| NPMM-O | | 0.779 | 0.779 | 0.805 | 0.706 |
| NPMM-G | | 0.832 | 0.832 | 0.854 | 0.772 |
| DMM | | 0.589 | 0.589 | 0.578 | 0.424 |
| GSDMM | | 0.773 | 0.773 | 0.681 | 0.513 |
| OSGM | | **0.854** | **0.854** | 0.912 | 0.869 |
| OSGM-ES | | 0.852 | 0.852 | **0.948** | **0.913** |
| OSDM | | 0.842 | 0.842 | 0.916 | 0.864 |
| MF-G | | 0.850 | 0.849 | 0.814 | 0.661 |
| MF-O | Tweets-T | 0.823 | 0.822 | 0.780 | 0.628 |
| DTM | | 0.814 | 0.814 | 0.840 | 0.776 |
| NPMM-O | | 0.774 | 0.774 | 0.766 | 0.645 |
| NPMM-G | | 0.826 | 0.826 | 0.819 | 0.711 |
| DMM | | 0.565 | 0.565 | 0.546 | 0.410 |
| GSDMM | | 0.785 | 0.785 | 0.709 | 0.560 |

News real-world dataset, by iterating the whole stream multiple times. However, the online process of the mentioned algorithm does not outperform while comparing it with the reported results of OSGM. Likewise, MF-G also iteratively processes each batch to infer the number of clusters, however due to the term ambiguity problem it is unable to achieve high performance. Besides, the crucial part of evaluating the cluster similarity is measured by the homogeneity measure. The significant difference in results on both synthetic data streams proves that OSGM can capture topic related terms and cluster evolution simultaneously.

### 4.3.6 Runtime Analysis

To compare the running time of competing algorithms, we performed all the experiments on a system with core i5-3470 and 8GB memory. The time consumption of all algorithms on News-T dataset is shown in Figure 4–5, where we split algorithms into three plots to increase the difference visibility. The first plot reflects algorithms having higher execution time than OSGM, which are NPMM-G and NPMM-O. The reason is NPMM exploit pre-trained word embeddings to calculate posterior probability. The second plot shows moderate execution time algorithms which are MF-G and DTM. As we can clearly observe that both algorithms execution time abruptly increase as topics in stream increase, which directly effects the runtime of their iterative process for topic inference. The third plot shows algorithm having comparative a bit lower execution time than OSGM. The visible difference is due to extra computation required by OSGM to calculate term co-occurrence matrix for semantic similarity, whereas MF-O, GSDMM, and DMM lack this property. The overall speed of OSGM is more efficient compared with most existing algorithms.

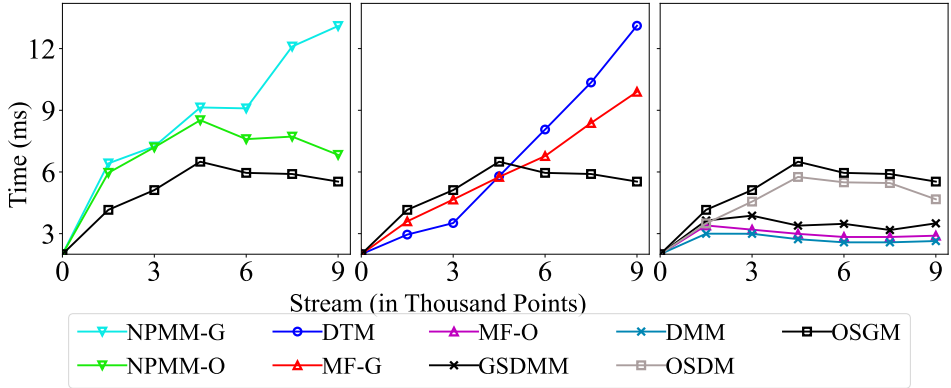

Figure 4–5 The runtime in milliseconds (ms) of different text stream clustering algorithms.

### 4.3.7 Parameter Sensitivity Analysis

We perform sensitivity analysis for OSGM with respect to four input parameters: $\alpha$, $\beta$, $\lambda$, and $\Gamma$ on the News dataset. Figure 4–6a shows the effect of $\alpha$ which ranges from $1e^{-3}$ to $1e^{-1}$. It can be observed that the performance of OSGM is quite stable over a large range of $\alpha$ values. The parameter $\alpha$ also contributes to infer the number of clusters, that is why the performance is stable. The multinomial distribution $\beta$ parameter is responsible while calculating the term distribution similarity. Figure 4–6b shows the sensitivity analysis over the same range of $\alpha$. We can observe that after a certain range, the relationship between NMI and Homogeneity becomes inverse. The reason behind this is that the increasing value of $\beta$ leads towards fine-grain clustering, which directly affects the NMI values. Our model follows the forgetting mechanism on decay factor $\lambda$ and the clusters are deleted from the model when the value approximately equals to zero. Figure 4–6c depicts the performance of OSGM on different decay factors ranging from $9e^{-7}$ to $9e^{-5}$. It can be observed that the behavior of a given evaluation measure is stable and does not show high variation. The triangular time threshold $\Gamma$ is responsible to capturing evolving features and extracting core terms. Figure 4–6d shows the sensitivity of a defined threshold ranging from 10 to 50. Interestingly, the performance is stable in terms of homogeneity. However, after a certain range it starts decreasing, particularly for NMI. This is due to the rapid disappearance of terms in clusters which lead towards highly fine-grained clustering.

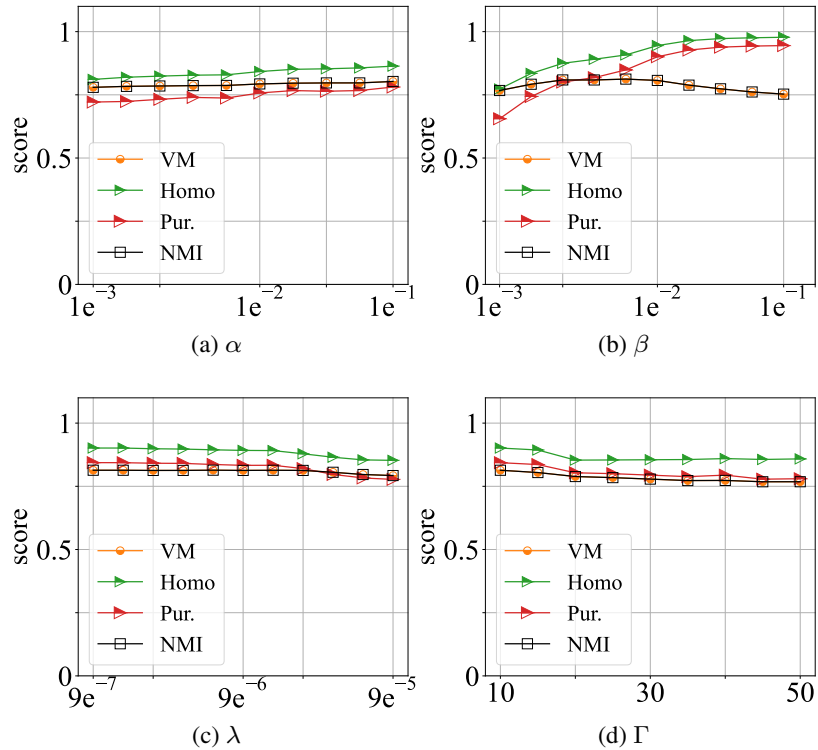

Figure 4–6 Parameter sensitivity analysis.

## 4.3.8 Semantic Smoothing Analysis

For the deep understanding of the significance of semantic smoothing, we compared our results of both variants with all state-of-the-art approaches. We can observe that both variants outperformed many state-of-the-art approaches. To explore the importance of introducing semantic smoothing, we demonstrate the parameter sensitivity analysis in terms of multinomial distribution parameter $\beta$. Figure 4–7 depicts the significant difference of semantic smoothing over a wide range of parameters of OSGM and OSGM-ES. It is clearly observable that the variation of NMI over different $\beta$ values is higher when we don't use semantic smoothing. Likewise, Figure 4–7a shows semantic smoothing leads towards a high NMI score by following coarse-grain clustering.

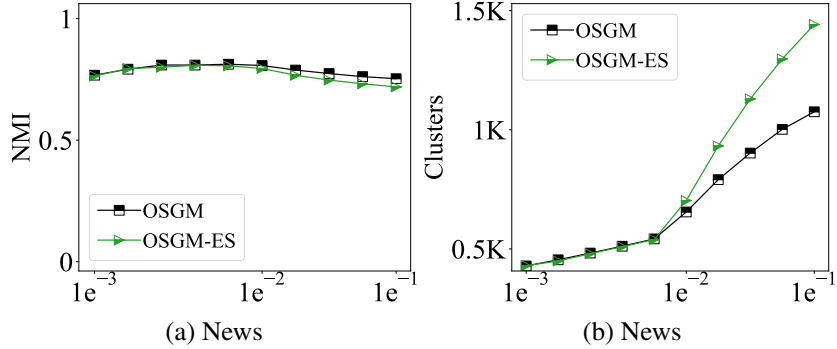

Figure 4–7 $\beta$ sensitivity of OSGM and OSGM-ES on real dataset in terms of NMI and cluster creation.

## 4.3.9 Topic Estimation Analysis

The OSGM automatically creates clusters based on calculated probability for evolving number of topics, thus it does not need number of topics as input. Figure 4–8 shows the active number of clusters and actual number of topics over time comparative to different algorithms including NPMM-O, OSDM, MStreamF-O (MF-O), and MStreamF-G. To interpret the given plots, suppose, our model has $|D|$ number of active documents at time-stamp $t$, and according to ground truth $D$ belongs to $\mathcal{L}$ topics. Consider, our model has grouped $D$ into $|M|$ number of clusters. Ideally, $|M| = |\mathcal{L}|$, therefore to evaluate here in Figure 4–8 the "*Actual*" represents the $|\mathcal{L}|$ and "OSGM" reflects the $|M|$ of the model over time. By observing the model cluster over time we can conclude that OSGM starts converging towards actual topics as clusters start merging with the passage of timestamps. However, OSDM and MF-O does not infer clusters in any way, that is the reason increase in clusters is observed continuously. In contrast, clusters of MF-G seem much closer due to batch-wise iterative process. Unlike other algorithms, NPMM uses pre-trained word embeddings (which already contains a closer distribution of words), therefore the number of clusters are much closer to actual number of clusters. Due to constant number of topics as input parameter, we do not include DDM, DTM, and GSDMM. Whereas, NPMM-G is not included as it iteratively processes whole stream multiple times to infer the number of topics.

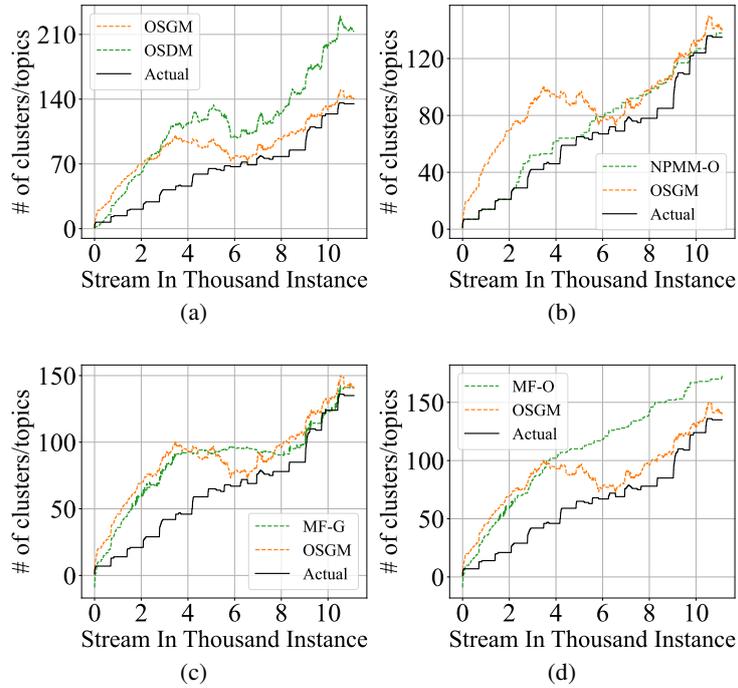

Figure 4–8 Predicted versus actual number of topics.

### 4.3.10 Statistical Tests

The series of previous state-of-the-art algorithms follows the similar non-parametric statistical model to assess the statistical significance between comparing approaches. As reported in [108], we perform Friedman rank test with 95% confidence level. We define the null hypothesis as no statistical difference among competing methods. If the null hypothesis is rejected, we further use Nemenyi post-hoc test to find these differences. The test is conducted on Table 4–3. On the standard threshold $p < 0.05$, for all average ranks of all algorithms on real and synthetic datasets, we get p-values $= 5.115e^{-15}$. This confirms that the proposed algorithm statistically differs from other methods. We apply the Nemenyi post-hoc test, which enables us to build a critical difference diagram shown in Figure 4–9. The closest critical difference score resembles with OSGM, which can deal with evolving topics but cannot deal with evolving term-subspace.

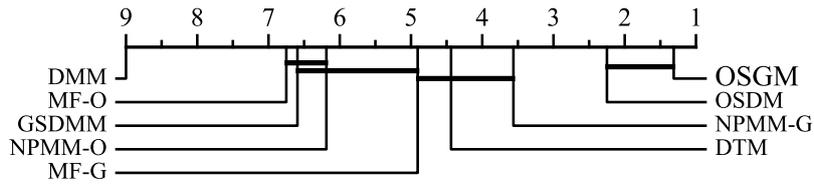

Figure 4–9 Nemenyi test on all data sets (real and synthetic) for OSGM.

### 4.3.11 Clustering Quality and Term Evolution Analysis

Here we discuss the performance of our proposed approach in terms of various evaluation measures and parameter sensitivity. To prove the superiority of OSGM, we conduct a deep analysis of clustering quality compared existing NPMM model which exploit Glove word-embedding. Figure 4–10 shows the frequency of some terms found in datasets that are ignored by previous Glove [109] word-embedding when clustering the documents. In contrast, the proposed approach does not rely on any external embedding, instead, it learns from the current stream of documents.

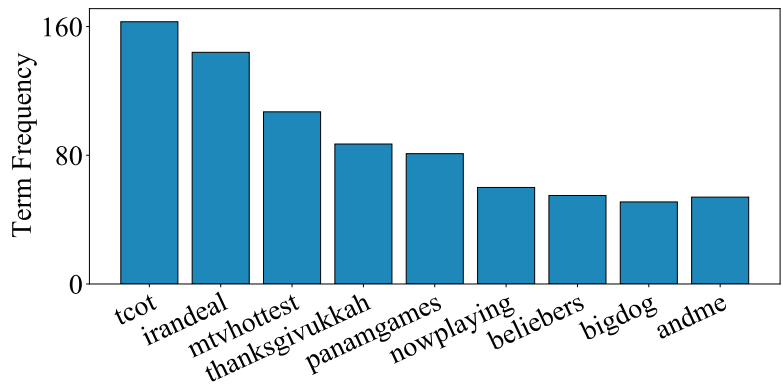

Figure 4–10 Term occurrence of few topic related terms ignored by word-embedding based approaches, which is successfully captured by OSGM.

Additionally, we choose a topic as a case study to exploit the cluster term-subspace evolution. Figure 4–4 shows the actual evolution of terms over time. We track the distribution of core-terms captured by OSGM related to the same topic at four different timestamps, as depicted in Figure 4–12. It is observed that OSGM successfully traces the term evolution and can extract the core terms.

## 4.4 Summary

In this chapter, we propose a new online semantic enhanced graphical model for evolving short text stream clustering. Compared to existing approaches, OSGM does not require to specify the batch size, the dynamic number evolving clusters, and can reduce cluster features to extract core terms automatically. It dynamically assigns each arriving document into an existing cluster or generating a new cluster based on the poly urn scheme. More importantly, OSGM incorporates semantic smoothing and term co-occurrence to deal with term ambiguity and coarse-grain clustering in the proposed graphical representation model. By exploiting the triangular time function, the proposed approach can track the change of term distribution over time. Moreover, we further investigate the importance of semantic smoothing in the proposed model. A deep conducted empirical study on synthetic and real-world datasets further demonstrates the benefits of OSGM compared to many state-of-the-art algorithms.

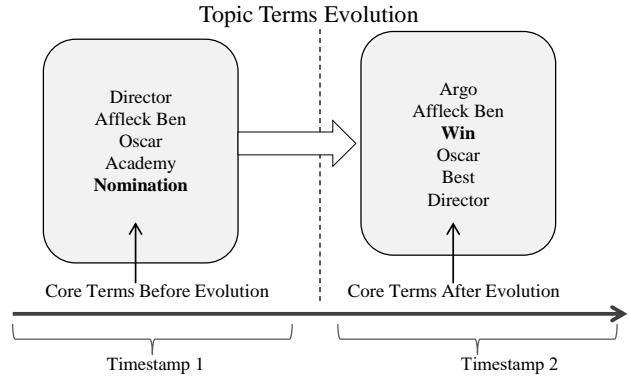

Figure 4–11 The changes of core terms for a topic.

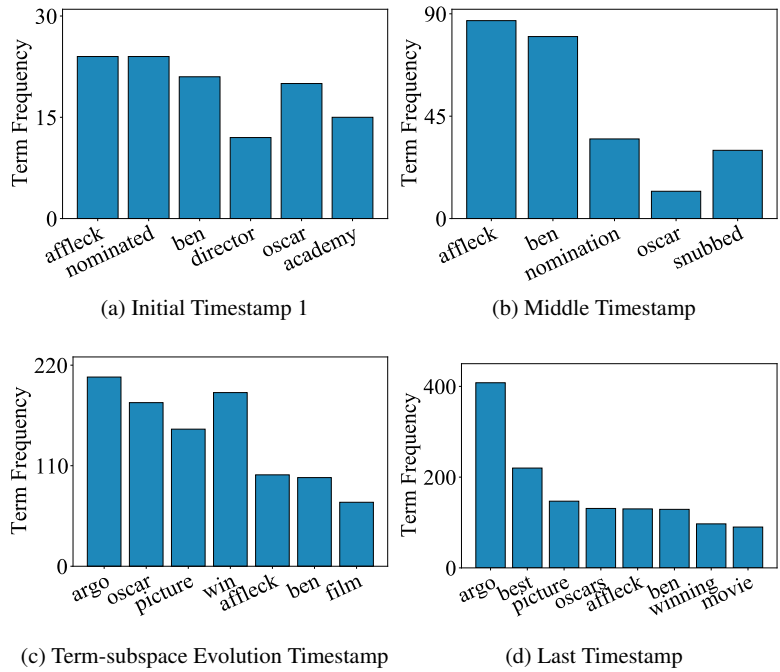

Figure 4–12 Term occurrence over different timestamps captured by OSGM.

# Chapter 5  A Semantic Co-occurrence Clustering Model with Episodic Inference for Short Text Streams

The topic and related term evolution is successfully captured by proposed model in the previous chapter. Whereas, contextual text representation is maintained by creating cluster level co-occurrence matrix. However, previous models assume relationship of every term in document with its each term to construct high dimensional co-occurrence matrix over time to capture semantic space which directly leads towards degrading cluster quality. Additionally, previous model adopted merging procedure to infer the number of cluster instead of adopting optimized inference process. This chapter highlights the importance of reduced semantic term-space and term specificity for clustering and proposes a novel episodic inference procedure for short text stream clustering task.

## 5.1 Introduction

During the past decade, huge volume of short text data has been generated on social media every day. Clustering of such continuous arriving short-length texts has gained intensive attention in recent years due to its diverse applications, such as topic tracking[96], event detection, news recommendation[95], and hot topic detection[97], etc. A series of clustering approaches have been introduced to tackle different problems during the past decade. Although each approach suffers from different limitations.

**Challenge 1: Reliable Term Weight.** Text representation of a document is an important task in natural language processing. The commonly obtained representation is based on selecting a set of suitable terms (vocabulary) and their relative importance (weighting scheme) to capture the document content[110]. For example, bag-of-words (BoW) representation and *TF-IDF* weighting scheme are widely used as term indexing and importance score, respectively[111]. However in streaming environment, such representation cannot be directly mapped since term distribution is unknown and may change over time. Recently, [42] and [107] represent streaming documents with micro-clusters containing overlapping term sub-space. However, the importance totally rely over cluster population of the model.

**Challenge 2: Reduced Dimension Semantic Representation.** Unlike long text documents, short-length documents suffers from the lack of supportive term occurrence to capture semantics[86] due to the sparse word representation. For most existing short text clustering algorithms like Dynamic cluster topic model (DCT)[39], SIF-AE[80], Sumblr[87], and MStreamF[42] exploiting independent word representation which lead towards "term ambiguity" issue[86]. The term ambiguity refers to a word which may have different meanings in different contexts. For this reason, previous works often enrich short text representations by incorporating features from external resources[102][84]. However, due to the context of terms changes over time, pre-trained word embeddings are thus not efficient in real-time streams[104]. Unlike previous external resource dependent models, recently OSDM[107] proposed a co-occurrence based semantic term space. However, due to high dimensional semantic term-space representation, not only the clustering quality is compromised but also consume more space and processing time.

**Challenge 3: Inference Procedure.** Another potentially problem of most existing approaches is the batch-wise processing. The batch-wise processing assumes that the instances are interchangeable within a batch① [97, 103]. Determining an optimal batch size is also a non-trivial task for different text streams. These algorithms also need to process each batch multiple times to infer the clusters.

## 5.2 Proposed Approach

To tackle the aforementioned problems, we propose a novel non-parametric Dirichlet model with episodic inference (EIDM) for short stream clustering. Compared to the state-of-the-art algorithms, the proposed model have (a) reduced semantic text representation by introducing window-based term co-occurrence matrix, (b) enhanced term weight by introducing word specificity and inverse document frequency, (c) can infer topics by new episodic inference procedure. The main contribution of this chapter are given as follows.

- We purpose a novel non-parametric Dirichlet model for text stream clustering with window based co-occurrence matrix capture the semantic space.
- A novel word specificity measure is introduced to compute the weight of terms
- For active cluster inference an episodic inference procedure is introduced.
- Extensive empirical analysis is conducted to prove the efficiency and robustness of the proposed model.

---

① A batch is a chunk of instances, objects or documents.

First, the overall flow of the model is described then elaborate its each component.

## 5.2.1 Model Overview

The probabilistic graphical model (PGM) is used to represent the stochastic generation of objects (short text document in our case) with Bayesian probabilistic modeling. To model the topic generation for the text stream clustering problem, the most crucial task is to define the relationship between documents and topics (clusters) in terms of a probability distribution. Each arriving document in the continuous flow of text stream, based on calculated probability, is either added in active maintained cluster by model or create a new cluster. If probability of existing cluster in the model (see Equation 5–8) is lesser than a pseudo probability (see Equation 4–8) then document is dealt as emergence of new topic thus a new cluster is created. With the arrival of documents, the model will check the old clusters (outdated topics) for deletion (see Section 4.2.2.4) or merging (see Section 4.2.2.5). In this way, the clusters of recent topics with current distribution is active in the model. To infer the number of clusters in the model, $\eta$ number of random documents from $\psi$ recent documents are resampled with the gap of $\rho$ time-unit. Before, describing each step in details, first we describe cluster feature set.

Table 5–1 Symbols and notations.

| Symbol | Definition |
|---|---|
| $D$ | total number of active documents |
| $V_d$ | unique terms / vocabulary in document $d$ |
| $V_z$ | unique terms / vocabulary of cluster $z$ |
| $V_{z \cup d}$ | vocabulary size of $V_z \cup V_d$ |
| $\mathcal{L}$ | average length of document in stream |
| $\bar{V}_z$ | average vocabulary size of active clusters ($V_z \cap V_d \neq \{\}$) |
| $N_d^w$ | occurrence of word $w$ in document $d$ |
| $N_z^w$ | occurrence of word $w$ in cluster $z$ |
| $N_d$ | total number of words in document $d$ |
| $N_z$ | total number of words in document $z$, $\sum_{d \in z} N_d$ |
| $cw_z^{w_i}$ | co-occurred neighbors of word $w_i$ in cluster $z$ |
| $m_z$ | number of documents in cluster $z$ |
| $n_z$ | number of words in cluster $z$, $\sum_{w \in z} N_z^w$ |
| $l_z$ | decay weight of cluster $z$ |
| $u_z$ | last updated time stamp of cluster $z$ |
| $\delta$ | co-occurrence window size |
| $\psi$ | buffer size |
| $\rho$ | episodic inference interval |

### 5.2.2 Cluster Feature Set

A micro-cluster in stream is represented by cluster feature set which contains different statistical values about the instances within a micro-cluster. Most similarity-based and model-based methods follow the vector space model (VSM) to represent the cluster feature space[92]. However, a topic needs to be represented in the term-changing subspaces. Here, same CF set is exploited as Chapter **3** defined as a 6-tuple $\{m_z, n_z^w, cw_z, N_z, l_z, u_z\}$. The description of each tuple is given in Table 4–1. The CF has two important addable and deletable property which allows a cluster to update incrementally over time. Addable property enables a cluster to update by adding a new document in it and defined in Section 3.3.1.

**Definition 5.2.1** A document $d$ can be removed from a cluster $z$ by using the *deletable property* to update cluster.

$$m_z = m_z - 1$$
$$n_z^w = n_z^w - N_d^w \quad \forall w \in d$$
$$cw_z = cw_z - cw_d$$
$$n_z = n_z - N_d$$

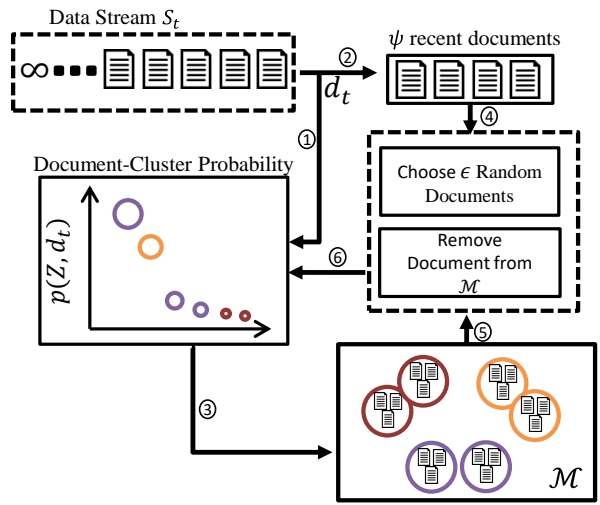

Figure 5–1 The overall flow of algorithm.

Here, $N_d$ represents the number of total words in the document. The $cw_d$ is co-occurrence matrix of document based on window size $\delta$ (see Section 5.2.3). A co-occurrence matrix $cw_d$ of a document contains the frequency ratio between two neighboring terms. Formally, the score between two $w_i$ and $w_j$ terms in a document $d$ are defined as

$$cw_{i,j}^d = \frac{N_d^{w_i}}{N_{d'}^{w_i} + N_{d'}^{w_j}}. \tag{5–1}$$

Here, $N_d^{w_i}$ is the term frequency of $w_i$ in document.

### 5.2.3 Window based Co-occurrence Matrix

The purpose of term co-occurrence is used for semantic understanding of the text which reflect the discriminative part of each term. In other words, the semantics of term is identified by accompanying terms. For static environment, $n$-gram technique is introduced however it increased the dimensionality of term space[112][113]. Thus to reduce the dimensionality many information gain scores are used to select the important $n$-gram features. Later on, text embeddings are introduced, learned by deep neural network, which transforms the actual very high dimensional term space into very low dimension[114][115][116]. However, both type of techniques requires the overall distribution of each term in the corpus. However, in streams, the distribution of each term change over time which directly effect the neighborhood of terms for semantic understanding. For example, [84] exploited static word embedding for short text clustering in stream environment, however, it is still unable to receive new terms with evolving distribution. The recent work [107] built co-occurrence matrix by considering pair of each term with every term of the same document. However, the higher number of paired co-occurrence not only increase the space complexity but also degrades the clustering quality with unimportant word co-occurrence pairs. For this reason, we introduce window-based term co-occurrence matrix, where each term can only make a pair with neighbors having $\delta$ distant while preserving sentence order. An example is depicted in Figure 5–2. In this way, a co-occurrence matrix of document $cw_d$ may at most have $O(\delta N_d)$ number of entries. Here, $N_d$ is the document length. The weight between two neighboring terms $w_i$ and $w_j$ in a cluster $z$ is defined in Equation (5–2) where $i \neq j$.

$$cw_{i,j}^z = \frac{\sum\limits_{d' \subseteq z} N_{d'}^{w_i}}{\sum\limits_{d' \subseteq z} N_{d'}^{w_i} + \sum\limits_{d' \subseteq z} N_{d'}^{w_j}} \quad \text{s.t. } (w_i, w_j) \in d' \tag{5-2}$$

### 5.2.4 Word Specificity

The weight of a term is important for extracting topic representative terms. For this reason, many information gain weight schemes have been introduced[117][118]. Previous techniques acquired the additional support of language dependent dictionary to extract the weight of term for specificity. However, social media and other short text platforms are indulged in generating new terms and phrases with event evolution. For this reason, we propose a novel weight to specify the specificity of a term. The basic idea is if a term is co-occurred with different terms every time it indicates that the term is less specific so that it can be used with variety of concepts. Whereas, high specific terms have less number of co-occurring neighbors. Additionally, the normal neighbors of a term depends on defined window size (as a model parameter), thus considering the co-occurrence window size constraint we define word specificity of $w_i$ as,

$$S_{w_i} = 1 + \left(\frac{e^{g(w_i)} - e^{-g(w_i)}}{e^{g(w_i)} - e^{-g(w_i)}}\right) + \delta \tag{5-3}$$

Here, $\delta$ is the neighboring window size, and $g(w_i)$ is defined as,

$$g(w_i) = \left(\frac{cw_{w_i}^{\mathcal{M}}}{n_{w_i}^{\mathcal{M}}}\right) \times \left(\frac{cw_{w_i}^{\mathcal{M}}}{n_{w_i}^{\mathcal{M}}} - (2\delta + 1)\right) \tag{5-4}$$

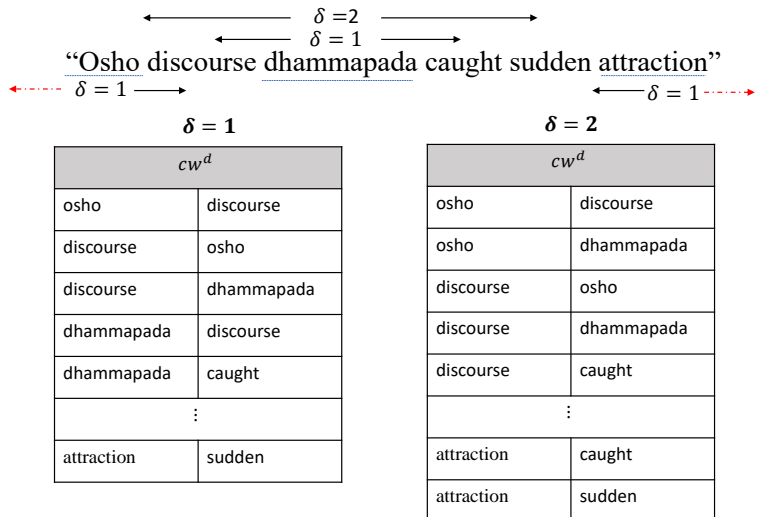

Figure 5–2 An example of word window.

$$cw_{w_i}^{\mathcal{M}} = \sum_{z \in \mathcal{M}} |cw_{w_i}^z| \tag{5-5}$$

$$n_{w_i}^{\mathcal{M}} = \sum_{z \in \mathcal{M}} n_z^{w_i} \tag{5-6}$$

Here, $cw_{w_i}^{\mathcal{M}}$ is the number of unique neighbors of term $w_i$ in the model and $n_{w_i}^{\mathcal{M}}$ is the total term frequency. The score is calculated by defining boundary of window size with hyperbolic tangent sigmoid function.

### 5.2.5 Document-Cluster Similarity

Initially the first arriving document creates its own cluster. For next arriving document, either it should be added in any active cluster of model ($\mathcal{M}$) or a new cluster is created for it. For calculating similarity between a document and an existing cluster, we transform Equation (2–5) and derive the probability of an arriving document $d$ to choose an existing cluster $z$ as we combine two probabilities (1) probability of topic popularity (defined by CRP) and (2) similarity between document and topic in terms of word distribution.

$$p(z_d|G_z) = p(G_z).p(d|G_z) \tag{5-7}$$

We derive the equation for calculating the probability of existing cluster of model, defined as,

$$p\left(z_d = z | \vec{z}, \vec{d}, \alpha, \beta\right) = \left(\frac{m_z}{D - 1 + \alpha D}\right) \cdot \left(\frac{\prod_{w \in d} \prod_{j=1}^{N_d^w} (N_z^w \cdot lCF_w \cdot S_w) + \beta + j - 1}{\prod_{i=1}^{N_d} n_z + (\beta V_{z \cup d}) + i - 1}\right) \cdot \left(1 + \sum_{\mathcal{V}_{z \cup d}} cw_{i,j}^z\right) \tag{5-8}$$

The first term of this Equation shows the cluster popularity $p(G_z)$, whereas the remaining two parts are derived to calculate homogeneity $p(d|G_z)$. The first part captures similarity in singular term space (the middle term of the Equation (4–5)) and second part captures similarity in semantic term space (co-occurrence of terms) to solve term-ambiguity problem. The middle term is based on the multinomial distribution (see Equation (2–4)) with a pseudo weight of words $\beta$ which defines the homogeneity between a cluster and a document. Previously, occurrence of a term in cluster $n_z^w$ is used while calculating the homogeneity (similarity). However, in the static environment inverse document frequency is used to calculate the term importance in global space. Here, in the dynamic environment we define similar weight measure called inverse cluster frequency $ICF_w$ to calculate term importance, defined as:

$$ICF_w = \log\left(\frac{\sum_{z \in \mathcal{M}} 1}{\sum_{z \in \mathcal{M}} 1[w \in V_z]}\right). \tag{5–9}$$

The denominator part of Equation (5–9) is the number of clusters containing the word $w$, and nominator is total number of active clusters of model. Occurrence of a term $w$ exists in more clusters, lesser the importance it has. If a term $w$ is found in very few clusters, it signifies its high importance. Here, we also calculate the word specificity $S_w$ of each term which is already defined in Equation (5–3).

### 5.2.6 Automatic Cluster Creation

Initially the first arriving document creates its own cluster. To identify new topics over time, we need to define the probability that can automatically create new cluster for the new topic if the incoming document in the stream does not relate to any existing active cluster. For automatic cluster creation, we transform Equation (2–5) and derive probability as follows.

$$p\left(z_d = z_{new} | \vec{z}_{\neg d}, \vec{d}, \alpha, \beta\right) = \left(\frac{\alpha D}{D - 1 + \alpha D}\right) \cdot \left(\frac{\prod_{w \in d} \prod_{j=1}^{N_d^w} \beta + j - 1}{\prod_{i=1}^{N_d} (\bar{V}_z \beta) + i - 1}\right) \tag{5–10}$$

Here, $\bar{V}_z$ is the average vocabulary size of active clusters. The $\alpha D$ represents the pseudo number of documents and $\beta$ helps to calculate pseudo term similarity with a new cluster. If this small probability of an arriving document is greater than the probability of existing clusters, then the model creates a new cluster containing the arrival document (see Line 10 Algorithm 5).

### 5.2.7 Episodic Inference

The inference procedure has proved useful for generative process to reduce the cluster scarcity and upgrade inverse clustering purity. Initially, GSDMM [82] extends the DMM [40] model to infer the number of clusters by integrating collapsed gibbs sampling. GSDMM split the stream into batch and collapsed gibbs sampling iteratively process each batch multiple times. Likewise, NPMM[84] also iterate the whole stream for cluster inference. However, there are two major flaws for their iterative process; (i) a batch may not capture the current distribution, (ii) it increases the processing time cost, thus not suitable for high velocity streams. In contrast, we propose episodic inference procedure. The basic idea is that the model stores $R$ number of recent documents in buffer, unlike previous approaches while process the whole batch of document, it randomly choose a small $\eta$ number of documents and resample it. The step by step procedure is given in Algorithm 5. The proposed inference procedure not only effectively decrease the processing cost but also able to cover current distribution.

**Algorithm 5:** EINDM: Episodic inference non-parametric Dirichlet model

1. $S_t : \{d_t\}_{t=1}^{\infty}$, $\alpha$ : concentration parameter, $\beta$ : pseudo weight of term in cluster, $\lambda$ : decay factor, $\psi$ : queue size, $\rho$ : episodic inference interval, $\delta$ : word window size Cluster assignments $z_d$
2. initialize $\mathcal{M} = \phi$     // empty model
3. $t_c = 0$     // model timestamp
4. $R = \phi$     // recent documents memory
5. **while** $d_t$ in $S_t$     // start streaming
6. **do**
7.     $t_c = t_c + 1$
8.     **if** $|R| == \psi$ **then**
9.         deQueue($R$)     // removing oldest document
10.     **end**
11.     enQueue($R, d_t$)     // inserting new document into queue
12.     sampleDocument($d_t, \mathcal{M}$)     // Algorithm 6
13.     $\mathcal{M} = updateActiveClusters(\lambda, \mathcal{M})$     // Algorithm 4
14.     **if** $t_c \% \rho == 0$ **then**
15.         $R_\eta$ = choose random $\eta$ number of documents from $R$
16.         **foreach** $d \in R_\eta$ **do**
17.             remove $d$ from $\mathcal{M}$ using Definition 12
18.             sampleDocument($d, \mathcal{M}$)
19.         **end**
20.     **end**
21. **end**

## 5.3 Experiments

This section provides the selection of datasets and state-of-the-art algorithm for experimental setup, results analysis and parameter sensitivity of proposed model.

**Algorithm 6:** Sample Document

1 **Function** *sampleDocument*$(d, \mathcal{M})$
2     **foreach** $z_i \in \mathcal{M}$ **do**
3         **if** $V_d \cap V_z \neq \phi$ **then**
4             $P_{Z_i} = p(z_d = z_i, d_t)$ using Equation (5–8)
5         **end**
6     **end**
7     $i = \arg\max_i (P_{Z_i})$
8     $P_{Z_n} = p(z_d = z_{new}, d_t)$ using Equation (5–10)
9     **if** $P_{Z_i} < P_{Z_n}$ **then**
        // create new cluster
10         $m_{z_n} = 1, n_{z_n}^w = N_{d_t}^w$
11         $cw_{z_n} = cw_{d_t}, n_{z_n} = N_{d_t}$
12         $l_{z_n} = 1, u_{z_n} = t_c, ta_{z_n} = ta_{d_t}$
13         $\mathcal{M} = \mathcal{M} \cup z_n$
14     **else**
15         update $z_i$ using Definition 3.3.1
16         $l_{z_i} = 1, u_{z_i} = t_c$
17     **end**
18 **return** $\mathcal{M}$
19 **End Function**

### 5.3.1 Dataset and Evaluation Measures

To evaluate the performance of the proposed algorithm, we conduct experiments on same four datasets given in Section 4.3.1. The statistics of the datasets are reported in Table 4–2.

We adopted four highly used evaluation metrics for deep analysis of all algorithms, which include Purity (Pur), Homogeneity (Homo), and Normalized Mutual Information (NMI). We utilized sklearn[①] API to compute the measures on overall clustering results[40]. The formal definition is provided for each measure in Section 2.4.1.

### 5.3.2 Baseline Algorithms

For the deep analysis, we have selected five state-of-the-art representative algorithms for text stream clustering to comparsion.

    (1) **DMM**[40] is an initial model based on multinomial Dirichlet modeling to simulate CRP.

---

① http://scikit-learn.org

(2) **DTM**[34] is an extension of Latent Dirichlet Allocation which traces the evolution of hidden topics from corpus over time. It was designed to deal with the sequential documents.

(3) **GSDMM**[40] is a Dirichlet multinomial mixture model for short text clustering, which does not consider the temporal dependency of instances. Whereas, the number of clusters is inferred with iterative Gibbs sampling procedure.

(4) **MStreamF**[42] is a recent model to deal with the infinite number of latent topics in short texts while processing one batch at a time. Two models of MStreamF were proposed, one with one-pass clustering process, and another with Gibbs sampling. We refer to the former algorithm as MStreamF-O (MF-O) and the latter as MStreamF-G (MF-G).

(5) **NPMM**[84] is the latest model to deal with the infinite number of latent topics in short texts while exploiting pre-trained Glove word-embeddings to capture semantics. Authors proposed two variations of their algorithm, one with iterative process referred as NPMM-G, and another without it which is NPMM-O.

(6) **OSGM**[119] is the latest model to deal with the infinite number of latent topics in short texts to capture evolving semantic representation.

### 5.3.3 Parameter Setting

We try to find the optimal parameter values for best results for most algorithms with grid search. Finally, we obtain best results on $\alpha = 0.01$ for DTM, $\alpha = 0.3$ and $\beta = 0.3$ for DMM. The DTM and DMM need fixed number of cluster as input therefore we set $K = 300$ and $K = 170$ for Tweets and News, respectively. For MF-O, and MF-G, we set $\alpha = 0.03$ and $\beta = 0.03$. By following [42], we set the number of iterations to 10 and *saved-batches* $= 2$ for MF-G. As defined in [84], we set $\alpha = 0.03$, $\beta = 0.03$, $\delta = 0.03$, and $\varepsilon = 3e^{-7}$ for NPMM. For OSGM, we set the same parameters defined in the paper. For EINDM, we choose $\alpha = 0.04$, $\beta = 5e^{-4}$, $\lambda = 6e^{-6}$, $\rho = 60$, $\delta = 1$, and $\psi = 500$. whereas we choose $\beta = 8e^{-4}$ for News-T and $\alpha = 5e^{-4}$ for Tweets-T dataset.

### 5.3.4 Performance Analysis

This section provides the performance comparison with state-of-the-art algorithms in terms of clustering quality and processing time complexity. Additionally, it shows the significance and robustness of proposed model with parameter sensitivity analysis.

### 5.3.4.1 Comparison with state-of-the-art approaches

Here, we compare the EINDM with chosen state-of-the-art five recent algorithms. Table 5–2 shows the overall clustering quality results on two real and two synthetic datasets in terms of NMI, Homogeneity and Purity. The given results show the superior performance of our proposed model. Compared with NPMM-G and MF-G which are batch-wise iterative processing models, our model outperformed with significant difference in all the evaluation measure. The NMI of NPMM-G for News dataset is equal to EINDM, however, NPMM-G exploit external word embeddings and require much higher processing cost (see Section 5.3.4.2) whereas due to their static distribution of terms it achieve very less score of homogeneity and purity.

### 5.3.4.2 Runtime Analysis

To investigate running time of EINDM with competing algorithms, we performed all experiments on a core i5 system with 16GB memory. Figure 5–3 shows time consumption of all algorithms on News-T dataset, where we split algorithms into three plots to increase the difference visibility. The first plot reflects algorithms comparative a bit lower execution time than proposed model. The visible difference is due to extra computation required by EINDM to calculate term co-occurrence matrix for semantic similarity and episodic inference, whereas MF-O, GSDMM, and DMM lack this property. The second plot shows moderate execution time algorithms which are MF-G and DTM. As we can clearly observe that both algorithms execution time abruptly increase as topics in stream increase, which directly effects the runtime of their iterative process for topic inference. The third plot shows algorithm having higher execution time which are NPMM-G and NPMM-O. The exploiting pre-trained word embeddings consume more computation time by NPMM-O and NPMM-G. The overall running time of EINDM is more efficient compared with most existing algorithms.

### 5.3.4.3 Parameter Sensitivity Analysis

To analyze the robustness of proposed model, we perform sensitivity parameter analysis with respect to six input parameters: $\alpha$, $\beta$, $\lambda$, $\psi$, $R$ and $\rho$ on the News dataset. Figure 5–4a shows the effect of $\alpha$ which ranges from $1e^{-3}$ to $1e^{-1}$. It can be observed that the performance of EINDM is quite stable over a large range of $\alpha$ values. The parameter $\alpha$ also contributes to infer the number of clusters, that is why the performance is stable. The slight increase of purity while increasing the value of $\alpha$ parameter cause increase in cluster population which leads towards fine-grain cluster modeling.

Table 5–2 The performance of different algorithms on four datasets in terms of different measures.

| Algorithm | Dataset | Evaluation Measures | | |
|---|---|---|---|---|
| | | NMI | Homogeneity | Purity |
| EINDM | | **0.843** | **0.926** | **0.869** |
| MF-G | | 0.780 | 0.751 | 0.653 |
| MF-O | | 0.685 | 0.654 | 0.552 |
| DTM | | 0.800 | 0.822 | 0.749 |
| NPMM-O | News | 0.809 | 0.802 | 0.710 |
| NPMM-G | | 0.843 | 0.834 | 0.745 |
| DMM | | 0.592 | 0.595 | 0.474 |
| GSDMM | | 0.821 | 0.751 | 0.605 |
| OSGM | | 0.815 | 0.893 | 0.824 |
| EINDM | | **0.844** | **0.961** | **0.932** |
| MF-G | | 0.795 | 0.738 | 0.801 |
| MF-O | | 0.746 | 0.695 | 0.529 |
| DTM | | 0.800 | 0.822 | 0.749 |
| NPMM-O | Tweets | 0.769 | 0.758 | 0.634 |
| NPMM-G | | 0.821 | 0.813 | 0.696 |
| DMM | | 0.392 | 0.549 | 0.141 |
| GSDMM | | 0.798 | 0.728 | 0.581 |
| OSGM | | 0.822 | 0.879 | 0.819 |
| EINDM | | **0.871** | **0.948** | **0.922** |
| MF-G | | 0.848 | 0.829 | 0.705 |
| MF-O | | 0.833 | 0.810 | 0.669 |
| DTM | | 0.819 | 0.844 | 0.773 |
| NPMM-O | News-T | 0.779 | 0.805 | 0.706 |
| NPMM-G | | 0.832 | 0.854 | 0.772 |
| DMM | | 0.589 | 0.578 | 0.424 |
| GSDMM | | 0.773 | 0.681 | 0.513 |
| OSGM | | 0.854 | 0.919 | 0.878 |
| EINDM | | **0.871** | **0.959** | **0.935** |
| MF-G | | 0.850 | 0.814 | 0.661 |
| MF-O | | 0.823 | 0.780 | 0.628 |
| DTM | | 0.814 | 0.840 | 0.776 |
| NPMM-O | Tweets-T | 0.774 | 0.766 | 0.645 |
| NPMM-G | | 0.826 | 0.819 | 0.711 |
| DMM | | 0.565 | 0.546 | 0.410 |
| GSDMM | | 0.785 | 0.709 | 0.560 |
| OSGM | | 0.854 | 0.912 | 0.869 |

The multinomial distribution $\beta$ parameter is responsible while calculating the term distribution similarity. Figure 5–4b shows the sensitivity analysis over the same range of $\alpha$. We can observe that after a certain range, the relationship between NMI and Homogeneity becomes inverse. The reason behind this is that the increasing value of $\beta$ also increase the threshold of homogeneity of a document to be added with existing clusters, which directly affects the NMI values.

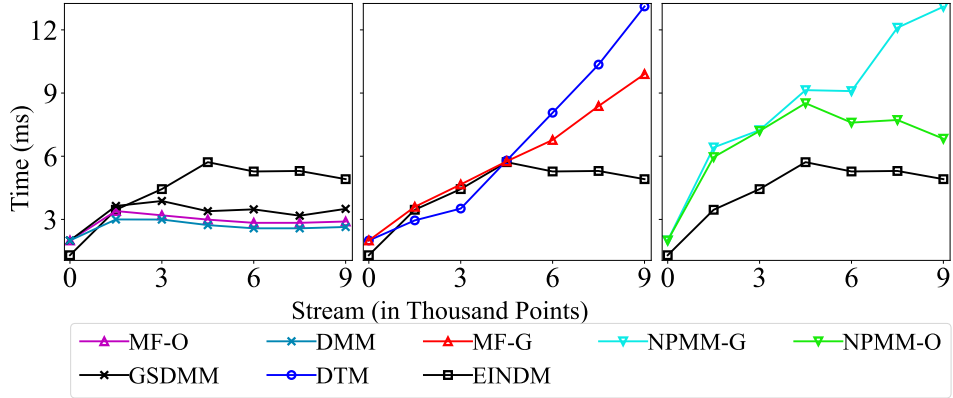

Figure 5–3 The runtime of selected algorithms in milliseconds (ms) on News-T dataset.

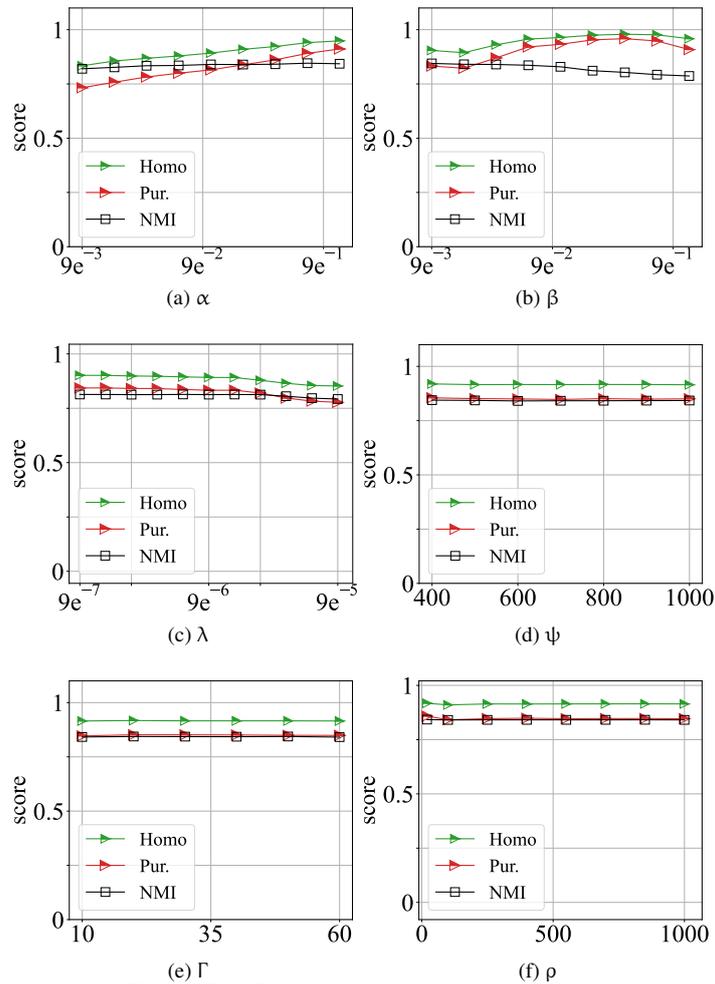

Figure 5–4 Parameter sensitivity analysis.

Our model follows the forgetting mechanism on decay factor $\lambda$ and the clusters are deleted from the model when $l_z$ approximately equals to zero. Figure 5–4c depicts the performance of EINDM on different decay factors ranging from $9e^{-7}$ to $9e-5$. More larger values enforce the model to delete clusters very rapidly as we assumed the stream time as instance population. For given range of parameter values, it can be observed that the behavior of a given evaluation measure is stable and does not show high variation.

The proposed episodic inference is dependent on three parameters including $\psi$, $R$ and $\rho$. The parameter $\psi$ is responsible to store population of documents related to current concept and we select the range from 400 to 1000. Whereas, the inference procedure starts after $\rho$ interval, for which we select range from 10 to 1000. The range of both parameters are selected by observing the size of dataset for analysis. After each interval $R$ number of samples from buffer are selected to infer the model thus smaller the number lesser the time complexity. Therefore we select range from 10 to 60 for $R$. Figure 5–4d, Figure 5–4e and Figure 5–4f show the performance on given range for these three parameters. We can clearly observe the stable performance for each of them.

## 5.4 Summary

In this chapter, we propose non-parametric Dirichlet model (EINDM) for evolving short text stream clustering. Compared to existing approaches, EINDM does not require to specify number of topics as it automatically detect new topics. It dynamically assigns each arriving document into an existing cluster or generating a new cluster based on the poly urn scheme. More importantly, EINDM incorporates word specificity as term importance and window-based term co-occurrence to deal with term ambiguity and context in the proposed graphical representation model. By exploiting the exponential decay function, the model deletes the outdated clusters from model as well as maintain active clusters. Moreover, we further introduce episodic inference procedure to infer the number of clusters. A deep conducted empirical study on synthetic and real-world datasets further demonstrates the benefits of EINDM compared to many state-of-the-art algorithms.

# Chapter 6  Semi-supervised Classification on Multi-label Evolving Text Streams

Previous chapter proposes an optimized Dirichlet model to cluster the text in streams. However, adopting such clustering approach is challenging while solving the multi-label classification problem where a document may associate with more than one topic. Additionally, in real time scenario, few number of labeled instances are available to learn classification model. This chapter gives a brief discussion about multi-label text data in streams with problem of label scarcity, label cardinality and proposes a new semi-supervised learning approach to classify text data.

## 6.1 Introduction

A massive amount of text data is continuously generated by many online sources over time. Classification of such streams has gained increasing research interest due to a wide scope of applications such as, information retrieval systems, news recommendation, rumor detection and many more. In the classic machine learning task, each document is associated with only single label. However, in many real world scenarios this assumption do not hold where each document can be associated with many mutually non-exclusive labels. For example, a news article may relate to *politics*, *election* and *government* themes simultaneously. The efficient and robust analysis of multi-label documents in data streams has become a complex task due to different types of challenges related to multi-label learning and concept evolution (drift)[21][120][121]. Additionally, representation of text for such machine learning algorithms is itself a non-trivial task while considering evolving stream environment[122].

**Challenge 1: High Dimensionality.** A stream is a sequence of infinite number of arriving documents where a document is represented by sequence of ordered terms. During the past decade, many multi-label learning models have been proposed to address the classification task from different perspective. Initially, classic multi-label algorithms were transformed for streams such as ML-HoeffdingTree[53], iSOUP-Tree[54], and ML-KNN[55], to mention a few. However, by considering global space for learning, these approaches are unable to deal with high dimensional data. For this reason, few research works focused on reducing the feature space to fit into the model (e.g., [56] exploit count-min sketch for dimension reduction in multi-label text streams). Also, later studies analyzed that these approaches still exploit global feature space whereas tracking label related features is important to improve classification performance. For this reason a variety of approaches such as ProSVM[57], FRS-LIFT[58], S-CLS[59], and ML-FSL[11] have been proposed to consider feature subspace. However, the main drawback of previous approaches is that they consider fixed number of dimensions (i.e., term space). Therefore, these techniques cannot be directly applicable in real-world scenarios where the number of dimensions may change over time.

**Challenge 2: Evolving Text Representation.** Text representation of a document is an important task in natural language processing. The commonly obtained representation is based on selecting set of suitable terms (vocabulary) and their relative importance (weighting scheme) to capture the document content[110]. For example, bag-of-words (BoW) representation and *TF-IDF* weighting scheme are widely used as term indexing and importance score, respectively[111]. However in streaming environment, such representation cannot be directly mapped since term distribution is unknown and may change over time. Recently, [42] and [107] represent streaming documents with micro-clusters containing overlapping term sub-space. However, their scope is limited for single labeled and short-length documents.

**Challenge 3: Label Scarity.** In the real-world scenarios, manually labeling all the data is impractical due to unbound volume of stream data arriving at high velocity[105]. In this case, the previous series of supervised algorithms are not feasible as they assume that labels are instantly available after prediction such as MLSAMPkNN[123], OZBAML[20], and Clasher[56], to mention a few. Recently, [124] and [73] is able to handle unlabeled data to update the model in stream. However, their technique does not deal with above mentioned problems instead solely deal with low dimensional multi-label data [53].

**Challenge 4: Online vs Batch processing.** Most previous state-of-the-art algorithms split the continuous sequence of documents into batch① while updating the classification model such as SSkC[72] and MLTM[125]. In the batch-wise processing, it is assumed that there exist no concept drift inside a single batch which directly leads to the problem of approximating the best size of a batch. Additionally, batch way processing

---

① chunk, batch or window are fixed number of instances either based on time or frequency scale.

lacks the capability of model to predict at any time as model needs to wait until a specific number of instances arrive to fill the batch-size.

**Challenge 5: Label Cardinality.** Apart from these issues, previous streaming algorithms predict static number of labels by exploiting known label cardinality [①][124][126]. Whereas, in stream environment the label cardinality is unknown. To solve this issue, recently [127] proposed a technique to predict $h$ number of labels for each instance in a batch, where the $h$ is approximated using hoeffding inequality. However, in real time data, the number of labels associated with an instance may differ from next instance in the same batch[128].

## 6.2 Related Work

The research work in this chapter is related to the task of multi-label learning and clustering unlabeled documents in streams. For this, first, we briefly reviewed recently proposed multi-label learning algorithms for streams. For the clustering task, our proposed non-parametric Dirichlet model is followed through a series of probabilistic models, thererby we describe evolution of these models to deal with streaming documents.

### 6.2.1 Multi-label Stream Classification

In static multi-label learning (ML), we have a set of $N$ documents $D = \{d_1, \ldots, d_N\}$ and a set of $M$ labels $\mathcal{L} = \{l_1, \ldots, l_M\}$ where each document $d_t$ is set of $N_d$ words, i.e., $\{w_1, \ldots, w_{N_d}\}$ and associated with one or multiple labels $(d_i \to L_{di})$, $L_{di} \subseteq \mathcal{L}$. In the literature[79], ML classification problem is solved through two methodologies, (i) problem transformation and (ii) algorithm adaptation. The former technique transforms the data to fit into single label classifier, such as 'label powerset' where subsets of labels are treated as individual class. Whereas, the latter approach transforms the algorithm for multi-label data such as ML-KNN [52]. Working on these methodologies over the past decade[129], a series of problems have been highlighted specifically for text documents of different domains which include label dependency [130], imbalance label distribution [131], label-feature dependency [59], and high dimensionality [132].

Unlike static environment, the streams have infinite sequence of documents arriving over time $D = \{d_t | t = 1, \ldots, \infty\}$. A concept drift in streams refers to change in

---

① Cardinality is the average number of labels per instance.

any kind of distribution over time which directly effect the decision boundaries created by the classifiers. The objective of classification task is to optimize the decision boundary over time for predicting the *L* for unknown documents. Unlike static environment, change in label distribution $p(L^{(i)})_t \neq p(L^{(i)})_{t+\Delta t}$, change in features $W_t \neq W_{t+\Delta t}$, change in label related features $p(W|L^{(i)})_t \neq p(W|L^{(i)})_{t+\Delta t}$, change in label dependency $p(L^{(i)}|L^{(j)})_t \neq p(L^{(i)}|L^{(j)})_{t+\Delta t}$, and change in feature dependency $p(w_i|w_j)_t \neq p(w_i|w_j)_{t+\Delta t}$ may occur over time *t* to time $t + \Delta t$. Here, *W* represents the feature space and $w_i \in W$. The challenging constraints while designing algorithm for stream are the limited processing time and memory. Over the past few years, a series of work on multi-label classification for real-valued data have been done. However, multi-label learning for text streams is less studied.

Among the initial approaches of algorithm adaptation, [53] proposed multi-label hoeffding tree for batch-wise processing of instances to detect distributional change in feature value. It tracks the log likelihood loss to change the tree nodes and leaves. However, the limitation lie towards the low dimensional applicability. In this regard, [120] introduced ensemble approach where micro-clusters are created for each data chunk to train the classifiers. Each classifier is updated after prediction of each batch. Within the same scope of fixed number of features to train the model, [56] proposed overlapping feature-space (clashing) based approach for text data. Using count-min sketch, it reduces the dimensionality for the model to detect change. Likewise, [122] tracks the change in topic distribution where each document is represented with pre-defined number of dimension. Dealing with evolving number of features is important while dealing with multi-label text in streams. In this regard, [133] proposed shelly neighborhood based model which calculates the redundancy score to adjust the number features. Recently, [124] proposed a label compression based technique to capture the high order label dependency, however, the constraint of pre-defined number of dimensions declare its non-applicability for evolving text data. To solve both issues together, [134] consider the first order label correlation for label prediction and adopt increasing feature by observing the fuzzy mutual information between batches of instances. In the same context, [135] proposed model based on evolving decision rules using Page-Hinkley, and it can preserve label dependency. However, its limitation tilt towards active window size and adoption of evolving features for text data.

In summary, for directly applying the approaches of real-value data for text data, pre-vious approaches do not consider evolving number of features, high dimensionality, and online processing together. Whereas, specifically for the text data, contextual dependency (dependency of a term over another term) is also not considered in these approaches.

## 6.2.2 Semi-supervised Multi-Label Text Classification

Apart from these issues, the classification for text data includes various problems such as term-weighting [56][111] and semantic-similarily [110]. Due to space limitation, we only report the closest approaches on mentioned problems.

Initially, different optimized variants of $k$-NN for multi-label data were proposed[136][55], however, including the high sensitivity of $k$ they also suffer from curse of dimensionality issue. To reduce the sensitivity of $k$, [55] proposed shelly structure based ML-KNN where negative and positive distances were considered with respect to each attribute. For the optimal neighbors selection, the ratio between minor and major occurring labels is also considered. Whereas, to deal with the high dimensionality problem, [59] proposed Laplacian score based feature selection method. Likewise, [132] proposed an ensemble learning based approach which chooses the different sub-space of dimensions to train base-classifiers. Then meta learning is used to select effective ones. Unlike previous fully supervised models, [137] proposed a semi-supervised approach which exploit different feature subsets for training the ensemble classifiers. In parallel, [58] implement label-specific feature reduction with fuzzy rough set which also find the relationship between features and labels. However, relationship among the labels is not considered in these approaches which we refer as the label dependency issue. To solve this problem, [125] and [138] proposed probability-based explicit and implicit relationship among labels while training the model.

## 6.3 Proposed Model

In literature, majority of research works on multi-label data do not consider streaming environment [138][132][72][55][130], and only a few works have contributed for multi-label streams [122][139][56]. Specifically, text data is very less studied in streaming environment[140]. In this regard, most of the proposed works are either fully supervised or consider static number of dimensions over streams. Whereas in real-time scenarios, these supposition do not hold. To address these issues, we propose an online semi-supervised classification model (OSMTC) to predict in online way for multi-label text streams. To work under semi-supervised setting, we propose a non-parametric Dirichlet model which is able to construct document clusters for streaming environment. Unlike previous Dirichlet models[42][105], our designed probabilistic model is insensitive to the size of document[①] and maintain semantic term space by storing cluster-level term co-occurrence matrix. For representing evolving term indexing, we propose a novel label specific micro-cluster representation (see Section 6.3) for multi-label documents. The defined addable-property (see Definition 6.3.1) can update model incrementally with evolving term-space over time. Specifically for online label specific feature reduction, we map triangular time decay factor over cluster features. Interestingly, unlike previous approaches where constant $h$ labels are predicted for each instance, we approximate $h$ (see Algorithm 8 Line 19) for each instance based on current label distribution of the nearest clusters. The advantages of proposed approach transcend over the limitations of previous approaches, which are: (i) it works with instance by instance processing instead of batch-wise processing; (ii) unlike fully supervised model, it works for limited number of labeled samples; (iii) it considers first-order label dependency for prediction; (iv) it is robust to long and short documents, and (v) it can capture semantic similarity for label-specific terms. The main contributions of this research can be summarized as follows.

- We propose an online semi-supervised classification algorithm, namely OSMTC, for multi-label text streams by introducing a novel non-parametric Dirichlet model.
- OSMTC introduces an evolving micro-cluster representation, which allows dynamically maintaining multiple subspaces of terms for each label.
- OSMTC is capable of handling both the gradual and sudden concept drifts in term space by employing the triangular time function and chinese restaurant process based on Dirichlet process.
- To capture the first-order label dependency, OSMTC embeds label co-occurrence probability with cluster similarity for next label prediction, which supports high prediction performance.

---

① In literature, previous graphic models based on dirchlet process either proposed to cluster the short text or for long text document

### 6.3.1 Model Overview

The proposed OSMTC model can be divided into three phases, 1) model initialization, 2) online prediction, and 3) model maintenance. Figure 6–1 shows the overall flow of OSMTC. In the first step, we take a small number of labeled documents ($D_{init}$) and create few micro-clusters ($Z_{min}$) for each label containing these documents. Now, the initial model is ready to predict for each arriving document in stream. For each arriving document the model calculates the cluster-document probability (using Equation 6–3). After that based on probability score, $k$ nearest micro-clusters are selected ($\mathcal{Z}_d$). At this stage, the number of labels to be predicted ($Y$) is equal to the number of micro-clusters ($l_{count}$) having probability greater than the mean of $\mathcal{Z}_d$ distribution. If the number of clusters greater than mean is one then we simply predict the label having high cluster population in nearest clusters. Otherwise, we compare the sum of cluster probabilities for each label and consider label co-occurrence score of nearest labels for prediction. After predicting the label(s), if the arriving document is not labeled then we simply add it to nearest micro-cluster of each predicted label. Otherwise, we add the arrived document into nearest micro-cluster of ground truth labels, whereas, a wrong prediction penalty, the score of common terms ($\mathcal{V}_{d \cap z}$) between each wrong predicted label-related cluster and document is slightly reduced. Before detailed discussion of each phase, we first describe the *micro-cluster*.

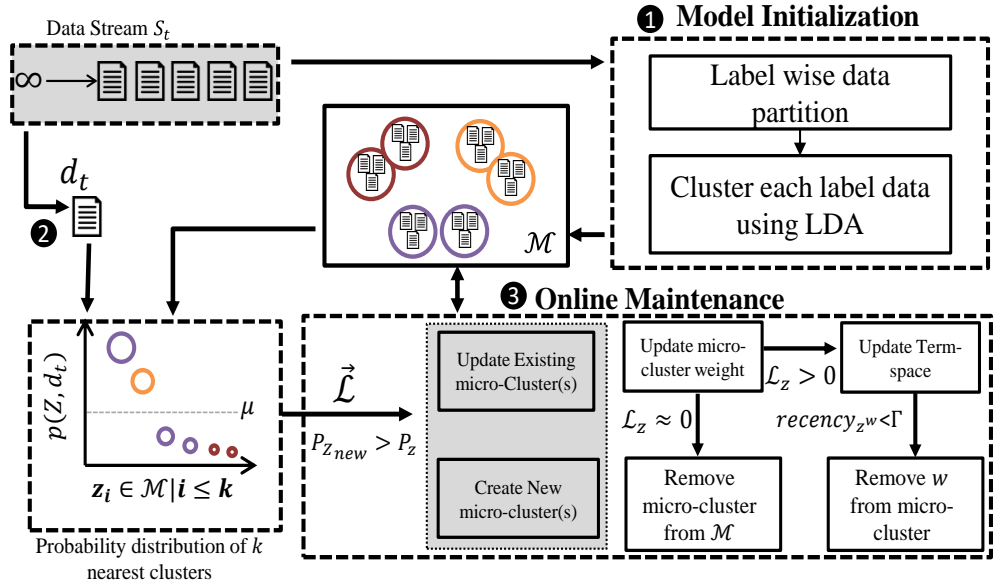

Figure 6–1 The overall flow of the proposed algorithm.

**Micro-cluster.** A micro-cluster (in stream) is represented by cluster feature set which contains different statistical values about the instances within it. In OSMTC the feature set of a micro-cluster $z$ is defined as a 8-tuple $\{m_z, n_z^w, n_z, cw_z, \mathcal{L}_z, r_z, u_z, ta_z\}$. Here, $m_z$ is the number of documents, $n_z^w$ represents term frequency of word $w$, $n_z$ is sum of frequency count of all words in cluster $\sum_{w \in z} n_z^w$, $\mathcal{L}_z$ stores assigned label of micro-cluster, $r_z$ contains the decay weight, $u_z$ is the last updated timestamp, and $ta_z$ is arriving timestamps of the words. The $cw_z$ is word to word co-occurrence score matrix and each entry is defined as follows.

Table 6–1 Definition of major notations and functions.

| Symbol | Definition |
| --- | --- |
| $S_t$ | stream of documents, i.e., $\{d_1, d_2, \dots\}$ |
| $N_d$ | total number of words in document $d$ |
| $V_d$ | unique terms of document $d$ |
| $Y_d$ | label associated with document $d$, $Y_d \subset \mathcal{L}$ |
| $\bar{N}_D$ | average length of documents in $S_t$ |
| $z$ | a 8-tuple micro-cluster |
| $\mathcal{M}$ | represent model containing set of active micro-clusters, i.e., $\{z_1, z_2, \dots, z_A\}$ |
| $Z_d$ | assigned micro-cluster of $d$ |
| $LC_{\mathcal{M}}$ | label co-occurrence weight matrix of model $LC_{\mathcal{M}} \in \mathbb{R}^{M \times M}$ |
| $Z^l$ | $\{z \in \mathcal{M} | \mathcal{L}_z = l, l \in \mathcal{L}\}$ |
| $X^l$ | a set of documents having label $l$ |
| $V_z$ | vocabulary of micro-cluster $z$ |
| $\bar{V}_Z$ | average vocabulary size of active clusters where $(V_z \cap V_d \neq \{\})$ |
| $\mathcal{V}_{z \cup d}$ | $\{w | w \in V_d \text{ or } w \in V_z\}$ |
| $\mathcal{V}_{d \notin z}$ | $\{w | w \in V_d \text{ and } w \notin V_z\}$ |
| $\mathcal{V}_{d \cap z}$ | $\{w | w \in V_d \text{ and } w \in V_z\}$ |
| $\mathcal{V}_{z \notin d}$ | $\{w | w \notin V_d \text{ and } w \in V_z\}$ |
| $\mathcal{Z}_d$ | $\mathcal{Z}_d : K \to V$ where $K \subseteq \mathcal{M}$ and $V = \{p(Z_d = K_i | d_t) | i = 1, \dots |K|\}$ |
| $\mathcal{Z}_d^l$ | $\{(K_i \to V_i) | \forall (K_i \to V_i) \in \mathcal{Z}_d, \mathcal{L}_{k_i} = l\}$ |
| $k\text{-}NC(k, \mathcal{Z}_d)$ | $\{f : K_i \to V_i | V_i > V_j, j \neq i, i = 1, \dots, k\}$, $k$ nearest micro-clusters |
| $\mu(\mathcal{Z}_d)$ | returns mean value of V in $\mathcal{Z}_d$ |
| $count\_g(\mathcal{Z}_d)$ | $|\{(K_i \to V_i) | \forall (K_i \to V_i) \in \mathcal{Z}_d, V_i > \mu(\mathcal{Z}_d)\}|$ |
| $extract(\mathcal{Z}_d, l)$ | returns a value pair $(p, q)$ where $q = \mathcal{Z}_d^l$ and $p = |q|$ |
| $extr\_mc(E)$ | $\{\mathcal{L}_k | \mathcal{L}_k \in \mathcal{L}, p = \max(p)\}$ where $E : \mathcal{L} \to O, O : p \to q$ |

$$cw_{w_i, w_j} = \sum_{d' \in z} \frac{n_{d'}^{w_i}}{n_{d'}^{w_i} + n_{d'}^{w_j}} \quad \text{s.t. } (w_i, w_j) \in d' \tag{6-1}$$

Here, $n_{d'}^w$ is frequency count of word $w$ in document $d'$. The ratio between $w_i$ and $w_j$ in a document must satisfy $cw_{w_i, w_j} + cw_{w_j, w_i} = 1$, where $i \neq j$. A document $d$ can be added into a micro-cluster as follows.

**Definition 6.3.1** A document $d$ can be added to a cluster $z$ by using the *addable*

*property* to update cluster.

$$m_z = m_z + 1$$
$$n_z^w = n_z^w + N_d^w \qquad \forall w \in d$$
$$cw_z = cw_z \cup cw_d$$
$$n_z = n_z + N_d$$
$$ta_z = ta_z \uplus ta_d \qquad \forall w \in d$$

Here, $N_d^w$ is frequency count of each word in document, $cw_d$ represents co-occurrence matrix of document, $\uplus ta_d$ merges the arrival time of document related terms, and $N_d$ is the length of document. The current time-stamp of the model is used to set $u_z$ when updated. The complexity of updating a cluster by adding a document is $O(\bar{n}_d^2)$, where $\bar{n}_d$ is the average length of the documents.

---

**Algorithm 7:** Initialize Learning Model

1   $S_t$ : Stream, $D_c$: number of instances for each label, $Z_{min}$: number of clusters $\mathcal{M}$ : Model

2   **Function** *InitializeModel*($D_c, S_t, Z_{min}$)
3     $\mathcal{M} = \phi$        // Empty model
4     $X_{train} = \phi$
5     **foreach** $l \in \mathcal{L}$ **do**
6        $X^l = \{d_i | d_i \in S_t \text{ and } d_i^l = l, i = 1,...,D_c\}$
7        $\mathcal{Z} = \text{LDA}(X^l, Z_{min})$        // partition instances using LDA
8        **foreach** $z \in \mathcal{Z}$ **do**
           // create new cluster
9           $Z = (m_z, n_z, n_z^w, cw_z, ta_z, \mathcal{L}_z = l)$
10          $\mathcal{M} = \mathcal{M} \cup Z$
11        **end**
12        $X_{train} = X_{train} \cup X^l$
13     **end**
14     $LC_\mathcal{M} = label\_cooccurence(X_{train})$ using Equation (6–2)
15     return $\mathcal{M}, LC_\mathcal{M}$

---

## 6.3.2 Model Initialization

The OSMTC model starts with initializing the learning model which takes $D_{init}$ number of labeled instances. First for each label $l \in \mathcal{L}$, we make a equal sized set of instances $S_i$ from given $D_{init}$. Formally, we can define as $S_{init} = \{S_1, \ldots, S_{|\mathcal{L}|}\}$ where $S_i$ may or may not be mutually exclusive because of multi-label data. We then split instances of each label $S_i$ into $Z_{min}$ number of partitions (clusters) using LDA [81]①. Unlike many clustering approaches for stream data where distance based k-means clustering is adopted to create

---

① We use LDA to make the process simple, whereas a variety of topic models can be applied for initialization process.

micro-clusters[92], LDA can effectively work on latent subspaces of text data. Algorithm 7 shows the step-wise procedure for model initialization. In this followed setting, each micro-cluster[①] can have only one assigned label, therefore the model can have maximum $|L| \times Z_{min}$ miro-clusters in initialization phase. Additionally, we compute first-level label co-occurrence weight matrix $LC_\mathcal{M}$ where each entry is calculated using heuristic probability between label $l_i$ and $l_j$ defined as

$$LC_{l_i,l_j} = P(l_i \cap l_j) = \frac{count(l_i \cap l_j)}{count(l_i)}. \tag{6–2}$$

After initializing the clusters and label co-occurrence score matrix, classification can be stated for incoming documents.

### 6.3.3 Classification Phase

The classification phase has two steps: (1) calculating the similarity score for each incoming instance in stream with all active clusters of the model and (2) predicting the labels by observing the *k* nearest clusters. We describe each step in detail as follows.

**Similarity calculation.** In the past decade, a series of algorithms have been proposed with either distance-based or probability-based similarity (between coming instance and existing clusters) to predict the labels[52]. The former type of measures suffer from curse of dimensionality where the model is represented on global feature space[56][120][127]. Whereas in latter approach, designing the probability-based similarity is still a challenging task where micro-clusters are represented into overlapping feature subspaces. For this reason most probability-based approaches dealt with either short text (instances only having low-dimensional representation) or long text (instances only having high-dimensional representation)[38]. Recently, [107] proposed multinomial distribution based probability to calculate instance-cluster similarity for short text document which also combines semantic similarity. However, their approach is confined for short text and single label documents. As OSMTC contains micro-clusters having different overlapping feature subspaces, therefore, to deal with short as well as long multi-label text document, we extend [107] and propose a novel probability score defined as follows.

---

① Micro-cluster and cluster will be interchangebly used in this chapter.

$$p\left(Z_d = Z | \vec{Z}, \vec{d}, \alpha, \beta\right) = \left(\frac{m_z}{D-1+\alpha D}\right) \cdot \left(\frac{\left(\prod_{w \in \mathcal{V}_{d \cap z}} \prod_{j=1}^{N_d^w} (n_z^w \cdot ICF_w) + \beta + j\right)\left(\frac{\mathcal{V}_{d \not\subset z}}{|V_d|}\right)}{\sum_{i=1}^{w \in \mathcal{V}_{z \not\subset d}} n_z^w + (\beta \mathcal{V}_{z \cup d}) + i}\right) \cdot \left(1 + \sum_{(w_i, w_j) \in \mathcal{V}_{d \cap z}} cw_{w_i, w_j}\right)$$

(6–3)

Here, $D$ is total number of active documents in the model $\mathcal{M}$. The $ICF_w$ is inverse cluster frequency to calculate the importance of weight, defined as

$$ICF_w = log\left(\frac{A}{|\{z \in \mathcal{M} | w \in V_z\}|}\right).$$

The definition of remaining terms are given in Table 6–1. The Equation (6–3) contains three prominent parts. The first part represents the cluster popularity which is the portion of cluster related documents out of active documents[①] of the model. The second part calculates similarity while considering single term space. In multi-label text document, a distribution of term sub-space represents the specific label. Therefore, for calculating the similarity we need to consider term length of the document $N_d^w$, term count of micro-cluster $n_z^w$, common term-space between micro-cluster and document $\mathcal{V}_{z \cup d}$, and non-overlapping term-space $\mathcal{V}_{d \not\subset z}$ and $\mathcal{V}_{z \not\subset d}$, respectively. Under the supposition of multinomial term distribution, $\beta$ is the given parameter. The third part captures the semantic space by analyzing co-occurred terms in the document. Each micro-cluster maintain a term co-occurrence matrix (see Equation (6–1)) through cluster related documents.

**Label Assignment.** After calculating the similarity of an arrived document with all active clusters of model using Equation (6–3), to predict the labels we then select $k$ nearest clusters having high probability score. Due to the unknown cardinality under the stream environment, (i) number of labels to be predicted and (ii) considering label dependency are still among non-trivial tasks for multi-label classification. Many previous lazy learning approaches, such as ML-KNN[52], do not consider these perspectives together using similarity score. OSMTC solves both problems by observing probability distribution of $k$ nearest clusters while considering basic label co-occurrence probability (using Equation (6–2)). The number of predicted labels is the number of clusters above mean of normalized probability distribution of $k$ nearest clusters. We already compute

---

① Active documents represents the current concept of model

**Algorithm 8:** OSMTC

1   $S_t : \{d_t\}_{t=1}^{\infty}$, $D_{init}$, $\alpha$, $\beta$, $\lambda$, $\Gamma$, $k$, $Z_{min}$ Label assignments: $Labs \subset \mathcal{L}$
2   $D_c = D_{init}/|\mathcal{L}|$
3   $\mathcal{M}, LC_{\mathcal{M}} = InitializeModel(D_c, S_t, Z_{min})$     // Alg. 7
4   $t_{\mathcal{M}} = 0$     // model time-stamp
5   **while** $d$ in $S_t$ **do** // start streaming
6      $t_{\mathcal{M}} = t_{\mathcal{M}} + 1$
7      $updateMCWeight(\lambda, Z_{min}, \mathcal{M})$     // Alg. 10
8      **foreach** $z \in \mathcal{M}$ **do**
9         remove outdated terms of $z$ using Equation (6–7)
10        **if** $V_d \cap V_z \neq \phi$ **then**
11           $P_z = p(z|d_t)$ using Equation (6–3)
12        **end**
13      **end**
14      $\mathcal{Z}_d = k\text{-}NC(k, P_Z)$     // Table 6–1
15      $l_{count} = count\_g(\mathcal{Z}_d)$     // Table 6–1
16      **foreach** $l \in \mathcal{L}$ **do**
17         $C^l = extract(\mathcal{Z}_d, l)$     // Table 6–1
18      **end**
19      $Y = extr\_mc(C^{\mathcal{L}})$     // Table 6–1
20      **if** $|Y| > 1$ **then**
21         **foreach** $l \in Y$ **do**
22           $P^l_{\mathcal{Z}} = \sum_{z \in \mathcal{Z}_d} LC_{l, \mathcal{L}_z}$     // Equation (6–2)
23         **end**
24         $Y' = \{\mathcal{L}_z | \forall z \in k\text{-}NC(l_{count}, P^Y_{\mathcal{Z}})\}$
25         $Y = Y'$
26      **end**
27      $P_{Z_{new}} = p(Z_{new}|d_t)$ using Equation (6–4)
28      $max\_prob = (P_Z)$
29      **if** $Y_d \neq \phi$ **then**     // instance is labeled
30         **foreach** $l \in (Y - Y_d)$ **do**
31           $z = k\text{-}NC(1, \mathcal{Z}^l_d)$
32           **foreach** $w \in \mathcal{V}_{d \cap z}$ **do**
33              $n^w_z = n^w_z \times (1 - (\Gamma/100))$
34           **end**
35         **end**
36         $Y = Y_d$
37      **end**
38      **foreach** $l \in Y$ **do**     // update model
39         **if** $max\_prob > P_{Z_{new}}$ **then**
40           $z = k\text{-}NC(1, \mathcal{Z}^l_d)$
41           add $d_t$ in $z$ using Definition 6.3.1
42           $u_z = t_{\mathcal{M}}$
43         **end**
44         **else**     // create new cluster
45           $z = (m_z, n_z, n^w_z, cw_z, ta_z, u_z = t_{\mathcal{M}}, \mathcal{L}_z = l)$
46           $\mathcal{M} = \mathcal{M} \cup z$
47         **end**
48      **end**
49   **end**

the label co-occurrence probability in initialization process at Line 14 of Algorithm 7 using Equation (6–2) which is the conditional probability between two labels. Finally, OSMTC predicts the label(s) with the combination of frequently occurred nearest pair of labels and their cluster count. The step-wise procedure is given in Algorithm 8. After predicting the labels, we then add the document into chosen micro-clusters. If we observe an abrupt concept drift in the label related feature space, then we create a new cluster(s) with this document (see Section 6.3.4.1). Otherwise we add document in existing cluster(s) (see Line 41 in Algorithm 8). The procedure of choosing (in Equation (6–5)) and creating new cluster (in Equation (6–4)) is described in Section 6.3.4.1. The main point to consider here is that a document can be added into more than one cluster and number of clusters is equal to the number of predicted labels (see Line 38 in Algorithm 8). Additionally, if the arriving instance is labeled then we add it to the closest cluster for each original label, whereas, for each wrong predicted label, we reduce the weight of common terms between closest cluster and document (see Line 33 in Algorithm 8). The purpose is to decrease the importance of term related to particular cluster label.

### 6.3.4 Model Maintenance

In this section, we describe the procedure of updating the model by adding an arrived document into cluster(s), removing the outdated terms (gradual concept drift) from the cluster, and remove outdated clusters (abrupt concept drift) from the model.

#### 6.3.4.1 Non-parametric Model Based Concept Evolution

In multi-label text streaming environment, concept evolution subsumes the change of label related terms. In other words the term subspace of model and each label changes over time. Whereas, concept drift is referred when the distribution of label related terms change over time [21]. To the best of our knowledge, concept evolution and concept drift are not studied specifically for multi-label text in streams [141][92]. Recently, to detect the concept evolution for text clustering, [42] and [107] exploit Dirichlet distribution based Chinese Restaurant Process (CRP). We adopt this approach for concept evolution in multi-label text streams.

As we already discussed that each active micro-cluster contains different term-space with different distribution. So, the basic idea is to check either new document belongs to existing nearest term-space of active micro-clusters ($k$ nearest micro-clusters) or belongs to a new term-space which leads for creating new micro-cluster in the model. Using Equation (6–3) we calculate the probability with existing micro-cluster. And the probability for creating new micro-cluster is defined as,

$$p\left(Z_d = Z_{new} | \vec{Z}_{\neg d}, \vec{d}, \alpha, \beta\right) = \left(\frac{\alpha D}{D - 1 + \alpha D}\right) \cdot \left(\frac{\prod_{w \in d} \prod_{j=1}^{N_d^w} \beta + j - 1}{\prod_{i=1}^{N_d} (\bar{V}_Z \beta) + i - 1}\right) \quad (6\text{–}4)$$

Here, we can clearly observe two parts, one for cluster popularity where $\alpha$ is a small pseudo value and $\beta$ is small pseudo occurrence of terms for new micro-cluster. Therefore, the condition of choosing a micro-cluster for an arrived document is

$$Z_d = \begin{cases} Z_i & \text{for } \exists p(Z_i \in \mathcal{M}) > p(Z_{new}) \\ Z_{new} & \text{otherwise} . \end{cases} \quad (6\text{–}5)$$

Along with creating micro-cluster for current concepts, the model also needs to consider the removal of outdated concepts. For this purpose, many approaches delete old batches of instance from model (such as MStreamF [42]). However, life-span of each document may vary in streaming environment, therefore, we adopt approach of deleting the outdated micro-clusters using a forgetting mechanism (e.g., decay rate) [9][29]. In decay mechanism, the importance weight ($u_z$) of a micro-cluster continuously reduces over time if the micro-cluster is not updated[①]. If a micro-cluster is updated over time which means it can capture the current concept. For simplicity, we adopt exponential decay function defined as,

$$r_z = r_z \times 2^{-\lambda \times (t_\mathcal{M} - u_z)}. \quad (6\text{–}6)$$

Here, $t_\mathcal{M}$ is current time of the model. Whenever, a document is added in the micro-cluster $z$ we set $r_z = 1$. If a micro-cluster is not receiving the documents over certain period of time then score of a micro-cluster approximately reaches to zero, which leads towards deletion of the micro-cluster. OSMTC maintains atleast $Z_{min}$ clusters for each label, therefore we check the active clusters before deleting (see Line 7 in Algorithm 10).

### 6.3.4.2 Removal of Outdated Label Related Terms

The previous studies [105] and [21] argue that in multi-label documents a label can be related with different terms. The label related core terms of each term-space may or may not be mutually exclusive, however, importance of these label related terms also

---

① Updating means instances are added in the micro-cluster.

change over time in text streams. To deal with this problem, we maintain the cluster feature $ta_z$ which store the arrival time sequence of each term in micro-cluster. Based on the triangular number time we calculate the cluster age and then compute the recency score of each term accordingly. If the *recency* is less than our defined threshold $\Gamma$ then we consider that this term cannot represent current concept of cluster. In other words if a term is frequently observed in cluster with the passage of cluster time-span then the term is useful to represent the current concept of terms. The triangular time decay is originally employed to fade out outdated micro-clusters. The function is defined as

$$\text{Triangular Number } \Delta f(T) = \left( \left( T^2 + T \right)/2 \right). \tag{6–7}$$

Here, $T$ is the timestamp number. The recency score of a term is measured by ratio of the sum of arrivals and age of cluster. For example, cluster $z$ contains 9 documents then the age of cluster $z$ will be:

$$Age_z = \Delta f(9) - \Delta f(1) = ((9^2 + 9)/2) - ((1^2 + 1)/2)$$

Suppose that a word $w_i$ occurred in three documents of cluster D1, D2 and D8 and arrival time of these documents are 1, 2 and 8, respectively. The arrival time of $w_i$ in cluster is stored as:

$$w_i = \{timestamp : 1, timestamp : 2, timestamp : 8\}$$

and the recency score of $w_i$ is calculated as

$$recency_{w_i} = ((1 + 2 + 8) \times 100)/Age_z.$$

It can be analyzed that more terms become less important when new documents are added in a micro-cluster. In this way, terms are filtered out gradually over time.

#### 6.3.4.3 Merging Micro-Clusters

The multinomial distribution parameter $\beta$ is responsible for calculating the homogeneity between a document and a micro-cluster. However, due to the exceeding noisy terms over time in a cluster, the probability of noisy terms becomes dominant over core terms. After concept drift of label related terms and gradual removal of noisy and outdated terms, two micro-clusters may share a high overlapping term-space which effects the cluster granularity. Previous approaches iterate the stream multiple times to infer the number of micro-clusters[42]. In contrast, we incorporate the clustering merging process

**Algorithm 9:** Update Micro-cluster Term-subspace

```
1  Function removeOldTerms(Γ, M)
2      S_z = Δf(1) using Equation (6–7)
3      foreach z ∈ M do
4          E_z = Δf(m_z) using Equation (6–7)
5          Age_z = E_z − S_z + 1                    // age of cluster
6          foreach (w, ta_w) ∈ ta_z do
7              sumta_w = ∑_{t∈ta_w} t
8              recency_w = (sumta_w × 100)/Age_z
9              if recency_w < Γ then
10                 remove(w, z)                     // remove word from cluster
11             end
12         end
13     end
```

using Equation (6–3) to calculate the probability between two clusters. With the passage of time, if a cluster is not updated then the value of $r_z$ reaches near to zero which indicate that a cluster is outdated however this particular cluster is created due to high noisy terms in the document. Therefore, before deleting the cluster we calculate the probability (using Equation (6–3)) with each active cluster of same label. In parallel we calculate the probability of this cluster for choosing a new cluster ($P_{z_{new}}$). If the new probability is greater than probability of active clusters then we simply delete the micro-cluster. Otherwise, we merge this micro-cluster with nearest micro-cluster containing same label. Algorithm 10 shows the step by step procedure to merge an outdated cluster with an active cluster.

### 6.3.5 Space and Time Complexity

The OSMTC always maintains an average $\bar{K}$ number of micro-clusters. Every cluster feature set stores an average $\bar{V}$ number of words in $n_z^w$, $2 \times |\bar{V}|$ timestamps of each word in $ta_z$, and at most $|\bar{V}_z| \times |\bar{V}_z|$ in $cw_z$. Thus the space complexity of OSGM is $O(\bar{K}(3\bar{V} + \bar{V}^2) + VD)$, where $V$ is the size of active vocabulary and $D$ is the number of active documents. On other side, OSMTC calculates the probability of arriving document with almost each active cluster (see Line 8 of Algorithm 8). Visiting and sorting cost is $O(\bar{K} + \bar{K}log(\bar{K}))$. Additionally, it also eliminates outdated terms by checking each micro-cluster. Therefore, the time complexity of OSMTC is $O(\bar{K}(\bar{d}\bar{V}) + \bar{K}log(\bar{K}))$, where $\bar{d}$ is the average size of arriving document.

**Algorithm 10:** Update Micro-cluster Weight

```
1  Function updateMCWeight(λ, Z_min, M)
2      Old = φ
3      foreach z ∈ M do
4          calculate r_z using Equation (6–6)
5          if r_z ≈ 0 and |Z_{Lz} ∈ M| > Z_min then
6              Old = Old ∪ z
7          end
8      end
9      foreach z_o ∈ Old do
10         foreach z_act ∈ (M − Old) do
11             if w_{z_act} ∩ w_{z_o} ≠ φ then
12                 P_{Z_i} = prob(z_act, z_o) using Equation (6–3)
13             end
14         end
15         i = arg max_i (P_{Z_i})
16         P_{Z_n} = calculate the probability of z_o choosing new cluster using Equation (6–4)
17         if P_{Z_i} < P_{Z_n} then
18             M = M − z_o
19         else
20             merge(z_o, Z_i) using Definition 6.3.1
21         end
22     end
23 End Function
```

## 6.4 Experiments

This section provides the selection of datasets and state-of-the-art algorithm for experimental setup, results analysis and parameter sensitivity of proposed model.

### 6.4.1 Datasets

To evaluate the performance of the proposed algorithm from different perspective, we conduct experiments on seven real and two synthetic datasets. The synthetic datasets are generated by TextGenerator library of MOA[20] tool and real datasets are obtained from publicly accessible repository①. The chosen real datasets were also used in [139], [53], [127], [54], [133], [132], [56] to evaluate the multi-label classification for streams. Table 6–2 shows the characteristics of all datasets. As we claim that our probability based similarity can deal with short as well as long text document, for this reason we mention the lowest dimension document ($W_{min}$) and highest dimension document ($W_{max}$) for each dataset. Additionally, most of previous algorithms have empirically evaluated their approach less than 5,000 dimensions datasets. However, most often in real time scenario, text data belongs to a very high dimensional space. Therefore, chosen 8 datasets having more than 20,000 dimensions to reflect actual constraint. Additionally, few datasets

---

① http://www.uco.es/kdis/mllresources/

suffers from high imbalance of instance distribution over labels, whereas, our prediction approach is based on initial created clusters for each label. For this reason, we consider labels having more than 30 instances in the stream.

Table 6–2 Dataset Statistics. dataset (Ds.), dimensionality ($W$), labels ($|\mathcal{L}|$), total Instances $|D|$, cardinality ($\bar{L}$), minimum document length($W_{min}$), maximum document length($W_{max}$).

| Ds. | $W$ | $|\mathcal{L}|$ | $|D|$ | $\bar{L}$ | $W_{min}$ | $W_{max}$ |
|---|---|---|---|---|---|---|
| Y-Art | 23,150 | 26 | 7,484 | 1.654 | 10 | 2773 |
| Y-Bus | 21,920 | 30 | 11,214 | 1.599 | 10 | 2376 |
| Y-Com | 34,100 | 33 | 12,444 | 1.507 | 10 | 7368 |
| Y-Edu | 27,530 | 33 | 12,030 | 1.463 | 10 | 5093 |
| Y-Ent | 32,000 | 21 | 12,730 | 1.414 | 10 | 4385 |
| Y-Soc | 52,350 | 39 | 12,111 | 1.280 | 10 | 9190 |
| 20NG | 1,006 | 20 | 19,279 | 1.029 | 1 | 497 |
| Syn-1 | 25,000 | 15 | 19,000 | 1.980 | 3 | 46 |
| Syn-2 | 45,000 | 20 | 30,000 | 2.420 | 4 | 41 |

## 6.4.2 Baselines and Evaluation Metrics

For comparative analysis of OSMTC, we choose nine supervised state of the art algorithms designed for streaming task which includes AMR[135], AHOT[142], OZBAML[20], SCD[143], OCB[144], ISOUPT[54], kNNPA[143], and SAMkNN[145]. We also selected two semi-supervised algorithms designed for static corpus which includes MASS[146] and TRANS[147]. The full names and main parameters of chosen algorithms are adopted according to their respective paper which can be seen in Table 6–3. All except few leading algorithms are specifically designed for multi-label streaming data. Some leading algorithms for single-label data have been used and adapted for multi-label data streams for more comprehensive analysis. All supervised algorithms were implemented in MOA[143] framework whereas source code of semi-supervised algorithms are publicly available. Our model is implemented for MOA and the code is publicly available[①].

We run experimentation on Intel Xeon CPU E5-2678v3 with 180 GB of memory on Linux system. For setting parameters, we performed population selection based genetic algorithm[148] to choose $\alpha$ and $\beta$ (ranging $[0.0001, 0.09)$) for each dataset, as their effect is more higher than other parameters, until convergence of hamming loss as objective function. We set $\Gamma = 5$, $k = 15$, $\lambda = 1e^{-5}$, $Z_{min} = 3$, and $D_{init} = 600$ by performing grid search on *20NG* dataset and use same values for all datasets.

Our comparative study between algorithms is based on choosing the well known and highly adopted evaluation measures which include Hamming loss, example-based

---

① https://github.com/JayKumarr/OSMTC

accuracy and micro average recall on stream setting for more strict performance evaluation. As hamming loss computes the instance wise label difference thus lesser the score better the algorithm is. The recall assures the purity of prediction label set. The reason of adopting recall is that many previous algorithms predict fixed number of labels either for whole dataset or for a single batch. However, this strategy decreases the quality of predicted label-set. Example-based accuracy is the average of overall precision and recall of the model.

Table 6–3 Algorithms and respective major parameter setting.

| Acronym | Parameters |
|---|---|
| AMR[135] | learner: MLNaiveBayes, changeDetector: DDM |
| AHOT[142] | splitCriterion: InfoGain, tieThreshold: 0.05 |
| OZBAML[20] | learner: HoeffdingTree, ensembleSize: 10 |
| SCD[143] | learner: HoeffdingTree, changeDetector: DDM |
| OCB[144] | learner: HoeffdingTree, ensembleSize: 10 |
| ISOUPT[54] | learningRate: 0.02, tieThreshold: 0.05 |
| kNNPA[143] | $k$=10, window: 1,000 |
| SAMkNN[145] | $k$: 5, $ratio_{LTM}$: 0.4, $m_{min}$: 50, $m_{max}$: 1,000 |
| MLSAMPkNN[123] | $k$: 3, $ratio_{penalty}$: 1, $m_{min}$: 50, $m_{max}$: 1,000 |
| MASS[146] | $k$: 3 |
| TRANS[147] | $\alpha = 0.01, \beta = 20, \tau = 1e^{-5}, \gamma = 50$ |

### 6.4.3 Results Analysis

The empirical score of the selected evaluation measures on all datasets can been seen in Table 6–4. For all algorithms, we calculate the sum of difference from mean of each evaluation measure to provide a clear understanding about conducted analysis. For hamming loss, lesser the score better the algorithm is. In other words, it calculates the distance between predicted and ground truth label set and the target is to minimize the distance. In contrast, greater the accuracy and recall better the model is. Additionally, to compare with static semi-supervised algorithms (i.e., MASS and TRANS), we analyze 10% and 20% labeled instances. In this context, we can analyze from Table 6–4 that in spite of very limited labeled data OSMTC acquired the highest score on most of datasets compared to all algorithms. The most significant different is observed in recall score, and the reason is certainly due to dynamic number of labels per instance which directly leads towards higher purity of predicted label-set whereas most of previous algorithm predict static number of labels for each instance according to dataset cardinality. Additionally, as OSMTC does not work well on *20NG* and the very reason lie behind the construction of the datasets, as global vector space of each dataset were transformed into low dimensional vectors thus losing the original contextual dependency of terms. To analyze the overall statistical significance, we use Nemenyi test on Table 6–4 with 95% confi-dence (the obtained p-value should be lesser than standard threshold which is 0.05) to plot critical difference diagram and the acquired p-value is 0.038. The Figure 6–2 clearly shows the our algorithm highly resembles with TRANS which is a designed for static data. Whereas, compared to streaming algorithms, OSMTC is statistical significant on just 20% labeled instances.

Table 6–4 Hamming loss, Example-based Accuracy, and Micro-average Recall comparison of all algorithms. Here bold represents the highest score. † shows the runtime failure due to memory capacity (180GB) exhausted. @*P* represents the percentage of labeled instances for particular algorithm. The *SDM* is sum of difference from mean of all values for each dataset.

| | Y-Art | Y-Bus | Y-Com | Y-Edu | Y-Ent | Y-Soc | 20NG | Syn-1 | Syn-2 | *SDM* |
|---|---|---|---|---|---|---|---|---|---|---|
| | | | | Hamming Loss | | | | | | |
| AMR | 0.063 | 0.053 | 0.045 | 0.044 | 0.067 | 0.032 | 0.046 | 0.132 | 0.125 | -0.009 |
| AHOT | 0.063 | 0.028 | † | 0.043 | 0.065 | † | 0.044 | 0.133 | † | -0.039 |
| OZBAML | 0.078 | 0.053 | 0.044 | 0.060 | 0.102 | 0.050 | 0.093 | 0.176 | 0.145 | 0.184 |
| SCD | 0.063 | 0.028 | 0.043 | 0.043 | 0.065 | **0.031** | 0.044 | 0.133 | 0.130 | -0.036 |
| OCB | 0.064 | 0.029 | 0.044 | 0.044 | 0.067 | † | **0.043** | 0.135 | † | -0.033 |
| ISOUPT | 0.068 | 0.059 | 0.050 | 0.052 | 0.074 | 0.039 | 0.135 | 0.135 | 0.126 | 0.121 |
| kNNPA | 0.064 | 0.029 | 0.043 | 0.046 | 0.068 | 0.037 | 0.051 | 0.133 | 0.121 | -0.024 |
| SAMkNN | 0.069 | 0.028 | † | 0.045 | † | † | 0.051 | 0.134 | † | -0.015 |
| MLSAMPkNN | 0.072 | 0.030 | † | † | † | † | 0.035 | 0.166 | † | 0.005 |
| MASS@10 | 0.066 | 0.033 | 0.044 | 0.046 | 0.100 | 0.033 | 0.065 | 0.142 | 0.125 | 0.037 |
| MASS@20 | 0.063 | 0.032 | 0.041 | 0.044 | 0.080 | 0.032 | 0.062 | 0.139 | 0.120 | -0.003 |
| TRANS@10 | 0.063 | 0.029 | 0.045 | 0.041 | 0.083 | 0.032 | 0.063 | 0.136 | 0.119 | -0.005 |
| TRANS@20 | 0.062 | 0.028 | 0.044 | **0.039** | 0.061 | **0.031** | 0.058 | 0.134 | 0.113 | -0.046 |
| OSMTC@10 | 0.063 | 0.029 | 0.044 | 0.042 | 0.060 | 0.032 | 0.051 | 0.129 | 0.118 | -0.048 |
| OSMTC@20 | **0.061** | **0.028** | **0.040** | 0.040 | **0.059** | 0.031 | 0.049 | **0.112** | **0.105** | **-0.092** |
| | | | | Example-based Accuracy | | | | | | |
| AMR | 0.203 | 0.010 | 0.001 | 0.004 | 0.003 | 0.001 | 0.251 | 0.001 | 0.001 | -1.853 |
| AHOT | 0.184 | 0.680 | † | 0.275 | 0.151 | † | 0.301 | 0.005 | † | 0.032 |
| OZBAML | 0.159 | 0.540 | 0.394 | 0.142 | 0.069 | 0.130 | 0.051 | 0.013 | 0.010 | -0.820 |
| SCD | 0.138 | 0.068 | 0.159 | 0.366 | 0.136 | 0.147 | 0.270 | 0.009 | 0.009 | -1.026 |
| OCB | 0.135 | 0.671 | 0.343 | **0.544** | 0.088 | † | 0.332 | 0.012 | † | 0.149 |
| ISOUPT | 0.163 | 0.053 | 0.151 | 0.447 | 0.067 | 0.033 | 0.052 | 0.013 | 0.123 | -1.226 |
| kNNPA | 0.143 | **0.677** | 0.560 | 0.057 | 0.508 | 0.414 | 0.036 | 0.003 | 0.002 | 0.071 |
| SAMkNN | 0.141 | 0.677 | † | 0.421 | † | † | 0.036 | 0.002 | † | -0.022 |
| MLSAMPkNN | 0.192 | 0.661 | † | † | † | † | **0.492** | 0.010 | † | 0.445 |
| MASS@10 | 0.180 | 0.651 | 0.562 | 0.517 | 0.391 | 0.332 | 0.044 | 0.101 | 0.103 | 0.552 |
| MASS@20 | 0.189 | 0.664 | 0.570 | 0.535 | 0.399 | 0.358 | 0.050 | 0.120 | 0.121 | 0.677 |
| TRANS@10 | 0.190 | 0.627 | 0.535 | 0.530 | 0.397 | 0.374 | 0.047 | 0.126 | 0.127 | 0.624 |
| TRANS@20 | 0.201 | 0.641 | 0.551 | 0.541 | 0.408 | 0.391 | 0.097 | 0.138 | 0.135 | 0.774 |
| OSMTC@10 | 0.209 | 0.662 | 0.550 | 0.529 | 0.407 | 0.398 | 0.050 | 0.134 | 0.121 | 0.731 |
| OSMTC@20 | **0.218** | 0.675 | **0.571** | 0.544 | **0.411** | **0.418** | 0.100 | **0.149** | **0.135** | **0.892** |
| | | | | Micro-Average Recall | | | | | | |
| AMR | 0.003 | 0.001 | 0.001 | 0.079 | 0.001 | 0.001 | 0.315 | 0.005 | 0.001 | -5.218 |
| AHOT | 0.040 | 0.568 | † | 0.080 | 0.145 | † | 0.327 | 0.050 | † | -2.299 |
| OZBAML | 0.906 | 0.988 | 0.863 | 0.931 | 0.990 | 0.985 | 0.953 | 0.930 | 0.889 | 2.809 |
| SCD | 0.031 | 0.564 | 0.348 | 0.069 | 0.122 | 0.136 | 0.280 | 0.009 | 0.005 | -4.061 |
| OCB | 0.026 | 0.550 | 0.298 | 0.060 | 0.076 | † | 0.355 | 0.013 | † | -2.823 |
| ISOUPT | 0.997 | 0.997 | 0.996 | 0.997 | 0.997 | 0.995 | 0.994 | 0.995 | 0.996 | 3.338 |
| kNNPA | 0.033 | 0.554 | 0.443 | 0.052 | 0.099 | 0.044 | 0.066 | 0.031 | 0.003 | -4.300 |
| SAMkNN | 0.010 | 0.548 | † | 0.452 | † | † | 0.036 | 0.002 | † | -1.828 |
| MLSAMPkNN | 0.170 | 0.569 | † | † | † | † | 0.491 | 0.110 | † | -0.933 |
| MASS@10 | 0.642 | 0.644 | 0.769 | 0.901 | 0.941 | 0.978 | 0.846 | 0.911 | 0.882 | 1.888 |
| MASS@20 | 0.737 | 0.651 | 0.821 | 0.930 | 0.957 | 0.984 | 0.867 | 0.949 | 0.927 | 2.197 |
| TRANS@10 | 0.651 | 0.609 | 0.855 | 0.925 | 0.949 | 0.981 | 0.894 | 0.943 | 0.917 | 2.098 |
| TRANS@20 | 0.743 | 0.621 | 0.913 | 0.957 | 0.963 | 0.985 | 0.928 | 0.972 | 0.954 | 2.410 |
| OSMTC@10 | 0.998 | 1.000 | 1.000 | 0.998 | 0.997 | 1.000 | 0.988 | 0.998 | 0.997 | 3.350 |
| OSMTC@20 | **0.999** | **1.000** | **1.000** | **1.000** | **1.000** | **1.000** | **1.000** | **0.999** | **0.999** | **3.371** |

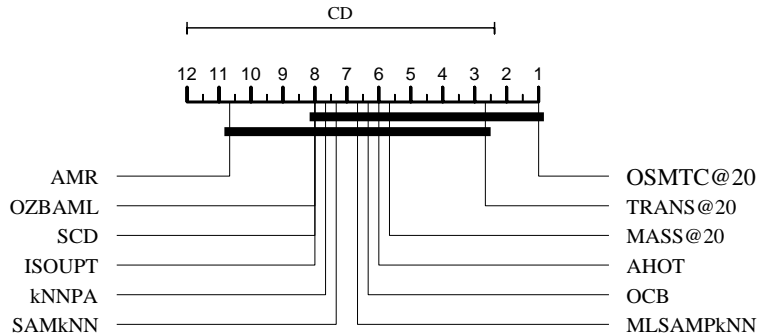

Figure 6–2 Critical difference diagram using Nemenyi test with a 95% confidence for all algorithms.

## 6.4.4 Running Time and Memory Analysis

The main challenges while designing a model for streaming task are limited processing time and memory. Less running time for processing instances empowers the model to work in low as well as high velocity data streams. Likewise, maintaining the active concepts in such environments bring the constraint of limited memory for the model. For this reason, we analyze the execution time and memory consumption of all streaming algorithms in Figure 6–3 and 6–4, respectively. MASS[146] and TRANS[147] are designed for static corpus thereby they iteratively process the whole data thus we do not perform the running time and memory comparative analysis. By observing the Figure 6–3, we can conclude that OSMTC consume less processing time compared to most of the previous algorithms. The next observation from given plots is that the proposed model comparatively takes less time if the average length of each document is less. And the reason is OSMTC has to maintain co-occurrence matrix to capture the semantics of each cluster therefore larger the document size is greater the time consumption to process the co-occurrence matrix.

In the context of memory utilization, Table 6–4 shows the failure of few recent algorithms to process the high dimensional datasets. In contrast, OSMTC consumed very less memory compared to other state-of-the-art algorithms over stream, as we show memory consumption in Figure 6–4 on *Y-Art* dataset. The reason for much less memory usage is flexibility of proposed model to store the current concept (active number of terms and clusters) by reducing cluster related terms and removing outdated clusters.

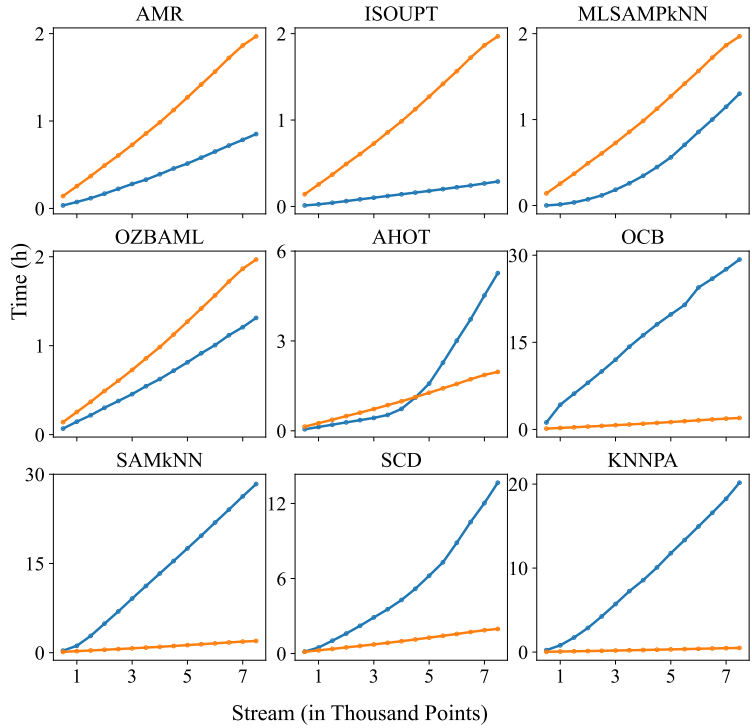

Figure 6–3 Run time analysis on Y-Art dataset. The orange line shows the execution time of OSMTC over stream whereas the blue line shows the corresponding algorithm entitled on each plot.

### 6.4.5 Term-space Analysis

As we discussed that previous approaches require pre-defined number of dimension before processing the documents in stream. Whereas, OSMTC does not hold such constraint and automatically updated the active term space over stream. To analyze the active terms of model, we show the number active terms for three datasets (see Figure 6–5) over the stream to signify the term reduction and maintaining concurrent feature-space strategy. By looking Figure 6–5 we can clearly observe that the model creates initial term space from $D_{init}$ documents and then it reduces the active terms by removing noisy terms from each cluster. The pattern of real datasets (i.e., Y-Art and Y-Bus) is comparatively homogeneous whereas synthetic dataset has low variation over stream. The reason is we intentionally inject such behavior to evaluate our algorithm while generating the dataset.

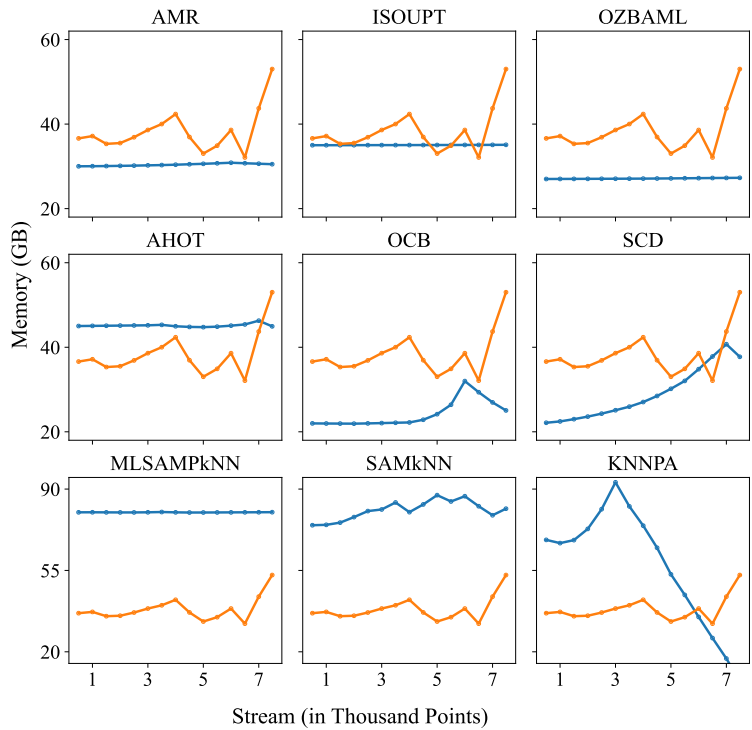

Figure 6–4 Memory analysis on Y-Art dataset. The orange line shows the execution time of OSMTC over stream whereas the blue line shows the corresponding algorithm entitled on each plot.

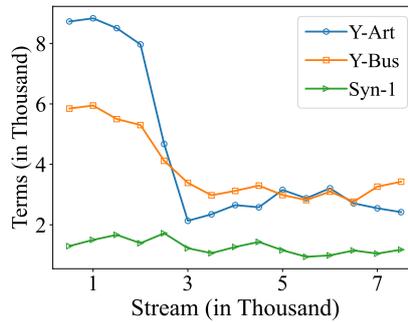

Figure 6–5 Active vocabulary size of the model on three datasets.

## 6.4.6 Parameter Sensitivity

In order to analyze the effect of different parameters and their relationship on prediction performance, we perform several experiments on *Syn-1* dataset.

The first step of model initialization is dependent on choosing initial number of documents ($D_{init}$) and clusters ($Z_{min}$). The main concept while setting the value for $D_{init}$ is choosing sufficient number of instances to represent basic label related term space, and if label related term space is consist of multiple subspaces then greater $Z_{min}$ value is needed to capture the subtle concepts. Figure 6–6 shows hamming loss of $D_{init}$ ranging from 400 to 1200 over four different values of $Z_{min}$. It can observed that for a large range of both values the model gradually reduces the hamming loss over stream.

For calculating the cluster probability, effect of $\alpha$ and $\beta$ needs to be considered as both parameters are responsible while predicting labels and creating new clusters. Figure 6–7 and 6–8 show the effect of both parameters ranging $9e^{-4}$ to $9e^{-2}$ on hamming loss and model clusters. From the figures, we can clearly analyze that greater value lead towards generating more number of clusters while maintaining the same hamming loss. The reason is increasing the value of any parameter lead towards high probability of creating new cluster. In other words, these two small values work as pseudo score for each term for creating new cluster, thus high value of $\alpha$ increases prior probability and high values of $\beta$ increase homogeneity of new cluster.

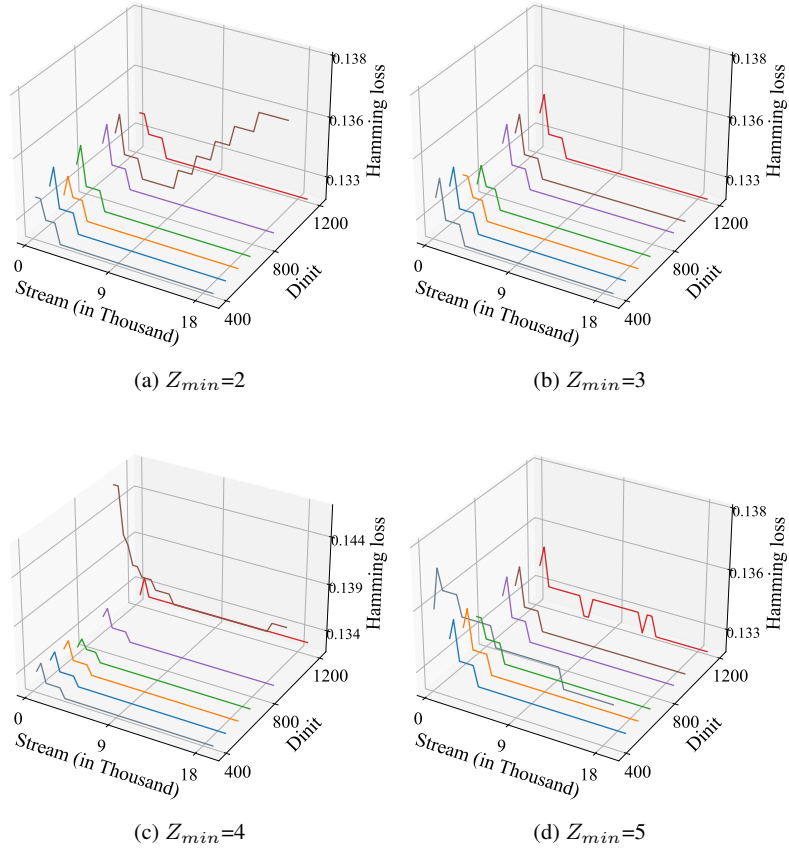

(a) $Z_{min}=2$  (b) $Z_{min}=3$

(c) $Z_{min}=4$  (d) $Z_{min}=5$

Figure 6–6 The effect of $D_{init}$ with different $Z_{min}$ on Syn-1 dataset.

For predicting the labels of an instance we observe the probability distribution and count number clusters for each label from $k$ nearest clusters, whereas lower bound of cluster population is controlled by $Z_{min}$. Thus, low value of $k$ with high value of $Z_{min}$ confines the model to predict more than one label for each instance. Figure 6–9 shows the effect of $k$ ranging from 3 to 15 on hamming loss with four different values of $Z_{min}$. From the given figure, we can analyze that variation in hamming loss is stabilized for any value of $k$ if $Z_{min} > 2$. The next parameter $\Gamma$ of our model is responsible for removing outdated terms with respect to cluster age. Greater the $\Gamma$ lesser the age of terms. As we already discussed that a global space related to a particular label may contain subtle subspaces and each subspace is captured through initial population of cluster using $Z_{min}$. Figure 6–10 shows the variation of hamming loss over $\Gamma$ ranging 3 to 15. From the given figures, we can analyzed that our approach is sensitive to $\Gamma$ while choosing a very less number of cluster population. For example, we can see high variation in Figure 6–10a compared to Figure 6–10d.

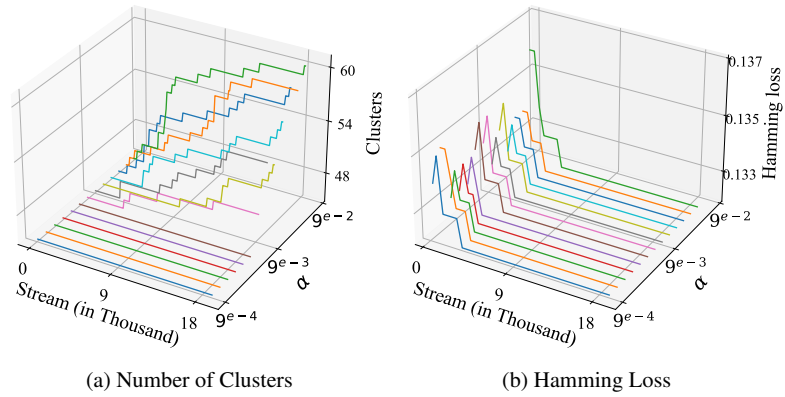

(a) Number of Clusters

(b) Hamming Loss

Figure 6–7 Model clusters and hamming loss over a range of $\alpha$ values on Syn-1 dataset.

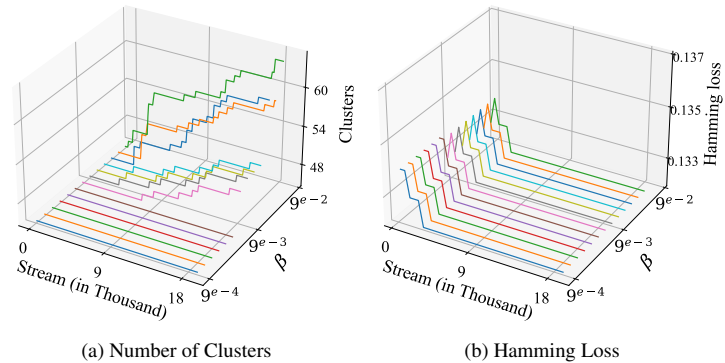

(a) Number of Clusters

(b) Hamming Loss

Figure 6–8 Model clusters and hamming loss over a range of $\beta$ values on Syn-1 dataset.

## 6.5 Summary

In this chapter, we have proposed OSMTC, a semi-supervised model to learn evolving multi-label text streams. To predict the label-set we design a non-parametric Dirichlet model which calculates the semantic and vector-based probability between clusters and document. The distribution of resulting probabilities then used to predict dynamic number of labels for each instance. The online processing of our model enables its applicability for both low and high velocity streaming environment. In addition, it can easily adopt the evolution of gradual change of label related term-space using term arrival time with respect to cluster time-span. As well as, it can adopt abrupt change by creating new clus-ters using Dirichlet process. We conducted an extensive empirical study on real as well as synthetic datasets. From acquired results we found our proposed semi-supervised algorithm to be highly effective for high dimensional text streams while consuming comparatively less execution time and memory with state-of-the-art fully supervised algorithms.

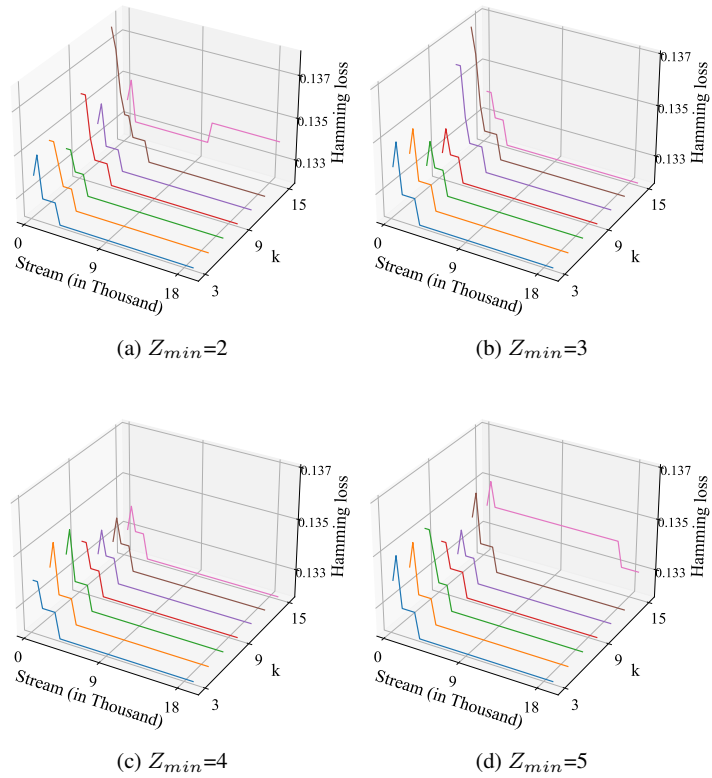

Figure 6–9 The effect of parameter *k* for different $Z_{min}$ on Syn-1 dataset.

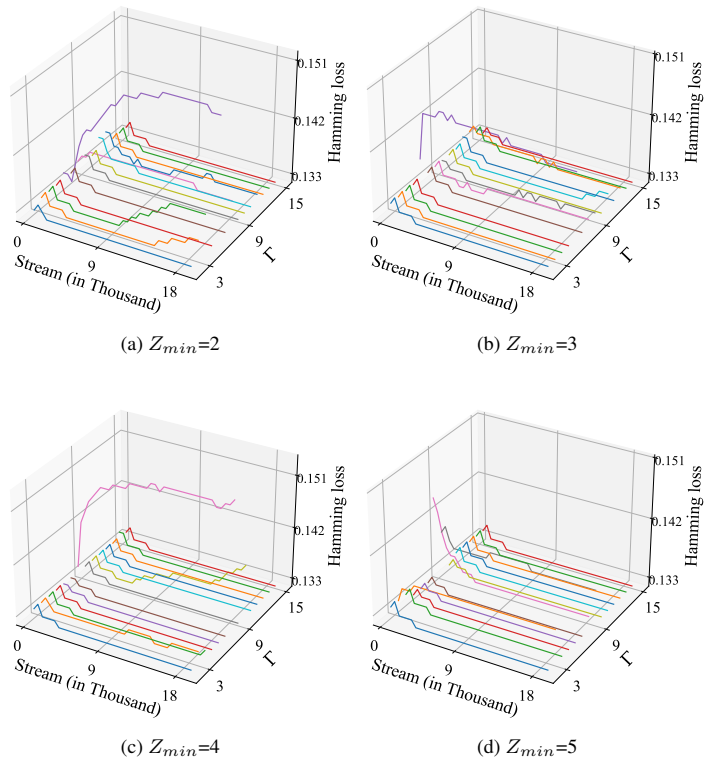

Figure 6–10 The effect of $\Gamma$ with different $Z_{min}$ on Syn-1 dataset.

# Chapter 7  Conclusion and Future Work

This chapter summarizes the scope of this research and provides an overview of each contribution in this thesis. Additionally, some major future directions to investigate the text mining problem are discussed.

## 7.1 Summary of Thesis

A text stream is an ordered sequence of text document arriving over time. Mining such text stream in an optimized way has become challenging task as it brings a variety of challenges which includes high dimensional text representation, sparsity, concept evolution, and label scarcity while consuming limited memory and very less processing time. The problem of high dimension occurs when representing text in semantic and contextual term space. The concept evolution enforce to maintain and track the active topics of the documents. To deal with both problem, a series of previous models work over subspace of the global term-space to detect small concepts (fine grain concept mining) inside stream. However, for many real world scenario, fine grain concept mining creates sparsity issue. Apart from these problems, availing fully labeled data to feed the learning model is nearly impossible as the velocity and volume of data is very high in stream thus creating label scarcity challenge. This dissertation proposes solutions of above mentioned problems with different assumptions.

In the first work, a new online semantic-enhanced dirichlet model is proposed for short text stream clustering. In contrast to existing approaches, proposed model dynamically maintains the active number clusters. It dynamically assigns each arriving document into an existing cluster or generating a new cluster based on the poly urn scheme. More importantly, it incorporates semantic information in the proposed graphical representation model to remove the term ambiguity problem in short-text clustering. Building upon the semantic embedding and online learning, proposed method allows finding high-quality evolving clusters.

In second work, an enhanced semantic enhanced graphical model is proposed for evolving short text stream clustering. Compared to existing approaches, it approximates the dynamic number evolving clusters, and can reduce cluster features to extract core terms automatically. It also dynamically assigns each arriving document into an existing cluster or generating a new cluster based on the poly urn scheme. More importantly, it incorporates semantic smoothing for coarse-grain clustering in the proposed graphical representation model. By exploiting the triangular time function, the proposed approach can track the change of term distribution over time.

In third work, a new dirichlet model is introduced on short text stream clustering. Compared with previous approaches, it incorporates a new word specificity score to increase the clustering quality. A reduced semantic window based co-occurrence matrix is introduced to capture the semantic similarity. Additionally, to infer the number of clusters of model, an episodic inference is proposed which resample random documents from buffer.

In the forth work, a semi-supervised model is proposed to learn evolving multi-label text streams. The advantages of proposed approach transcend over the limitations of previous approaches, which are: (i) it works with instance by instance processing instead of batch-wise processing; (ii) unlike fully supervised model, it works for limited number of labeled samples; (iii) it considers first-order label dependency for prediction; (iv) it is robust to long and short documents, and (v) it can capture semantic similarity for label-specific terms. To deal with text representation challenge, the proposed model can (i) deal with evolving number of dimensions, and (ii) predict well with a few set of labeled data in text streams.

## 7.2 Future Work

Although a variety of algorithms are proposed for learning from text data streams. However, still there are various scenarios, drawbacks of existing algorithms and related challenges in real-world applications which need to be focused on for algorithm design. This section gives some main future directions to continue the research in text stream mining.

(1) **Dirichlet Parameter Estimation:** In the last decade, many topic modeling algorithms are proposed, which then transformed to work in the streaming environment. However, automation of parameter value selection is considered in parallel with enhancing the topic quality. The performance of many algorithms rely on fine tunned parameter values as document-length plays an important role to calculate the probability. Therefore, it is necessary to reduce the parameters or automate the parameter values selection.

(2) **Label Correlation Evolution:** The relationship of a label with another label in multi-label dataset is referred as label correlation. In the multi-label text stream, a little work have been studied by the researchers to deal with different type of concept drift. Including the evolving label cardinality, label correlation may also change over time. Therefore, both label related concept drift need to studied for multi-label classification problem.

(3) **Novel Class Detection:** The concept evolution in stream may also refer to emergence of new class or new label. The research work in this thesis is mainly focused to deal with evolving number of features. Whereas, in many real world applications, new classes or labels may come with the passage of time. Therefore, novel class detection for multi-class and multi-label learning need to be considered for designing the effective algorithm.

# **Acknowledgements**

I would like to thank many people who contributed during my Ph. D. study. First, I would like to thank my supervisor, Prof. Dr. Junming Shao, for taking me as a PhD student when I had very little research experience. His guidance, advice, and good ideas has provided throughout my time as his student as well as I thank for all lunch and dinner parties. With his experience, knowledge, and ability to explain everything clearly, I learned to produce such a quality research work.

I express my humble gratitude to my mother, father and brother for their every possible support and encouragement. I would also like to thank all teachers of UESTC who taught different courses to me and shared their knowledge. I would like to extend my thankfulness for everybody in Data Mining Lab, especially Prof. Dr. Yang Qingli, Dr. Salah-Ud-Din, Mawuli Benard, Zhong Zhang, and Qin Zhili for their discussions and technical support. Additionally, I would like to thank all my friends, whose social and motivational support was also very necessary, including Nisha Kunwar, Junaid Bhatti, Wazeer Ali, NoorBakhsh, Zakria, Abdullah Khan, Naveed Ali, Punal Sehto, Asif Raza, Wang Yanqun, Afzal, Jenette, Abbas, Mutahir, Nasir Ilyas, Pirah Memon, Safeer Ali and other friends which I have not mentioned here. I want to thank our coordinator Miss Liu Yi. I am also grateful to reviewers for their valuable comments, helpful advice, and constructive suggestions for my research. Finally, I would like to thank the UESTC, for awarding me scholarship for all my study period. A overwhelmed gratitude to the existence.

# Research Results Obtained During the Study for Doctoral Degree